\documentclass[preprint,12pt,3p]{elsarticle}

\usepackage{isabelle,isabellesym}

\usepackage{pdfsetup}

\usepackage{graphicx}
\usepackage{amsfonts,amssymb,amsmath,alltt}
\usepackage{array}
\usepackage{mdwmath}
\usepackage{mdwtab}
\usepackage{fixltx2e}
\usepackage{stfloats}
\usepackage{url}
\usepackage{hyperref}
\hypersetup{breaklinks=true}
\usepackage{flushend}
\usepackage{xspace}
\usepackage{color}
\usepackage{stmaryrd}
\usepackage{tikz} 
\usepackage[textsize=tiny]{todonotes} 
\newcommand{\mycomment}[1]{}

\definecolor{darkgreen}{rgb}{0.0, 0.5, 0.0}
\def\mcolor#1#2{\rule{0ex}{0ex}\color{#1}#2\color{black}{}}
\newcommand{\new}[1]{\mcolor{blue}{}\mcolor{black}{#1}}

\newenvironment{ttbox}{\begin{alltt}\ttbraces\small\tt}%
                      {\end{alltt}}
\def\ttbraces{\let\.=\nobreak\chardef\{=`\{\chardef\}=`\}\chardef\|=`\\}

\newcommand{\TODO}[1]{}
 
\newcommand\ie{i.e.\!\,, }
\newcommand\ttand{\mbox{{$\land$}}}

\newcommand\ttfun{\mbox{{$\Rightarrow$}}}
\newcommand\ttcap{\mbox{{$\cap$}}}

\newcommand\ttimp{\mbox{{$\longrightarrow$}}}
\newcommand\ttequiv{\mbox{{$\equiv$}}}
\newcommand\ttexists{\mbox{{$\exists$}}}
\newcommand\ttforall{\mbox{{$\forall$}}}
\newcommand\ttneg{\mbox{{$\neg$}}}
\newcommand\ttneq{\mbox{{$\neq$}}}
\newcommand\ttin{\mbox{{$\in$}}}
\newcommand\ttnin{\mbox{{$\notin$}}}
\newcommand\ttImp{\mbox{{$\Longrightarrow$}}}
\newcommand\ttlam{\mbox{\( \lambda \)}}
\newcommand\tttimes{\mbox{\( \times \)}}
\newcommand\ttlbrack{\mbox{\(\llbracket\)}}
\newcommand\ttrbrack{\mbox{\( \rrbracket \)}}

\newcommand\ttatI{\mbox{\( @_G \)}}
\newcommand\ttat[1]{\mbox{\( @_{#1} \)}}
\newcommand\ttleq{\mbox{{$\le$}}}
\newcommand\ttrelI{\mbox{{$\to_{i}$}}}
\newcommand\ttrelN{\mbox{{$\to_{n}$}}}

\newcommand\ttsubseteq{\mbox{{$\subseteq$}}}
\newcommand\ttf{\mbox{{$f$}}}
\newcommand\ttvdash{\mbox{{$\vdash$}}}

\begin{document}

\begin{frontmatter}

\title{Applying the Isabelle Insider Framework to Airplane Security
}

\author{Florian Kamm\"uller}  
\ead{f.kammueller@mdx.ac.uk}
\address{Middlesex University London and Technische Universit\"at Berlin} 
\author{Manfred Kerber}  
\ead{M.Kerber@cs.bham.uk}
\address{University of Birmingham, UK}


\begin{abstract}
\new{Avionics is one of the fields in which verification methods have been pioneered and brought a new level of reliability to systems used in safety critical environments.}
  Tragedies, like the 2015 insider attack on a German airplane, in
  which all 150 people on board died, show that safety and security
  \new{crucially depend not only on the well functioning of systems but also on the way how humans interact with the systems. Policies are a way to describe how humans should behave in their interactions with technical systems, formal reasoning about such policies requires integrating the human factor} into the
  verification process.  In this paper, we report on our work on using
  logical modelling and analysis of infrastructure models and policies
  with actors to scrutinize security policies in the presence of
  insiders.
  
We model insider attacks on airplanes in the Isabelle Insider framework. 
This application motivates the use of an extension of the framework with 
Kripke structures and the temporal logic CTL to enable reasoning on dynamic 
system states. Furthermore, we illustrate that Isabelle modelling and 
invariant reasoning reveal subtle security assumptions. We summarize by 
providing a methodology \new{for the development of policies that satisfy stated properties.}
\end{abstract}
\begin{keyword} 
Airplane safety and security, Insider threats, Interactive theorem proving, Security policies, Verification
\end{keyword}
\end{frontmatter}

\section{Introduction}
Airplanes offer a very safe way of travelling. Accidents and terror
attacks are extremely rare. After the 2001-09-11 attacks stringent
measures were taken and have been to the day of writing
successful. The most recent major incident was an insider attack in
which the copilot of Germanwings Flight 9525 on 2015-03-24 hijacked
the aircraft by locking out the captain, who had left the cabin, and
subsequently brought the aircraft to a crash in which all 150 persons
on board died. As a consequence, airlines introduced a two-person rule
that a pilot must never be on their own in the cockpit. The two-person
rule has been rescinded in 2017 only two years after it was
introduced. The 2015-03-24 incident shows that insider attacks are an
important issue and it motivated earlier work \cite{kk:16} of applying
the existing Isabelle Insider framework \cite{kp:16} to verify
airplane policies in the presence of insider attacks.

This earlier work has revealed some major challenges for the Isabelle
Insider framework:
\begin{itemize}
\item Since the policies are dealing with actors and their possibilities of moving
     within the infrastructure, for example an airplane, a fixed association of actors with 
     locations, roles, and credentials in the model must be extended to enable
     representing dynamic change. 
\item We need to integrate dedicated logics into the framework enabling the expression of 
      security and safety guarantees over the dynamically changing infrastructure state. 
      We need to express global validity of logical properties of policies 
      over all reachable states; for example, we want to express ``for all states
      reachable from an acceptable initial state, a suicidal copilot cannot
      crash the plane''. 
\end{itemize}

In the current paper, we provide solutions to
these challenges and demonstrate them on the airplane case study.
The main contribution of this paper are:
\begin{itemize}
\item State transitions as well as rules for 
     expressing changes to the state of infrastructures including locations, actors, 
     their roles, credentials and behaviours are provided by Kripke structures. This
     allows modelling state change and state transition.
\item Temporal logic CTL is provided within the framework to formalize and prove logical properties. 
      This enables (a) detecting attack paths through 
      the graph of infrastructure state evolution and (b) from there identifying additional 
      security assumptions that when met guarantee that the attack is not possible any more 
      on any path.
\end{itemize}
Another contributions of this work is to identify an improved
methodology for policy invalidation and model refinement.


After discussing related work in Section \ref{sec:related}, 
we present in Section \ref{sec:air} a retrospective of the development of safety and security
regulations for airplanes.
We then present the existing Isabelle Insider framework in Section \ref{sec:isains}.
Next, we use this framework to model an airplane scenario
including an Insider attacker. We integrate Kripke structures into the model and 
express and interactively prove central security properties 
using the branching time temporal logic CTL (Section \ref{sec:model}).
Section \ref{sec:ana} presents the analysis of those properties on the airplane 
scenario showing how the framework can be used to scrutinize the security 
policies and thereby reveal existing loopholes within their formal specifications.
This procedure is summarized into our methodology before Section \ref{sec:concl}
concludes.

\new{The full Isabelle sources are available online \cite{kam:19isa}. In
order to give an impression of the kind of formalization the most
important definitions and theorems can be found in the Appendix.}

\section{Related Work}\label{sec:related}

In this section, we present some related work from the field of
insider threats and work in which reasoning approaches similar to the
one applied in our work are applied. \new{Furthermore we discuss work
related to the verification in avionics.}

The insider threat patterns provided by CERT~\cite{cmt:12} use the System 
Dynamics model, which can express dependencies between variables. The
System Dynamics approach is also successfully being applied in other
approaches to insider threats, for example, in the modelling of unintentional 
insider threats \cite{gscmmc:14}.
Axelrad et al.~\cite{10.1109/SPW.2013.35} have used Bayesian
networks for modelling insider threats in particular the human disposition.
In comparison, the model we rely on for modelling the human disposition in 
the Isabelle Insider framework is a simplified classification 
following the taxonomy provided in \cite{nblgcww:14}. In contrast to all these approaches, our work provides an additional model of infrastructures
and policies allowing reasoning at the individual and organizational level.

\new{A major field of application of formal methods is avionics.  Companies
(such as Airbus and Boeing) and organizations (such as NASA) use
formal methods to prove formal properties of aircrafts and
spacecrafts.  There is a large body of work, including work based on
model checking and theorem proving, which we cannot give justice in
this paper. We will mention only a few. \cite{Mo-et-al-2013} is mainly
concerned with the relationship between software testing and formal
verification, and Moy et al.\ argue that in many application areas formal
verification outperforms testing, firstly in that the proofs show the
correctness on all inputs and not just the ones tested, but secondly
also in the person power required. \cite{OHalloran-2013} shows how a
Z-based toolset is used to prove the correctness of embedded real time
safety critical software for Eurofighter Typhoon. Khan et
al. \cite{Khan-et-al-2012} argue that complexity of avionics has
increased to a level that verification and validation of the systems
need computer based approaches. They use model abstraction to simulate
hardware and software interactions.}

In the domain of rigorous analysis of airplane systems, work often follows
for practical and economic reasons a philosophy of using a mix of formal and
systematic informal methods. An example from airplane maintenance procedures 
\cite{ICAS08-Siemens-Boeing} uses a security evaluation methodology following 
the Common Criteria and a formal model and verification with the model checker 
AVISPA. In comparison, we use a more expressive logical model in the Isabelle 
Insider framework than the AVISPA specification. \new{To our knowledge, the focus of work on formal methods in avionics is directed towards the
correct functioning of the hardware and the software.
However, it is very important to consider the human factor.}\footnote{\new{Quote by Chesley B. Sullenberger [\href{http://www.sullysullenberger.com/my-testimony-today-before-the-house-subcommittee-on-aviation/}{http://www.sullysullenberger.com/my-testimony-today-before-the-house-subcommittee-on-aviation/}]:}\\
{\sl \new{Pilots must be able to handle an unexpected emergency and still keep their passengers and crew safe, but we should first design aircraft for them to fly that do not have inadvertent traps set for them.}

\new{We must also consider the human factors of these accidents.}

\new{From my 52 years of flying experience, and my many decades of safety work – I know that nothing happens in a vacuum, and we must find out how design issues, training, policies, procedures, safety culture, pilot experience and other factors affected the pilots’ ability to handle these sudden emergencies, especially in this global aviation industry.}

\new{Dr. Nancy Leveson, of the Massachusetts Institute of Technology, has a
quote that succinctly encapsulates much of what I have learned over
many years: `Human error is a symptom of a system that needs to be
redesigned.'}}} \new{We assume that our} \new{work is the first to consider
insider threats within airplane safety and security in a formal way.}

Logical modelling and analysis of insider threats has started off by investigating 
insider threats with invalidation of  security policies in connection with 
model checking by one of us in \cite{kp:13,kp:14}. This early approach also uses infrastructure models 
of organizations, actors and policies but was more restricted than the 
Isabelle Insider framework discussed in Section \ref{sec:isains}.
The use of sociological explanation has been pioneered in \cite{bikp:14} by one of us already 
with first formal experiments in Isabelle. Finally, one of us has established the Isabelle Insider framework 
in \cite{kp:16}. It has been validated on two of the main three 
insider patterns the Entitled Independent and Ambitious Leader. 
Relevant in the context of this application are other applications of the 
Isabelle Insider framework, and been applied to IoT Insiders \cite{knp:16,kam:17b} by using in addition
the extension of the framework to attack trees. 
Attack trees 
provide the possibility to refine attacks once they have been identified. 
This refinement is formalized together with the notion of attack trees as first 
introduced for insider models in general in \cite{iphk:15}. 
In other work, we applied the insider framework to auction protocols \cite{kkp:16}. 
In the CHIST-ERA project SUCCESS \cite{suc:16} we use the framework in
combination with attack trees and the Behaviour Interaction Priority (BIP) component 
architecture model to develop security and privacy enhanced IoT solutions.

In \cite{kamali-et-al-2017} Kamali et al.\ present reasoning that
integrates deduction based reasoning and model checking for the formal
verification of vehicle platooning. The idea is that vehicles move in
platoons and can join and leave them under certain safety
conditions. In order to model the hybrid aspects of the real-time
system a hybrid system is used that makes use of discrete decision
making (such as, initiating joining a platoon) and continuous control
(of actually driving the vehicle). The formal discrete reasoning is
translated to a timed automaton which can then be used to produce
actual running code (in a simulator). The right level of abstraction
is important in order to deal with complexity issues.

\section{Development of Airplane Safety and Security}
\label{sec:air}
On 2001-09-11, four terrorist attacks took place in the USA, two on
the two towers of the World Trade Center, one on the Pentagon, and in
a fourth attack the airplane crashed when passengers tried to overcome
the hijackers.\footnote{For a description of the events,
  see~\cite{Wikipedia-September_11_attacks}, including more than 300
  further pointers. A detailed account of the events of 9/11 and
  recommendations can be found in a 585 page report by the 9/11
  commission~\cite{911CommissionReport}. A list of aircraft hijackings
  can be found
  as~\cite{Wikipedia-List_of_aircraft_hijackings}.} Before these
  attacks, aircraft hijacking typically meant that the hijackers had
  some negotiable demands. Because of the risk to life for the people
  on board the aircraft, the standard approach was to enter
  negotiations and to avoid a resolution by force while the aircraft
  was in the air.

In particular, also there was no secured door between the passenger
compartment and the cockpit in airplanes; actually the door was
occasionally open, even allowing passengers to get a glimpse of the
cockpit during the flight. In Western countries there were no airplane hijackings with
major loss of life between the 1970s and the 2001-09-11 attacks. This
may have created in the USA and other countries a
false sense of security. In the wake of the attacks a serious rethink
of the security provision has
happened. In
particular, the cockpit doors were reinforced and made bullet-proof, making it
nearly impossible to open by intruders \cite{theStar}.

These (and other) changes seem to have had the wanted effect, since in
the time since the introduction of secured cockpit doors there
were only 17 airplane hijackings or attempted airplane hijackings\footnote{Note however that there
  were other attacks on flights which did not originate from
  passengers, such as the Malaysia Airline Flight MH17 which was
  brought down by a missile over Ukraine on 2014-07-17.}  (as listed on
\cite{Wikipedia-List_of_aircraft_hijackings}), all but one of them
could be prevented from causing fatalities, and the one that did
result in fatalities was an insider attack. One nearly successful
airplane hijacking has been caused by the copilot who forced
Ethiopian Airlines Flight 702 to land at Zurich airport in an attempt
to blackmail asylum for himself in Switzerland. Also
this airplane hijacking can be characterized as an insider attack
since the attacker was part of the crew.

The one major exception to the rule was Germanwings Flight 9525 on
2015-03-24, which was on the way from Barcelona to D\"usseldorf. The
aircraft was hijacked by the copilot who locked out the captain who
had left the cabin. The pilot tried to regain access to
the cockpit but did not succeed. Subsequently, the copilot brought
the aircraft to a crash in which all 150 people on board died.

Let us now look more closely into the door and its release
mechanism.\footnote{The information is extracted from a 5:32
  film by Airbus \cite{airbus2002}.}
The door is operated by a switch from inside the cockpit (with three
positions: ``unlock'', ``norm'', ``lock'') and a keypad outside the
cockpit. In order to gain access to the cockpit normally a crew member
would use the inter-phone to contact a pilot in the cockpit to request
access, then presses the hash key on the keypad, which triggers a
buzzer in the cockpit, and the pilot releases the door using the
switch to open the door (by keeping it in the ``unlock'' position). In
case the pilot(s) is/are incapacitated the crew member outside the
cockpit can enter an emergency code to open the door. After 30 seconds
(during which the buzzer sounds in the cockpit) of no reaction by the
pilots the crew member can open the door for five seconds. 

\begin{figure}
\begin{center}
\input{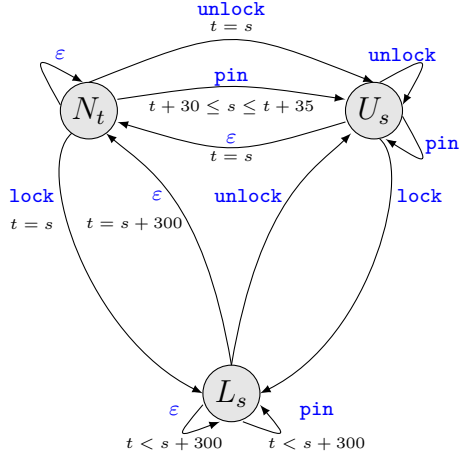}

\vspace{-.3cm}
\caption{A finite timed automaton to describe the lock mechanism of
  the door. In the three states, $N_t$, $U_s$, and $L_s$ (for normal,
  unlocked, locked at times $t$ or $s$, respectively), the pilots can
  \texttt{\color{blue}lock} the door, or \texttt{\color{blue}unlock}
  the door at any time with immediate effect, or do nothing with
  respect to the door -- indicated by
  $\color{blue}\tt\varepsilon$. Cabin crew can enter the
  \texttt{\color{blue}pin} of the door; entering an incorrect PIN
  corresponds to the empty action
  $\color{blue}\tt\varepsilon$. Entering the correct PIN has an effect
  only in the state $N_t$ after 30 seconds in a time window of five
  seconds unless the pilots take the \texttt{\color{blue}lock}
  action. After no action for 300 seconds the $L_s$ state is
  transformed to the $N_t$ state.}\label{fig:automaton}
\end{center}
\vspace{-.7cm}
\end{figure}
Since this access method could be used by a hijacker to force a crew
member to open the door from outside the cockpit, the pilots can, within
the 30 seconds between entering the emergency code and the release of
the door, lock the cockpit door by putting the toggle button into the
``lock'' mode. In that case the keypad is disabled for five minutes
and the door can be opened during this time only from inside the
cockpit by putting the button in the position ``unlock''.

The mechanism can be described on different levels and each level
requires certain assumptions (for instance, that the door itself will
withstand any physical force that may be exerted by an
attacker). According to Occam's razor, we try to give a representation
that is as easy as possible and still describes the situation in
sufficient detail that the important aspects are modelled. A first
approximation can be given by the timed finite state machine in
Figure~\ref{fig:automaton} with three states ``$N$'', ``$U$'', and ``$L$''
for ``normal'', ``unlocked'', and ``locked'',
respectively. While time
plays a role and it makes a difference for humans whether the door is
locked for 300ms, 300s, or 300 minutes, we will abstract from this in
the following formalization.  During the fatal flight, the copilot
used this locking mechanism to lock out the captain from the
cockpit. While the mechanism has been successful so far from
preventing any fatal attempt by an outsider to hijack an aircraft, the
same mechanism prevented the captain from re-entering the cockpit and
take action to rescue the aircraft in this case.


\section{Isabelle Insider Framework}
\label{sec:isains}

Before we formalize the airplane scenario in the
section~\ref{sec:model}, we give first a brief introduction to
Isabelle in this section; describe the Isabelle Insider framework with
infrastructures, policies, actors, and insiders; and describe how
Kripke Structures and CTL are modelled.

\subsection{Isabelle and Modular Reasoning}
Isabelle/HOL is an interactive proof assistant based on Higher Order Logic (HOL). 
Application specific logics are formalized into new theories extending HOL.
They are called object-logics. Although HOL is undecidable and therefore proving
needs human interaction, the reasoning capabilities are very sophisticated
supporting ``simple'', i.e., repetitive, tedious proof tasks to a level of
complete automation. The use of HOL has the advantage that it enables expressing
even the most complex application scenarios, conditions, and logical
requirements and HOL simultaneously enables the analysis of the meta-theory. 
That is, repeating patterns specific to an application can be abstracted and
proved once and for all. As an example, we will see how general preservation
theorems of the state transition relation over the system graph and over policies
can be proved as part of the insider framework and applied in concrete
applications like the airplane scenario (see Section \ref{sec:kripke}).

An object-logic contains new types, constants, and definitions. These items 
reside in a theory file, for instance, the file \texttt{Insider.thy} contains the object-logic 
for insider threats described in the following paragraphs. 
This Isabelle Insider framework is a {\it conservative extension} of HOL. This means
that our object logic does not introduce new axioms and hence guarantees consistency.
Conceptually, new types are defined as subsets of existing types and properties are proved
using a one-to-one relationship to the new type from properties of the existing type.
This process of conservative extension has been greatly facilitated by the 
datatype package that offers a restricted sort of simple recursive type definitions.
Inductive definitions are a similar tool to define new predicates by a set of
rules. Both extension features offer the specification of model elements with
a theory of induction and exhaustion properties necessary for the proof of
theorems over the model.

Besides datatypes and inductive definitions, we make also use of local assumptions 
within {\it locales}.
This is the reasoning process we propose as part of our methodology: 
the insider condition in Section \ref{sec:att} is not an axiom but is locally assumed 
to analyze the infrastructure's policies.

This process has been conceived as {\it Modular Reasoning} in Isabelle \cite{kam:00}
and implemented in the locales mechanism. Locales have been motivated by case studies
from abstract algebra where proofs about algebraic structures -- like groups, rings,
or fields -- frequently use assumptions -- like $\forall x. x \circ 1 = x$ --
that are valid within these algebraic structures but not outside. Rather
than repeating those local assumptions continuously in large numbers of 
property statements and proofs, locales realize contexts in which those
assumption can be used.
Insider threat modelling and analysis using logics shows the same needs, since 
assumptions about actors are specific to a certain application's infrastructure.
Moreover, the definition and the assumption of a locale are accessible later on,
whenever the locale is invoked. But since they are local assumptions and
definitions they do not endanger HOL's principle of conservative extension.

We are going to use Isabelle syntax and concepts in this paper and will
explain them when they are used.

\subsection{Infrastructures, Policies, Actors, and Insiders}
\label{sec:infra}

In the Isabelle/HOL theory for Insiders, one expresses policies over
actions \texttt{get}, \texttt{move}, \texttt{eval}, and \texttt{put}.
An actor may be enabled to 
\begin{itemize}
\item \texttt{get} data or physical items, like keys,
\item \texttt{move} to a location,
\item \texttt{eval} a program,
\item \texttt{put} data at locations or physical items -- like airplanes --
``to the ground''.
\end{itemize}
The precise semantics of these actions is refined in the state transition 
rules for the concrete infrastructure.
The framework abstracts from concrete data -- actions have no parameters:
\begin{ttbox}
{\bf datatype} action = get | move | eval | put
\end{ttbox}
The human component is the {\it Actor} which is represented by an abstract
type \texttt{actor} and a function \texttt{Actor}
that creates elements of that type from identities (of type \texttt{string}):
\begin{ttbox}
{\bf {typedecl}} actor 
{\bf {type}}_{\bf{synonym}} identity = string
{\bf {consts}} Actor :: string \ttfun actor
\end{ttbox}
Note that it would seem more natural and simpler to just 
define \texttt{actor} as a datatype over identities with a constructor \texttt{Actor}
instead of a simple constant together with a type declaration like, for example,
in the Isabelle inductive package \cite{pau:97}. 
This would, however, make the constructor \texttt{Actor} an injective function
by the underlying foundation of datatypes therefore excluding the fine
grained modelling that is at the core of the insider definition:
In fact, it defines the function \texttt{Actor} to be injective
for all except insiders and explicitly enables insiders to have
different roles by identifying \texttt{Actor} images.

Atomic policies of type \texttt{apolicy}
describe prerequisites for actions to be granted to actors given by
pairs of predicates (conditions) and sets of (enabled) actions:
\begin{ttbox}
{\bf type}_{\bf{synonym}} apolicy = ((actor \ttfun bool) \tttimes action set)
\end{ttbox}
For example, the \texttt{apolicy} pair \texttt{(\ttlam x. True, \{move\})}
specifies that all actors are enabled to perform action \texttt{move}.
To represent the macro level view seeing the actor within an infrastructure,
we define a graph datatype \texttt{igraph} for infrastructures containing:
a set of location pairs -- the actual ``map'' of the infrastructure
and a list of actor identities associated with each node (location) in that graph.\footnote{Note that $\lambda$ is the usual lambda-operator 
of higher order logic that describes functions. For instance, the square function 
can be defined -- without giving it a name -- as $\ttlam x. x*x$.}
Moreover, an \texttt{igraph} contains a function associating actors with a pair
of string lists: the first list describes the credentials an actor has while the
second list defines the roles that an actor can take on. Finally, an 
\texttt{igraph} has a component assigning locations to a string list describing
the state of the component. Slightly adapting the original insider
framework, we needed to integrate the credentials, roles, and location state 
into the infrastructure graph to enable the dynamic view of state 
transition and Kripke structures (see Section \ref{sec:kripke}). 
For each of the components there exist corresponding projection functions and
predicates \texttt{has} and \texttt{role} to express that actors have
credentials or that they can perform in specified roles, respectively, 
and \texttt{isin} to express that locations are in a specified state 
(see Appendix). 
\begin{ttbox}
{\bf datatype} igraph = Lgraph (location \tttimes location)set 
                          location \ttfun identity list
                          actor \ttfun (string list \tttimes string list)  
                          location \ttfun string list
\end{ttbox}
Infrastructures combine an infrastructure graph of type \texttt{igraph}
with a policy function that assigns local policies 
over a graph to each location of the graph, that is, it is a function
mapping an \texttt{igraph} to a function from \texttt{location} to 
\texttt{apolicy set}. The Isabelle type \texttt{[igraph, location] \ttfun\ apolicy set}
abbreviates \texttt{igraph \ttfun\ (location \ttfun\ apolicy set)} hence the 
stepwise application to \texttt{igraph} to return a function is possible.
\begin{ttbox}
{\bf datatype} infrastructure = Infrastructure igraph 
                                          [igraph, location] \ttfun apolicy set
\end{ttbox}
Elements of the datatype infrastructure can
thus be constructed using the constructor \texttt{Infrastructure}, which
is a higher order function, because it takes as (second) input a policy 
valued function.
This higher order parameter represents local policies, that is, maps from graph
locations to policies for that location. In the following section, we will see
how this higher order function enables proof of general preservation properties.

Policies specify the expected behaviour of actors of an infrastructure. 
We define the behaviour of actors using a predicate \texttt{enables}:
within infrastructure \texttt{I}, at location \texttt{l},
an actor \texttt{h} is enabled to perform an action \texttt{a} if there
is a pair \texttt{(p,e)} in the local policy of \texttt{l} --  \texttt{delta I l} 
projects to the local policy -- such that action \texttt{a} is in the action set 
\texttt{e} and the policy predicate \texttt{p} holds for actor \texttt{h}.
\begin{ttbox}
enables I l h a \ttequiv \ttexists (p,e) \ttin delta I l. a \ttin e \ttand p h
\end{ttbox} 
For example, the statement 
\texttt{enables I l (Actor''Bob'') move} is true if 
the atomic policy \texttt{(\ttlam x. True, \{move\})} is in the set
of atomic policies \texttt{delta I l} at location \texttt{l} in infrastructure
\texttt{I}. \new{Double quotes as in \texttt{''Bob''} create a string in Isabelle/HOL.}

The human actor's level is modelled in the Isabelle Insider framework by assigning
the individual actor's psychological disposition\footnote{Note that the determination of the psychological state of an actor is not done using the formal system. It is up to a psychologist to determine this. However, if for instance, an actor is classified as \texttt{disgruntled} then this may have an influence on what they are allowed to do according to a company policy and this can be formally described and reasoned about in Isabelle.} \texttt{actor\_state}
to each actor's identity.
\begin{ttbox}
{\bf datatype} actor_state = State psy_state motivations
\end{ttbox}
The values used for the definition of the types
\texttt{motivations} and \texttt{psy\_state} (see Appendix)
are based on a taxonomy from psychological insider research \cite{nblgcww:14}.
The transition 
to become an insider is represented by a {\it Catalyst} that tips the insider 
over the edge so he acts as an insider formalized as a ``tipping point'' 
predicate. To embed the fact that the attacker is an insider, the actor can then
impersonate other actors. In the Isabelle Insider framework, the 
predicate \texttt{Insider} must be used as a {\it locale} assumption
to enable impersonation for the insider:
this assumption entails that an insider \texttt{Actor ''Eve''} can act like 
their alter ego, say \texttt{Actor ''Charly''} within the context of the locale.
This is realized by the predicate  \texttt{UasI}:
\begin{ttbox}
UasI a b \ttequiv (Actor a = Actor b) \ttand
           \ttforall x y. x \ttneq a \ttand y \ttneq a \ttand Actor x = Actor y \ttimp x = y
\end{ttbox}
Note that this predicate also stipulates that the function \texttt{Actor} 
is injective for any other than the identities \texttt{a} and \texttt{b}.
This completion of the Actor function to an ``almost everywhere injective function''
is needed in some proofs (for an example see Section \ref{sec:foe}).
\new{We generalize here from other approaches on formal security analysis used in particular 
in security protocol verification known as the Dolev-Yao attacker model \cite{DolevYao1981}.
Our approach is more flexible because it addresses not just one specific attacker with a set 
range of abilities (eavesdrop, intercept, fake in Dolev-Yao) but more generally an insider, 
that is, someone who can impersonate any other actor and thereby attain any ability or access 
rights that exist in the system. This flexibility also allows modeling an attacker that 
``impersonates'' more than one actor to analyze collusions of insiders. In an earlier application 
of the Isabelle Insider framework \cite{kkp:16}, we illustrated this by a ``ringing attack'' on 
the Cocaine Auction protocol.}

\subsection{Kripke Structures and CTL}
\label{sec:kripke}
The expressiveness of Higher Order Logic allows formalizing the notion of Kripke 
structures as sets of states and a transition relation over those in Isabelle. 
Moreover, temporal logic can be directly encoded using Isabelle's fixpoint definitions 
for each of the CTL operators \cite{kam:16b}. Combining the two, we can then
apply them as generic tools to analyze dynamically changing infrastructures
with Insiders: we consider snapshots of infrastructures as states, use the 
actors and their action based behaviour definition to define a state transition,
to then use temporal logic to express safety and security properties over dynamically 
changing infrastructures. This application will be demonstrated on our case study
in Section \ref{sec:forair}. We briefly introduce here the necessary facts of 
Kripke structures and CTL showing how they are instantiated for Insiders.

The transition relation on system states is defined as an 
inductive predicate called \texttt{state\_transition\_in}. It introduces the syntactic 
infix notation \texttt{I \ttrelN\, I'} to denote that system state \texttt{I} and 
\texttt{I'} are in this relation.
\begin{ttbox}
{\bf inductive} state_transition_in :: [state, state] \ttfun bool  ("_  \ttrelN _")
\end{ttbox}

The specification of the behaviour of actors in the Insider framework allows
defining the rules for the state transition relation of the Kripke
structure for infrastructures for each of the actions. 
Here is the rule for put. 
The expression \texttt{h \ttatI\ l}
says that \texttt{h} is at location \texttt{l} in the graph \texttt{G}. 
The next state construction \texttt{I'} 
uses the projections \texttt{gra}, \texttt{agra}, \texttt{cgra},
\texttt{\texttt{lgra}} to select the graph itself, the actors-location association,
the credentials and roles, and the location state map, respectively. 
The rule expresses that an actor -- who is at location \texttt{l}
and is ``put''-enabled in the infrastructure \texttt{I} by its 
policy at location \texttt{l} -- can ``put'' the location into a 
state \texttt{z} in the successor state \texttt{I'} of the state 
transition for infrastructures.
The double brackets enclose the preconditions of the meta-implication $\ttImp$
in Isabelle. A proposition $\ttlbrack A ; B \ttrbrack \ttImp C$ simply abbreviates
$A \ttImp (B \ttImp C)$.

\begin{ttbox}
put: \ttlbrack G = graphI I; h \ttatI l;  
      enables I l (Actor a) put; 
      I' = Infrastructure 
            (Lgraph (gra G)(agra G)(cgra G)((lgra G)(l := [z])))
            (delta I)
     \ttrbrack \ttImp I \ttrelN I' 
\end{ttbox}
We illustrate this particular rule here because we use it
in the case study to express that an actor can put the airplane
to the ground (see Section \ref{sec:attack}). 

We can already develop some very useful theorems for the state transition
relation and Kripke structures. For example, the following lemma motivates
why we define infrastructures as higher order functions where
the local policies map the graph to a function over its locations: 
precisely because of that generality of the infrastructure constructor
we can prove that state transitions do not change the policy \texttt{delta}
-- as one would expect.
\begin{ttbox}
{\bf lemma} init_state_policy: I \ttrelN\^{}* I' \ttImp  delta I = delta I'
\end{ttbox}
The relation \texttt{\ttrelN\^{}*} is the reflexive transitive closure -- an operator
supplied by the Isabelle theory library -- applied to the relation \texttt{\ttrelN}.

The proof of this invariant illustrates why for policy verification
as we show here a deductive framework like Isabelle is  well suited.
To deduce the above theorem, we first prove that single step state
transitions preserve the policy. 
\begin{ttbox} 
 \ttforall I I'. I \ttrelN I' \ttimp  delta I = delta I'
\end{ttbox}
Then we use this lemma within an application of the induction for
reflexive transitive closure  of relations that is provided in the
Isabelle theory library to infer the above lemma \texttt{init\_state\_policy}.
Note that it is the specification in HOL of the state transition relation
that provides the case analysis rule and the induction scheme as sound
rules automatically generated from the definition.

Branching time temporal logic CTL has been integrated by one of us as part of the
Isabelle Insider framework \cite{kam:16b} built over Kripke structures.
A generic type \texttt{state} including a transition $\ttrelI$ is defined there 
using the concept of type classes in Isabelle. This type class \texttt{state} is 
then instantiated to the type of \texttt{infrastructures} thereby instantiating 
the state transition relation to $\ttrelN$ defined in the insider theory presented 
above (see Appendix).
Thereby, the theory constructed and proved for this state transition $\ttrelI$ over 
a generic type \texttt{state} are transferred automatically to infrastructures and their
transition relation $\ttrelN$.

Summarizing, the CTL-operators \texttt{EX} and \texttt{AX} express that property $f$ 
holds in some or all next states, respectively.
\begin{ttbox}
 AX \ttf \ttequiv \{ s. \{f0. s \ttrelI f0 \} \ttsubseteq \ttf \}
 EX \ttf \ttequiv \{ s. \ttexists f0 \ttin \ttf. s \ttrelI f0 \}
\end{ttbox}
The CTL formula \texttt{AG} $f$ means that on all paths branching from a state $s$
the formula $f$ is always true (\texttt{G} stands for `globally'). It can be defined
using the Tarski fixpoint theory by applying the greatest fixpoint operator.
\begin{ttbox}
 AG \ttf \ttequiv gfp(\ttlam Z. \ttf \ttcap AX Z)
\end{ttbox}
In a similar way, the other CTL operators are defined. 
The formal Isabelle definition of what it means that formula 
$f$ holds in a Kripke structure \texttt{M} for insiders can be
stated as: the initial states of the Kripke structure \texttt{init M} 
need to be contained in  the set of all states \texttt{states M} 
that imply $f$.
\begin{ttbox}
 M \ttvdash \ttf \ttequiv  init M \ttsubseteq \{ s \ttin states M. s \ttin \ttf  \}
\end{ttbox}
In an application, the set of states of the Kripke structure will be defined 
as the set of states reachable by the infrastructure state transition from 
some initial state, say \texttt{example\_scenario}.
\begin{ttbox}
  example\_states \ttequiv \{ I. example\_scenario \ttrelI\^{}*  I \}
\end{ttbox}
The \texttt{Kripke} constructor combines the constituents initial state, state set, and
state transition relation \texttt{\ttrelI}. 
\begin{ttbox}
 example\_Kripke \ttequiv Kripke example\_states \{example\_scenario\} \ttrelI 
\end{ttbox}
Given some \texttt{example\_policy} -- a predicate over an infrastructure using
actors, actions, and their behaviours -- we can then for example try to 
prove that this property holds generally by attempting the following proof
in Isabelle.
\begin{ttbox}
  example\_Kripke \ttvdash AG example_policy 
\end{ttbox}
If the proof fails, the failed attempt will reveal conditions describing a
state in the Kripke structure as well as actions leading to this state that identify
an attack possibility. In the example in Section \ref{sec:attack}, this will be 
illustrated. Additionally, the failed attempts to prove the global validity
also lead to identifying invariants of the system helping to establish
decisive side conditions as well as identifying loopholes. The loopholes
lead to a deeper insight into problems with the policy. 
By defining new locale assumptions and re-proving global properties, the
newly found assumptions can be refined until the proof succeeds.
This procedure will be illustrated on the airplane case study as well in 
Section \ref{sec:foe}.

Note that all the definitions in the locale \texttt{airplane}
that we use in Section \ref{sec:model} have been implemented as {\it locale} 
definitions using the locale keywords \texttt{\bf fixes}  and 
\texttt{\bf defines}~\cite{kpw:99}. 
Thus they are accessible whenever the locale \texttt{airplane} is invoked.
But since definitions are essentially abbreviations, they adhere to the principle
of conservative extension of HOL not endangering consistency.


\section{Formalizing the Airplane Scenario}
\label{sec:model}

In this section we first provide the necessary infrastructure, then
specify global and local policies, and finally formalize insider
attacks and safety and security.
\TODO{M@F: Please check this
  paragraph. It's new. I felt that something is needed here before the
  subsection starts. F@M: checked ok}

\subsection{Formalization of Airplane Infrastructure and Properties}
\label{sec:forair}
We restrict the Airplane scenario to four identities: Bob, Charly, Alice, and Eve.
Bob acts as the pilot, Charly as the copilot, and Alice as the flight attendant. 
Eve is an identity representing the malicious agent that can act as the copilot
although not officially acting as an airplane actor. The identities that act
legally inside the airplane infrastructure are listed in the set
of airplane actors.
\begin{ttbox}
{\bf fixes} airplane_actors :: identity set
{\bf defines} airplane_actors_def: airplane_actors \ttequiv \{''Bob'', ''Charly'', ''Alice''\}
\end{ttbox}
In the above locale definition we use the \texttt{\bf fixes} keyword to introduce
a locale constant with its type which is then specified by \texttt{\bf defines}. 
In the following, we drop all these elements but the actual definition to make
the exposition shorter and clearer.

To represent the layout of the airplane, a simple architecture is best suited
for the purpose of security policy verification. The locations we consider for
the graph are \texttt{cockpit}, \texttt{door}, and \texttt{cabin}. They are
defined as locale definitions and assembled in a set \texttt{airplane\_locations}.
\begin{ttbox}
cockpit \ttequiv Location 2
door \ttequiv Location 1
cabin \ttequiv Location 0
airplane_locations \ttequiv \{ cabin, door, cockpit \}
\end{ttbox}

The actual layout and the initial distribution of the actors in the 
airplane infrastructure is defined by the following graph in which 
the actors Bob and Charly are in the cockpit and Alice is in the 
cabin.
\begin{ttbox}
ex_graph \ttequiv Lgraph
          \{(cockpit, door),(door,cabin)\}
           (\ttlam x. if x = cockpit then [''Bob'', ''Charly''] 
                 else (if x = door then [] 
                       else (if x = cabin then [''Alice''] else [])))
          ex_creds ex_locs
\end{ttbox}
The two additional inputs \texttt{ex\_creds} and \texttt{ex\_locs} 
for the constructor \texttt{Lgraph} are the credential and role assignment 
to actors and the state function for locations introduced in Section 
\ref{sec:infra}, respectively. For the airplane scenario, we use 
the function \texttt{ex\_creds} to assign the roles and credentials 
to actors. For example, for \texttt{Actor ''Bob''} the following 
function returns the pair of lists \texttt{([''PIN''], [''pilot''])}
assigning the credential \texttt{PIN} to this actor and 
designating the role \texttt{pilot} to him.
\begin{ttbox}
ex_creds \ttequiv (\ttlam x.
  (if x = Actor ''Bob'' then ([''PIN''], [''pilot''])        
   else (if x = Actor ''Charly'' then ([''PIN''],[''copilot''])
         else (if x = Actor ''Alice'' then ([''PIN''],[''flightattendant''])
               else ([],[])))))"
\end{ttbox}
Locations  of the infrastructure graph have specific states.
For example, the door can be in state \texttt{locked}.
Similar to the previous function \texttt{ex\_creds}, 
the function \texttt{ex\_locs} assigns these states to the locations
of the infrastructure graph.
\begin{ttbox}
ex_locs \ttequiv \ttlam x. if x = door then [''norm''] 
                else (if x = cockpit then [''air''] else [])
\end{ttbox}

\subsection{Initial Global and Local Policies}
\label{sec:attack}
In the Isabelle Insider framework, we define a global policy reflecting
the global safety and security goal and then break that down into local 
policies on the infrastructure. The verification will then analyze whether
the infrastructure's local policies yield the global policy.

Globally, we want to exclude attackers to ground the plane. In the 
formal model, landing the airplane results from an actor performing a
\texttt{put} action (see Section \ref{sec:kripke}) in the cockpit and 
thereby changing the state from \texttt{air} to \texttt{ground}.

Therefore, we specify the global policy as ``no one except
airplane actors can perform \texttt{put} actions at location cockpit''
by the following predicate over infrastructures \texttt{I} and
actor identities \texttt{a}.
\begin{ttbox}
global_policy I a \ttequiv  a \ttnin airplane\_actors \ttimp \ttneg(enables I cockpit (Actor a) put)
\end{ttbox}
We next attempt to define the local policies for each location as a function
mapping locations to sets of pairs: the first element of each pair for a location 
\texttt{l} is a predicate over actors specifying the conditions necessary for an actor
to be able to perform the actions specified in the set of actions which is the
second element of that pair.
The local policy functions are additionally parameterized over an infrastructure
graph \texttt{G} since this may dynamically change through the state transition. 
\begin{ttbox}
local_policies G \ttequiv
(\ttlam y. if y = cockpit then
      \{(\ttlam x. (\ttexists n. (n \ttatI cockpit) \ttand Actor n = x), \{put\}),
       (\ttlam x. (\ttexists n. (n \ttatI cabin) \ttand Actor n = x 
             \ttand has (x, ''PIN'')\ttand isin G door ''norm''), \{move\}) \}
      else (if y = door then \{(\ttlam x. True, \{move\})\}
            else (if y = cabin then \{(\ttlam x. True, \{move\})\} else \{\})))
\end{ttbox}
This policy expresses that any actor can move to door and cabin but
places the following restrictions on cockpit.
\begin{description}
\item[\texttt{put}:] to perform a \texttt{put} action, that is, put the plane into a new position 
            or put the lock, an actor must be at position cockpit, \ie in the cockpit;
\item[\texttt{move}:] to perform a move action at location cockpit, that is, move into it,
             an actor must be at the position cabin, must be in possession of 
             PIN, and door must be in state norm.
\end{description}
Although this policy abstracts from the buzzer, the 30 sec delay, and a few
other technical details, it captures the essential features of the cockpit door.

The graph, credentials, and features are plugged together with the policy 
into the infrastructure \texttt{Airplane\_scenario}. 
\begin{ttbox}
 Airplane_scenario \ttequiv Infrastructure ex_graph local_policies
\end{ttbox}

\subsection{Insider Attack, Safety, and Security}
\label{sec:att}
We now first stage the insider attack and introduce
basic definitions of safety and security for the airplane scenario.
To invoke the insider within an application of the Isabelle
Insider framework, we assume in the locale \texttt{airplane} 
as a locale assumption with \texttt{assumes} that the tipping point 
has been reached for \texttt{Eve} which manifests itself in her
\texttt{actor\_state} assigned by the locale function \texttt{astate}
\begin{ttbox}
astate x =  (case x of 
           ''Eve'' \ttfun Actor_state depressed \{revenge, peer_recognition\}
          | _ \ttfun Actor_state happy \{\})
\end{ttbox}
In addition, we state that she is an insider being able to
impersonate \texttt{Charly} by locally assuming the \texttt{Insider}
predicate. This predicate allows an insider to impersonate a set of
other actor identities; in this case the set is singleton.
\begin{ttbox}
{\bf{assumes}} Eve_precipitating_event: tipping_point(astate ''Eve'')
{\bf{assumes}} Insider_Eve : Insider ''Eve'' \{''Charly''\}
\end{ttbox}
Next, the process of analysis uses this assumption as well as the definitions
of the previous section to prove security properties interactively as theorems
in Isabelle. We use the strong insider assumption here up front to provide a 
first sanity check on the model by validating the infrastructure
for the ``normal'' case. We prove that the global policy holds for the 
pilot Bob. To illustrate a proof in Isabelle, we show the statement of the
theorem including the Isabelle proof script. The system replies of the
interaction with Isabelle are omitted but can be simply recreated by
running that script.
\begin{ttbox}
{\bf lemma} ex\_inv: global_policy Airplane_scenario ''Bob''
{\bf by} ({\bf{simp add}}: Airplane\_scenario\_def global\_policy\_def airplane\_actors\_def)
\end{ttbox}
The proof is finished with one complex step: 
unfold the definitions of the scenario given by
\texttt{Airplane\_scenario\_def} and two other definitions and then apply the simplifier, 
an automated technique that applies equational (including conditional) rewriting
to solve a goal. 

We can prove the same theorem for \texttt{Charly} who 
is the copilot in the scenario (omitting the proof and accompanying
Isabelle commands).
\begin{ttbox}
 global_policy Airplane_scenario ''Charly''
\end{ttbox}
But \texttt{Eve} is an insider and is able to impersonate \texttt{Charly}.
She will ignore the global policy.
This insider threat can now be formalized as an invalidation 
of the global company policy for \texttt{''Eve''} in the following ``attack'' theorem 
named \texttt{ex\_inv3}:
\begin{ttbox}
{\bf theorem} ex_inv3: \ttneg global_policy Airplane_scenario ''Eve''
\end{ttbox}
This theorem can be proved by first invoking the above insider assumption 
about Eve unfolding the corresponding underlying definitions provided in the
Isabelle Insider framework but finally then again using the powerful simplification
tactic \texttt{simp}.
The attack theorem is proved in Isabelle: it says that \texttt{Eve} can get access 
to the cockpit and put the position to \texttt{ground}. In other words, Eve can 
crash the plane. 
The proof is very similar to proofs of comparable theorems in other
applications of the Isabelle Insider framework, for instance, 
for the IoT \cite{kam:17b} or for auctions \cite{knp:16}, and can basically
be copied from there just replacing local definition names.
\new{Summarizing, the insider assumption allows modeling that actors may 
be the same as other actors. Policies that are expressed according to roles
thus apply to those insiders which -- given that they are attackers -- are harmful.}

Safety and security are sometimes introduced in textbooks as complementary
properties, see, e.g., \cite{gol:08}. Safety expresses that humans and goods should 
be protected from negative effects caused by machines while security is the inverse
direction: machines (computers) should be protected from malicious humans.
Similarly, the following descriptions of safety and security in the airplane
scenario also illustrate this complementarity: 
one says that the door must stay closed to the outside; the other that there must 
be a possibility to open it from the outside.
\begin{description}
\item[\it Safety:] if the actors in the cockpit are out of action, there must be a possibility
to get into the cockpit from the cabin, and 
\item[\it Security:] \hspace*{2ex}if the actors in the 
cockpit fear an attack from the cabin, they can lock the door.
\end{description}
In the formal translation of these properties into HOL, this complementarity
manifests itself even more clearly: the conclusions of the two formalizations of 
the properties are negations of each other.
Safety is quite concisely described by stating that airplane actors can move 
into the cockpit.
\begin{ttbox}
Safety I a \ttequiv a \ttin airplane_actors \ttimp (enables I cockpit (Actor a) move)
\end{ttbox}
Security can also be defined in a simple manner as the property that
no actor can move into the cockpit if the door is on lock.
\begin{ttbox}
Security I a \ttequiv isin (graphI I) door ''locked''
                \ttimp \ttneg(enables I cockpit (Actor a) move)
\end{ttbox}
These two properties are defined for any infrastructure \texttt{I}
so we can apply them to the initial airplane scenario we have defined
in the previous section.
For this \texttt{Airplane\_scenario}, we can show safety, for example, for
\texttt{Alice} because she is in the cabin.
\begin{ttbox}
{\bf{lemma}} Safety: Safety Airplane_scenario ''Alice''
\end{ttbox}
In general, we could prove safety for any airplane actor who is in the cabin
for this state of the infrastructure.

In a slightly more complex proof, we can prove security for any
other identity which can be simply instantiated to \texttt{''Bob''}.
\begin{ttbox}
{\bf{lemma}} Security: Security Airplane_scenario ''Bob''
\end{ttbox}

The simple formalizations of safety and security enable 
proofs only over a particular state of the airplane infrastructure at a time 
but this is not enough since the general airplane structure is subject
to state changes.
For a general verification, we need to prove that the properties of interest
are preserved under potential changes. Since the airplane infrastructure
permits, for example, that actors move about inside the airplane, we need to 
verify safety and security properties in a dynamic setting.
After all, the insider attack \new{on Germanwings Flight 9525} appeared when the pilot had moved out of the
cockpit. Furthermore, we want to redefine the policy into the two-person
policy and examine whether safety and security are improved.
For these reasons, we next apply the general Kripke structure mechanism
introduced in Section \ref{sec:kripke} to the airplane scenario.

\section{Analysis of Safety and Security Properties}
\label{sec:ana}
In this section we first introduce a Kripke structure to model state
transitions in the airplane scenario. Then we formalize the two-person
rule and look how this rule is related to the property that the
airplane is not in danger with respect to an insider attack. We show
that an additional assumption is necessary to prove this property. We
conclude the section by summarizing the methodology. 
\TODO{M@F: Again,
  please check this paragraph. It's new. I felt that something is
  needed here before the subsection starts.
   F@M: checked ok}

\subsection{Kripke Structure for Airplane Scenario}
\label{sec:airkripke}
The state transition relation $\ttrelI$ introduced in Section \ref{sec:kripke}
is generally defined for a type class \texttt{state}. Therefore, we can 
instantiate the state transition for the type \texttt{infrastructure} as $\ttrelN$.
Consequently, we can define the set of all states that are in the reflexive
transitive closure of the infrastructure transition relation when starting
in the infrastructure \texttt{Airplane\_scenario} as a locale definition
\texttt{Air\_states}.
\begin{ttbox}
 Air_states \ttequiv \{ I. Airplane_scenario \ttrelN\^{}* I \}
\end{ttbox}
From there, we can define a corresponding Kripke structure 
by applying the constructor \texttt{Kripke}
to the above state set and the singleton set of \texttt{Airplane\_scenario} 
as the (only) initial state.
\begin{ttbox}
 Air\_Kripke \ttequiv Kripke Air\_states \{Airplane\_scenario\}
\end{ttbox}

We now illustrate how we can use this Kripke structure to 
explore and potentially invalidate the policy. The state of the infrastructure
that represents the fatal state is when the pilot has moved out and the 
door is locked. We introduce a locale definition \texttt{aid\_graph}
to represent the graph for this infrastructure.
\begin{ttbox}
aid_graph \ttequiv Lgraph
            \{(cockpit, door),(door,cabin)\}
             (\ttlam x. if x = cockpit then [''Charly''] 
                 else (if x = door then [] 
                       else (if x = cabin then [''Bob'', ''Alice''] else [])))
            ex_creds ex_locs'
\end{ttbox}
The function \texttt{ex\_locs'} encodes the state of the airplane 
where the door is now locked.
\begin{ttbox}
ex_locs' \ttequiv \ttlam x. if x = door then [''locked''] 
                 else (if x = cockpit then [''air''] else [])
\end{ttbox}
We finally define a new infrastructure state that takes this graph 
and the same \texttt{local\_policies} as \texttt{Airplane\_scenario}.
\begin{ttbox}
 Airplane_in_danger \ttequiv Infrastructure aid_graph local_policies
\end{ttbox}
For the analysis of security, we need to ask whether this new
infrastructure state \texttt{Airplane\_in\_}\linebreak[2]\texttt{danger} is reachable
via the state transition relation from the initial state.
It is. We can prove the following as a theorem in the locale \texttt{airplane}.
\begin{ttbox}
 {\bf theorem} step_allr: Airplane_scenario \ttrelN\^{}* Airplane_in_danger
\end{ttbox}
As the name of this theorem suggests it is the result of lining up a 
sequence of steps that lead from the initial \texttt{Airplane\_scenario}
to that \texttt{Airplane\_in\_danger} state.
In fact there are three steps via two intermediary infrastructure
states \texttt{Airplane\_getting\_in\_danger0} and 
\texttt{Airplane\_getting\_in\_danger} (see Appendix).
The former encodes the state where \texttt{Bob} has moved to the cabin and
the latter encodes the successor state in which additionally the lock state 
has changed to \texttt{locked}. 
The definitions of these states are very similar to the above definition
of \texttt{Airplane\_in\_danger} (see Appendix).
The proof of the theorem \texttt{step\_allr} correspondingly lines
up lemmas for each of the state transitions between the involved states.
Once provided with these lemmas, the main proof is just one simplification 
with the underlying definition of the reflexive transitive closure of a 
relation. This is the advantage of using a richly equipped proof assistant:
the theory library is well equipped with standard mathematics and the tactics
work well on this basis. The only real work has to be done to prove the
individual steps. However, although the proof scripts are a bit lengthy, this
is just simple step by step unfolding of definitions and simplification.
The only reason why it is not done in one step fully automatically is that
some instantiations under existential quantifiers have to be inserted
in the application of the state transition rules, like for example the
rule \texttt{put} we have seen in Section \ref{sec:kripke}.

Using the formalization of CTL over Kripke structures introduced in Section
\ref{sec:kripke}, we can now transform the attack sequence
represented implicitly by the above theorem \texttt{step\_allr} into a 
temporal logic statement. This attack theorem states that there is a 
path from the initial state of the Kripke structure \texttt{Air\_Kripke}
on which eventually the global policy is violated by the attacker.
\begin{ttbox}
{\bf theorem} aid_attack: Air_Kripke \ttvdash EF (\{x. \ttneg global_policy x ''Eve''\})
\end{ttbox}
The proof uses the underlying formalization of CTL and the lemmas that
are provided to evaluate the \texttt{EF} statement on the Kripke structure.
However, the attack sequence is already provided by the previous theorem.
So the proof just consists in supplying the step lemmas for each step and
finally proving that for the state at the end of the attack path, i.e.,
for \texttt{Airplane\_in\_danger}, the global policy is violated.
This proof corresponds precisely to the proof of the attack theorem 
\texttt{ex\_inv3}. It is not surprising that the security attack is 
possible in the reachable state \texttt{Airplane\_in\_danger} when it
was already possible in the initial state. However, this statement is 
not satisfactory since the model does not take into account whether 
the copilot is on his own when he launches the attack.
This is the purpose of the two-person rule which we want to investigate in
more detail in this paper. Therefore, we next address how to add the 
two-person role to the model.

\subsection{Introduce Two-Person Rule}
To express the rule that two authorized
personnel must be present at all times in the cockpit, we define a second set of
local policies. The following function realizes the two-person constraint.
It requests that the number of actors at the location \texttt{cockpit} in the
graph \texttt{G} given as input must be at least two to enable actors at
the location to perform the action \texttt{put}.  Formally, we can express 
this here as \texttt{2 \ttleq length(agra G cockpit)} since we have all of
arithmetic available (remember \texttt{agra G y} is the list of 
actors at location \texttt{y} in \texttt{G} introduced in Section 
\ref{sec:kripke}).
\begin{ttbox}
local_policies_four_eyes G \ttequiv
(\ttlam y. if y = cockpit then
       \{(\ttlam x. (\ttexists n. n \ttatI cockpit \ttand Actor n = x) \ttand 2 \ttleq length(agra G y) \ttand 
             \ttforall h \ttin set(agra G y). h \ttin airplane_actors), \{put\}),
       (\ttlam x. (\ttexists n. n \ttatI cabin \ttand Actor n = x) \ttand has (x, ''PIN'')\ttand 
                   isin G door ''norm''), \{move\})\}
       else (if y = door then 
             \{(\ttlam x. ((\ttexists n. n \ttatI cockpit \ttand Actor n = x) 
                     \ttand 3 \ttleq length(agra G cockpit)), \{move\})\}
            else (if y = cabin then 
                  \{(\ttlam x. \ttexists n. n \ttatI door \ttand Actor n = x), \{move\})\} 
                 else \{\})))
\end{ttbox}
Note that the two-person rule requires three people to be at the cockpit
before one of them can leave. This is formalized as a condition on the 
\texttt{move} action of location \texttt{door}. A move of an actor
\texttt{x} in the cockpit to \texttt{door} is only allowed if three people are in 
the cockpit.
Practically, it enforces a person, say Alice to first enter 
the cockpit before the pilot Bob can leave.
However, this condition is necessary to guarantee that the two-person
requirement for \texttt{cockpit} is sustained by the dynamic changes to 
the infrastructure state caused by actors' moves.
A move to location \texttt{cabin} is only allowed from \texttt{door}
so no additional condition is necessary here.

What is stated informally above seems intuitive and quite easy to believe.
However, comparing to the earlier formalization of this two-person rule \cite{kk:16}, 
it appears that the earlier version did not have the additional condition on the 
action \texttt{move} to \texttt{door}. 
One may argue that in the earlier version the authors did not consider
this because they had neither state transitions, Kripke structures, nor CTL
to consider dynamic changes. However, in the current paper this additional
side condition only occurred to us when we tried to prove the following 
invariant which is needed in a subsequent security proof.
\begin{ttbox}
 {\bf lemma} two_person_inv1: 
  Airplane_not_in_danger_init \ttrelN\^{}* I \ttImp 2 \ttleq length (agra (graphI I) cockpit)
\end{ttbox}
This proof requires an induction over the state transition relation
starting in the infrastructure state \texttt{Airplane\_not\_in\_danger\_init}
with Charly and Bob in the cockpit and the two-person policy in place.
\begin{ttbox}
Airplane_not_in_danger_init \ttequiv Infrastructure ex_graph local_policies_four_eyes
\end{ttbox}
The corresponding Kripke structure of all states originating in this 
infrastructure state is defined as \texttt{Air\_tp\_Kripke}.
Within the induction for the proof of the above \texttt{two\_person\_inv1}, 
a preservation lemma is required that proves that if the condition 
\texttt{2 \ttleq\ length (agra (graphI I) cockpit)}
holds for \texttt{I} and \texttt{I \ttrelN\ I'} then it also 
holds for \texttt{I'}. The preservation lemma is actually trickier to
prove. It uses a case analysis over all the transition rules for each action. 
The rules for \texttt{put} and \texttt{get} are easy to prove for the user 
as they are solved by the simplification tactic automatically. The case for
action \texttt{move} is the difficult case. Here we actually need to 
use the precondition of the policy for location \texttt{door} in 
order to prove that the two-person invariant is preserved
by an actor moving out of the cockpit.
In this case, we need for example, invariants like the following
lemma that shows that in any infrastructure state originating
from \texttt{Airplane\_not\_in\_danger\_init} actors only ever appear
in one location and they do not appear more than once in a location --
which is expressed in a predicate \texttt{nodup} (see Appendix).
The following lemma is an instantiation of a similar general lemma proved
for all Kripke structures -- similar to the lemma \texttt{init\_state\_policy}
mentioned in Section \ref{sec:kripke}.
\begin{ttbox}
{\bf lemma} actors_unique_loc_aid_step: 
  Airplane_not_in_danger_init \ttrelN\^{}* I
  \ttImp \ttforall a. (\ttforall l l'. a \ttat{\texttt{graphI I}}) l \ttand a \ttat{\texttt{graphI I}} l' \ttimp l = l'
         \ttand (\ttforall l. nodup a (agra (graphI I) l))
\end{ttbox}

\subsection{Revealing Necessary Assumption by Proof Failure}
We would expect -- and this has in fact been presented in \cite{kk:16} --
that the two-person rule guarantees the absence of the insider attack.
This is indeed a provable fact in the following state 
\texttt{Airplane\_not\_in\_danger} defined similar to 
\texttt{Airplane\_in\_danger} from Section \ref{sec:airkripke} but
using the two-person policy.
\begin{ttbox}
 Airplane_not_in_danger \ttequiv Infrastructure aid_graph local_policies_four_eyes
\end{ttbox}
For this state, it can be proved  \cite{kk:16} that for any actor 
identity \texttt{a} the global policy holds.
\begin{ttbox}
  global_policy Airplane_not_in_danger a
\end{ttbox}
So, in the state \texttt{Airplane\_not\_in\_danger} with the two-person rule,
there seems to be no danger. But this is precisely the scenario of the suicide 
attack! Charly is on his own in the cockpit -- why then does the two-person rule
imply he cannot act?
The state \texttt{Airplane\_not\_in\_danger} defined in the earlier
formalization is mis-named: it uses the graph \texttt{aid\_graph} to define
a state in which Bob has left the cockpit and the door is locked. 
Since there is only one actor present, the precondition of the local policy 
for \texttt{cockpit} is not met and hence the action \texttt{put} is not 
enabled for actor Charly.
Thus, the policy rule for cockpit is true because the precondition of this 
implication -- two people in the cockpit -- is false, 
and false implies anything: seemingly a disastrous failure of logic.

Fortunately, the above theorem has been derived in a preliminary model only \cite{kk:16} 
in which  state changes were not integrated yet and which has been precisely
for this reason recognized as inadequate.
Now, with state changes in the improved model, we have proved the two-person invariant 
\texttt{two\_person\_inv1}.
Thus, we can see that the system -- if started in \texttt{Airplane\_not\_in\_danger\_init} 
-- cannot reach the mis-named state \texttt{Airplane\_not\_in\_danger} in which  Charly is on his own in the cockpit.


However, so far, no such general theorem has been proved yet. We 
only used CTL to discover attacks using \texttt{EF} formulas.
What we need for general security and what we consider next is to prove a global 
property with the temporal operator \texttt{AG} that proves that from a given 
initial state the global policy holds in all (\texttt{A}) states globally (\texttt{G}).

As we have seen in the previous section when looking at
the proof of \texttt{two\_person\_inv1}, it is not evident and trivial 
to prove that all state changes preserve security properties.
However, even this invariant does not suffice.
Even if the two-person rule is successfully enforced 
in a state, it is on its own still not sufficient. 
When we try to prove
\begin{ttbox}
  Air_tp_Kripke \ttvdash AG \{x. global_policy x ''Eve''\}
\end{ttbox}
for the Kripke structure \texttt{Air\_tp\_Kripke} of all states
originating in \texttt{Airplane\_not\_in\_danger\_init},
we cannot succeed. In fact, in that Kripke structure
there are infrastructure states 
where the insider attack is possible.
Despite the fact that we have stipulated the two-person rule as
part of the new policy and despite the fact that we can prove that
this policy is preserved by all state changes, the rule has no
consequence on the insider. Since Eve can impersonate the copilot
Charly, whether two people are in the cockpit or not, the attack can happen.

What we realize through this failed attempt to prove a global property
is that the policy formulation does not entail that the presence of
two people in itself actually disables an attacker. 

This insight 
reveals a hidden assumption. \new{Formal reasoning systems have the advantage that hidden assumptions must be made explicit. In human reasoning they occur when people assume a common understanding, which may or may not be actually the case. In the case of the rule above, its purpose may lead to an assumption that humans accept but which is not warranted.}

We use again a locale definition to encode this intentional understanding
of the two-person rule. The formula \texttt{foe\_control} encodes for any 
action \texttt{c} at a location \texttt{l} that if there is an \texttt{Actor x}
that is not an insider, that is, is not impersonated by Eve, then the insider is
disabled for that action \texttt{c}.
\begin{ttbox}
foe_control l c \ttequiv (\ttforall I. (\ttexists x. x \ttat{I} l \ttand Actor x \ttneq Actor ''Eve'')
                   \ttimp \ttneg(enables I l (Actor ''Eve'') c))
\end{ttbox}

\subsection{Proving Security in Refined Model}
\label{sec:foe}
Having identified the missing formulation of the intentional effects of
the two-person rule, we can now finally prove the general security property 
using the above locale definition.
We assume in the locale \texttt{airplane} an instance of \texttt{foe\_control}
for the cockpit and the action \texttt{put}.
\begin{ttbox}
{\bf assumes} cockpit_foe_control: foe_control cockpit put
\end{ttbox}
With this assumption, we are now able to prove that 
for all infrastructure states of the system \texttt{airplane} originating 
in state \texttt{Airplane\_not\_in\_danger\_init} Eve cannot put the 
airplane to the ground. 
\begin{ttbox}
{\bf theorem} Four_eyes_no_danger: Air_tp_Kripke \ttvdash AG \{x. global_policy x ''Eve''\}
\end{ttbox}
The proof uses as a key lemma that within Kripke structure
\texttt{Air\_tp\_Kripke} there is always someone in the cockpit 
who is not the insider.
\begin{ttbox}
{\bf lemma} tp_imp_control: Airplane_not_in_danger_init \ttrelN\^{}* I
                        \ttImp \ttexists x. x \ttat{I} cockpit \ttand Actor x \ttneq Actor ''Eve''
\end{ttbox}
This lemma can be proved by using the invariant that always two people are
in the cockpit. 
However, the invariant \texttt{two\_person\_inv1} cannot be used directly
since it is a lemma over lists rather than sets. Instead of re-formulating
the model with sets, we use a simple fact about sets and lists.
\begin{ttbox}
 (\ttforall a. nodup a l) \ttimp card (set l) = length l
\end{ttbox}
This general lemma enables together with the invariant 
\texttt{actors\_unique\_step\_loc\_aid\_step}
the proof of the more suitable invariant \texttt{two\_person\_set\_inv}.
\begin{ttbox}
{\bf lemma} two_person_set_inv: Airplane_not_in_danger_init \ttrelN\^{}* I
                           \ttImp 2 \ttleq card (set (agra (graphI z) cockpit))
\end{ttbox}
Using the assumption \texttt{foe\_control}, we can now mainly by applying
modus ponens derive that Eve is not enabled in \texttt{cockpit} to
perform \texttt{put} for any infrastructure state originating from
\texttt{Airplane\_not\_in\_danger\_init}. 
\begin{ttbox}
  Airplane_not_in_danger_init \ttrelN\^{}* I \ttImp \ttneg enables I cockpit (Actor ''Eve'') put
\end{ttbox}
Now, the proof of theorem \texttt{Four\_eyes\_no\_danger} (see Appendix) uses
simplification on basic lemmas for Kripke structures and CTL
to reduce to the above fact which finishes the proof.

\subsection{Summarizing Methodology}
We propose an informal methodology by summarizing the steps for the development of
secure policies in the presence of insiders using the Isabelle Insider
framework.
\begin{enumerate}
\item Build a model of the infrastructure, its actors, and local policies
with roles and credentials and define the security property of interest as global policy.
\item Identify initial state(s) and define Kripke structure.
\item Use the tipping point and insider assumptions to specify the potential
insider(s).
\item Invalidate the global policy, that is, negate the property to specify
an infrastructure state in which the insider can violate it.
\item Explore the state transition function to find a path from the initial 
state(s) to this state in which the global policy is violated.
For the invalidation and exploration, CTL can be used: first attempt to prove
\texttt{AG \{x. global\_policy x ''Eve''\}}; failure produces potentially a candidate
for an attack; next prove \texttt{EF \ttneg\{x. global\_policy x ''Eve''\}} to establish
the attack path.
\item Repeat the previous two steps to improve the policy, until the proof
of \texttt{AG \{x. global\_}\linebreak[2]\texttt{policy x ''Eve''\}} succeeds. 
\item If after repeated cycles in the previous 3 steps the proof of the \texttt{AG} property
   is still not successful, try to identify a missing global assumption (like
   \texttt{foe\_control}). Going back to step 4, add the assumption as a locale
   assumption and re-iterate. 
\end{enumerate}

\section{Discussion and Conclusions}
\label{sec:concl}

In this section, we briefly discuss limitations and approaches to
developing airplane policies, summarize the contributions of the
paper, and present some concluding remarks.

\subsection{Aspects of Airplane Policies}

In order to prove consequences of policies certain assumptions have to
be made and it is important to analyze the assumptions, since any
consequences hold only with respect to the assumptions. An important
assumption is that the airplane is initially not in danger,
\texttt{Airplane\_not\_in\_danger\_init}. That is, if the assumption
is violated initially (before the airplane leaves the ground) then we
cannot conclude that the airplane will not be in danger
later. Current policies do not assume that the cockpit door must be
locked before passengers board the airplane. Actually, often it is
still open and closed only later. This means that an attack by an
outsider during this phase cannot be ruled out.

For airlines it is an important question whether they should follow a two-person rule and as a consequence of the events on 2015-03-24 with the Germanwings flight 9525 a number of countries recommended the rule and a number of airlines\footnote{This is reported, for instance, in an article of 2015-03-26 by Reuters, \protect\url{http://www.reuters.com/article/france-crash-cockpits-idUSL6N0WS6GR20150326}{.}} introduced them -- without consideration of possible negative consequences.
In a more recent development, some German airlines have rescinded the
two-person
rule,\footnote{See \url{https://phys.org/news/2017-04-german-airlines-scrap-two-person-cockpit.html} and
  \url{https://www.swiss.com/corporate/EN/media/newsroom/press-releases/media-release-20170428}.}
since the introduction has also the disadvantage that it takes
considerably longer for one person to leave and another to enter the
cockpit than just for one person to leave. This means that with the
two-person policy, each time a pilot/co-pilot leaves the cockpit the
door is open for much longer than without the policy, hence increasing
the risk of a hostile attack. Up to now no good improvement on the
protocol for the door has been found, since any change seems to be
paired with substantial disadvantage as well.

We have not formally modelled the situation and the reasoning behind
this. We do this informally here. If we assume $p_0$, the probability
that one pilot is an insider; $p_1$, the probability that a terrorist
can use the time the door is open to enter the cockpit following the
one-person rule and take over the plane; and $p_2$, the corresponding
probability that a terrorist can enter the cockpit following the
two-person rule.

Fortunately all these probabilities are very small. This means, however, that there is no reliable way to determine their values. It
  seems obvious that $p_2>p_1$, it can be assumed that
  $p_2$ is considerably bigger than $p_1$.\footnote{See,
\href{https://www.easa.europa.eu/newsroom-and-events/news/minimum-cockpit-occupancy-easa-issues-revised-safety-information-bulletin}{\tt https://www.easa.europa.eu/newsroom-and-events/news/minimum-cockpit-occupancy-easa\-issues-revised-safety-information-bulletin}}

With these probabilities we get that an aircraft is in danger according to the one-person rule:\\
${\it probability}({\it insider}\ \textrm{OR}\ {\it terrorist}) = 
p_0 + p_1 - {\it probability}({\it insider}\ \textrm{AND}\ {\it terrorist}) \approx p_0 + p_1$

With the two-person rule: \\
${\it probability}({\it insider}\ \textrm{OR}\ {\it terrorist}) =
   0 + p_2 - 0 \cdot p_2 = p_2$

The second equation of the first case assumes that the events that a
pilot is an insider and that a terrorist can use the one-person rule to
enter the cockpit are independent. The approximate equality follows
since both $p_0$ and $p_1$ are very small, that is, the size of
$p_0\cdot p_1$ is negligible compared to either $p_0$ or $p_1$. In the
second case it is assumed that the probability that an insider can
harm the plane if not on their own is $0$.

In order to follow a rational policy, an airline should look at the
relationship of the probabilities in the two cases, that is, between
$p_2$ and $p_0 + p_1$. It should go for the smaller probability. If
the probability of a terrorist getting in following the two-person rule
is greater than that of getting in following the one-person rule plus
the probability of an insider doing harm then follow the one-person
rule, else the two-person rule.

However, as we have mentioned above it is very difficult to determine
these probabilities. Hence, when it comes to defining policies, it
looks much more fruitful to consider possibilistic specifications of
systems, actors, and their possible behaviours in order to understand
better the shortcomings and possible glitches when imposing policies
as security rules than to apply probabilistic reasoning.

\subsection{Advantages and drawbacks of the approach}

While the detection of attacks is a very useful feature, the use of a 
heavier, that is, a more labour intensive analysis, like interactive theorem
proving with Isabelle may seem to be an academic exercise. Particularly in the 
light of related logical analysis techniques like model checking or SMT \new{(Satisfiability Modulo Theory)} solving,
the interaction might appear like an unnecessary limitation. However, as the 
foundations of  logic and computation theory teach us, properties 
may become undecidable as soon as higher order elements are in the models.
And this is the case when we want to express policies over infrastructures,
and prove properties that often necessitate proofs of invariants which can
only be proved by induction.

Invalidation of policies of infrastructures to detect insider threats
\cite{kp:13} uses model checking to discover paths to system states in which
the security policy is violated. 
However, the restrictions on the description of 
infrastructure models in model checkers renders them insufficient for our purposes:
we need to consider a variety of actors and restrictions like the number
of people in locations and changing configurations created by actors moving about
between them. 
Model checking explores the entire state space of systems for all possible 
instantiations of all state variables. This process -- if implemented as
a decision procedure -- requires finite models and is exponential in 
the number of state components -- a problem known as state explosion. Due to 
the resulting restrictions on the state specification it is not possible to 
use general arithmetic expressions -- for example using state variables over infinite 
data domains like $x < 2$ for $x$ being an integer -- nor to describe security policies 
using higher order predicates -- for example using expressions like ``$a$ is at location $l$'' 
as an input for a graph based policy model generalizing over actors $a$ and locations $l$ 
of an infrastructure.
Similarly, SMT solvers use a complete enumeration of all 
possible interpretations of logical formulas and satisfiability checking \new{has only recently been extended to higher order logic in an efficient way~\cite{barbosa:19}. The application of model checkers would require to apply abstraction and considerable work would need to be done to find a suitable level of abstraction. The formalization in the rich language of Isabelle/HOL looks cognitively more adequate and allows to more easily experiment with different policies.}


\subsection{Summary of Contributions}
The current paper presents a 
complete formalization and analysis of 
preliminary work previously presented as a workshop paper \cite{kk:16} 
on examining insider attacks on airplanes. The main improvements and 
additional contributions over this and other previous works with the 
Isabelle Insider framework are:
\begin{itemize}
\item We have improved the Isabelle Insider framework by integrating the
  credentials, roles, and location state into the infra\-structure graph.
  This is necessary when using infra\-structures as states in Kripke structures
  but also generally improves the infrastructure model.
\item We have identified a crucial implicit condition intentional in the 
  two-person rule formalizing it as \texttt{foe\_control} in our model.
\item We have shown for the first time how invariant reasoning and induction
   can be used to prove that a global policy holds over a Kripke structure
   in the Isabelle Insider framework. By using an instance of 
   \texttt{foe\_control} we showed that the two-person rule provides 
   insider security.
\item We have summarized the procedure as an informal methodology.
\end{itemize}


Isabelle and other HOL tools support a rich set of type definitions and
inductive predicates. This work has shown the benefits of using these
definition tools as a natural match for concepts in the application.
Without such well-founded definitions, proof rules that are used on
features of a model cannot be considered as mathematically sound.
Datatypes and induction on predicates are derived from first principles
like fixpoint induction and datatype isomorphism in HOL. 
This is known as the principle of conservative extension. It is this principle
that adds a special quality of mathematical soundness to Isabelle formalizations.

The complexity of the application domain of infrastructures including 
actors and policies necessitated the use of higher order functions to 
represent policies. We have illustrated this necessity by showing some
meta-level invariants for the insider framework.
The proof of invariants needs induction.


\new{As discussed in Section \ref{sec:related}, the Isabelle Insider framework has been 
initially designed and validated on the insider threat patterns identified by 
CERT \cite{cmt:12}.
The present application of the Isabelle Insider framework is based on the same insider model 
but greatly enhances it by the generic state transition model based on Kripke structures
and CTL. The definition of the airplane application  uses earlier insider
applications as a blueprint. Hence basic proofs can be reused.
The current application additionally provides reusable proofs at the level
of the insider theory itself (for example, preservation of the local policies by
the state transition) and shows how proofs about the dynamic behaviour of the application
are conducted. This can similarly inspire future applications allowing reusability of the
Isabelle Insider framework. }

\subsection{Conclusions}
The current work has picked up on the challenging earlier application \cite{kk:16}
on investigating airplane safety and security in the presence of insiders.
We have successfully proved the major observation of that earlier paper: a
thorough logical analysis of the airplane scenario requires the exploration
of the state space for all possible changes to the state. Integrating the extensions
to Kripke structures and CTL in our model we were now able to explore the
airplane scenario thoroughly and completely. The analysis in the interactive 
theorem prover Isabelle has shown that earlier results were partly misleading 
because security results were only relating statically to one specific state 
at a time. 
In the current version, the use of an inductive state transition relation enables us
to prove invariants and most prominently revealed a missing assumption 
when clarifying the policy specification.

\new{This shows that a rigorous validation as part of the process in the
development of new airplane policy is very import.}

\new{Finally, we were able to establish the proof 
of the global security property in presence of an insider.} As a by-product, the 
extensive study has provided general improvements to the Isabelle Insider framework.

\mycomment{
In section~\ref{sec:air}, we have also discussed how the policy
makers's decisions should rationally follow from the relationships
between the probabilities of attacks, which depend on the policies
adopted.
}

\makeatletter
\g@addto@macro{\UrlBreaks}{\UrlOrds}
\g@addto@macro{\UrlBreaks}{%
\do\/\do\d%
}
\makeatother

\appendix
\color{blue}
\section{Isabelle Code Extracts}
\newcommand\prf{{$\langle$proof$\rangle$}}
This section contains a subset of the Isabelle formalization of the 
Insider framework and the airplane case study showing
all relevant definitions, most interesting lemmas and theorems without proofs
(proofs are replaced by the tag {\it \prf}), and some proof examples. The following code
has been abridged from the latex generated  from the Isabelle sources 
available online \cite{kam:19isa}. In this repository there 
is also a directory \texttt{latex} that contains the latex-generated pdf 
outputs of the formalization in full (\texttt{document.pdf}, 61 pages) as well as the 
outline (\texttt{outline.pdf}, 25 pages).

\subsection{Kripke Structures and CTL}
\urlstyle{rm}
\isabellestyle{it}
\parindent 0pt\parskip 0.5ex
\begin{isabellebody}%
\setisabellecontext{MC}%
\isadelimtheory
\endisadelimtheory
\isatagtheory
\isacommand{theory}\isamarkupfalse%
\ MC\ \isanewline
\isakeyword{imports}\ Main\isanewline
\isakeyword{begin}%
\endisatagtheory
{\isafoldtheory}%
\isadelimtheory
\isanewline
\endisadelimtheory
\isanewline
\isacommand{definition}\isamarkupfalse%
\ monotone\ {\isacharcolon}{\isacharcolon}\ {\isachardoublequoteopen}{\isacharparenleft}{\isacharprime}a\ set\ {\isasymRightarrow}\ {\isacharprime}a\ set{\isacharparenright}\ {\isasymRightarrow}\ bool{\isachardoublequoteclose}\isanewline
\isakeyword{where}\ {\isachardoublequoteopen}monotone\ {\isasymtau}\ {\isasymequiv}\ {\isacharparenleft}{\isasymforall}\ p\ q{\isachardot}\ p\ {\isasymsubseteq}\ q\ {\isasymlongrightarrow}\ {\isasymtau}\ p\ {\isasymsubseteq}\ {\isasymtau}\ q\ {\isacharparenright}{\isachardoublequoteclose}\isanewline
\isanewline
\isacommand{lemma}\isamarkupfalse%
\ monotoneE{\isacharcolon}\ {\isachardoublequoteopen}monotone\ {\isasymtau}\ {\isasymLongrightarrow}\ p\ {\isasymsubseteq}\ q\ {\isasymLongrightarrow}\ {\isasymtau}\ p\ {\isasymsubseteq}\ {\isasymtau}\ q{\isachardoublequoteclose}
\isanewline\prf\isanewline\isanewline
\isacommand{lemma}\isamarkupfalse%
\ lfp{\isadigit{1}}{\isacharcolon}\ {\isachardoublequoteopen}monotone\ {\isasymtau}\ {\isasymlongrightarrow}\ {\isacharparenleft}lfp\ {\isasymtau}\ {\isacharequal}\ {\isasymInter}\ {\isacharbraceleft}Z{\isachardot}\ {\isasymtau}\ Z\ {\isasymsubseteq}\ Z{\isacharbraceright}{\isacharparenright}{\isachardoublequoteclose}
\isanewline\prf\isanewline\isanewline
\isacommand{lemma}\isamarkupfalse%
\ gfp{\isadigit{1}}{\isacharcolon}\ {\isachardoublequoteopen}monotone\ {\isasymtau}\ {\isasymlongrightarrow}\ {\isacharparenleft}gfp\ {\isasymtau}\ {\isacharequal}\ {\isasymUnion}\ {\isacharbraceleft}Z{\isachardot}\ Z\ {\isasymsubseteq}\ {\isasymtau}\ Z{\isacharbraceright}{\isacharparenright}{\isachardoublequoteclose}\isanewline\prf
\isanewline\isanewline
\isacommand{primrec}\isamarkupfalse%
\ power\ {\isacharcolon}{\isacharcolon}\ {\isachardoublequoteopen}{\isacharbrackleft}{\isacharprime}a\ {\isasymRightarrow}\ {\isacharprime}a{\isacharcomma}\ nat{\isacharbrackright}\ {\isasymRightarrow}\ {\isacharparenleft}{\isacharprime}a\ {\isasymRightarrow}\ {\isacharprime}a{\isacharparenright}{\isachardoublequoteclose}\ {\isacharparenleft}{\isachardoublequoteopen}{\isacharparenleft}{\isacharunderscore}\ {\isacharcircum}\ {\isacharunderscore}{\isacharparenright}{\isachardoublequoteclose}\ {\isadigit{4}}{\isadigit{0}}{\isacharparenright}\isanewline
\isakeyword{where}\ \isanewline
power{\isacharunderscore}zero{\isacharcolon}\ {\isachardoublequoteopen}{\isacharparenleft}f\ {\isacharcircum}\ {\isadigit{0}}{\isacharparenright}\ {\isacharequal}\ {\isacharparenleft}{\isasymlambda}\ x{\isachardot}\ x{\isacharparenright}{\isachardoublequoteclose}\ {\isacharbar}\isanewline
power{\isacharunderscore}suc{\isacharcolon}\ {\isachardoublequoteopen}{\isacharparenleft}f\ {\isacharcircum}\ {\isacharparenleft}Suc\ n{\isacharparenright}{\isacharparenright}\ {\isacharequal}\ {\isacharparenleft}f\ o\ {\isacharparenleft}f\ {\isacharcircum}\ n{\isacharparenright}{\isacharparenright}{\isachardoublequoteclose}\isanewline
\isanewline
\isacommand{lemma}\isamarkupfalse%
\ predtrans{\isacharunderscore}empty{\isacharcolon}\ \isanewline
\ \ \isakeyword{assumes}\ {\isachardoublequoteopen}monotone\ {\isasymtau}{\isachardoublequoteclose}\isanewline
\ \ \isakeyword{shows}\ {\isachardoublequoteopen}{\isasymforall}\ i{\isachardot}\ {\isacharparenleft}{\isasymtau}\ {\isacharcircum}\ i{\isacharparenright}\ {\isacharparenleft}{\isacharbraceleft}{\isacharbraceright}{\isacharparenright}\ {\isasymsubseteq}\ {\isacharparenleft}{\isasymtau}\ {\isacharcircum}{\isacharparenleft}i\ {\isacharplus}\ {\isadigit{1}}{\isacharparenright}{\isacharparenright}{\isacharparenleft}{\isacharbraceleft}{\isacharbraceright}{\isacharparenright}{\isachardoublequoteclose}\isanewline
\isadelimproof
\endisadelimproof
\isatagproof
\isacommand{proof}\isamarkupfalse%
\ {\isacharparenleft}rule\ allI{\isacharcomma}\ induct{\isacharunderscore}tac\ i{\isacharparenright}\isanewline
\ \ \isacommand{show}\isamarkupfalse%
\ {\isachardoublequoteopen}{\isacharparenleft}{\isasymtau}\ {\isacharcircum}\ {\isadigit{0}}{\isacharcolon}{\isacharcolon}nat{\isacharparenright}\ {\isacharbraceleft}{\isacharbraceright}\ {\isasymsubseteq}\ {\isacharparenleft}{\isasymtau}\ {\isacharcircum}\ {\isacharparenleft}{\isadigit{0}}{\isacharcolon}{\isacharcolon}nat{\isacharparenright}\ {\isacharplus}\ {\isacharparenleft}{\isadigit{1}}{\isacharcolon}{\isacharcolon}nat{\isacharparenright}{\isacharparenright}\ {\isacharbraceleft}{\isacharbraceright}{\isachardoublequoteclose}\ \isacommand{by}\isamarkupfalse%
\ simp\isanewline
\isacommand{next}\isamarkupfalse%
\ \isacommand{show}\isamarkupfalse%
\ {\isachardoublequoteopen}{\isasymAnd}{\isacharparenleft}i{\isacharcolon}{\isacharcolon}nat{\isacharparenright}\ n{\isacharcolon}{\isacharcolon}nat{\isachardot}\ {\isacharparenleft}{\isasymtau}\ {\isacharcircum}\ n{\isacharparenright}\ {\isacharbraceleft}{\isacharbraceright}\ {\isasymsubseteq}\ {\isacharparenleft}{\isasymtau}\ {\isacharcircum}\ n\ {\isacharplus}\ {\isacharparenleft}{\isadigit{1}}{\isacharcolon}{\isacharcolon}nat{\isacharparenright}{\isacharparenright}\ {\isacharbraceleft}{\isacharbraceright}\ \isanewline
\ \ \ \ \ \ {\isasymLongrightarrow}\ {\isacharparenleft}{\isasymtau}\ {\isacharcircum}\ Suc\ n{\isacharparenright}\ {\isacharbraceleft}{\isacharbraceright}\ {\isasymsubseteq}\ {\isacharparenleft}{\isasymtau}\ {\isacharcircum}\ Suc\ n\ {\isacharplus}\ {\isacharparenleft}{\isadigit{1}}{\isacharcolon}{\isacharcolon}nat{\isacharparenright}{\isacharparenright}\ {\isacharbraceleft}{\isacharbraceright}{\isachardoublequoteclose}\isanewline
\ \ \isacommand{proof}\isamarkupfalse%
\ {\isacharminus}\isanewline
\ \ \ \ \isacommand{fix}\isamarkupfalse%
\ i\ n\isanewline
\ \ \ \ \isacommand{assume}\isamarkupfalse%
\ a\ {\isacharcolon}\ {\isachardoublequoteopen}\ {\isacharparenleft}{\isasymtau}\ {\isacharcircum}\ n{\isacharparenright}\ {\isacharbraceleft}{\isacharbraceright}\ {\isasymsubseteq}\ {\isacharparenleft}{\isasymtau}\ {\isacharcircum}\ n\ {\isacharplus}\ {\isacharparenleft}{\isadigit{1}}{\isacharcolon}{\isacharcolon}nat{\isacharparenright}{\isacharparenright}\ {\isacharbraceleft}{\isacharbraceright}{\isachardoublequoteclose}\isanewline
\ \ \ \ \isacommand{have}\isamarkupfalse%
\ {\isachardoublequoteopen}{\isacharparenleft}{\isasymtau}\ {\isacharparenleft}{\isacharparenleft}{\isasymtau}\ {\isacharcircum}\ n{\isacharparenright}\ {\isacharbraceleft}{\isacharbraceright}{\isacharparenright}{\isacharparenright}\ {\isasymsubseteq}\ {\isacharparenleft}{\isasymtau}\ {\isacharparenleft}{\isacharparenleft}{\isasymtau}\ {\isacharcircum}\ {\isacharparenleft}n\ {\isacharplus}\ {\isacharparenleft}{\isadigit{1}}\ {\isacharcolon}{\isacharcolon}\ nat{\isacharparenright}{\isacharparenright}{\isacharparenright}\ {\isacharbraceleft}{\isacharbraceright}{\isacharparenright}{\isacharparenright}{\isachardoublequoteclose}\ \isacommand{using}\isamarkupfalse%
\ assms\isanewline
\ \ \ \ \ \ \isacommand{apply}\isamarkupfalse%
\ {\isacharparenleft}rule\ monotoneE{\isacharparenright}\isanewline
\ \ \ \ \ \ \isacommand{by}\isamarkupfalse%
\ {\isacharparenleft}rule\ a{\isacharparenright}\isanewline
\ \ \ \ \isacommand{thus}\isamarkupfalse%
\ {\isachardoublequoteopen}{\isacharparenleft}{\isasymtau}\ {\isacharcircum}\ Suc\ n{\isacharparenright}\ {\isacharbraceleft}{\isacharbraceright}\ {\isasymsubseteq}\ {\isacharparenleft}{\isasymtau}\ {\isacharcircum}\ Suc\ n\ {\isacharplus}\ {\isacharparenleft}{\isadigit{1}}{\isacharcolon}{\isacharcolon}nat{\isacharparenright}{\isacharparenright}\ {\isacharbraceleft}{\isacharbraceright}{\isachardoublequoteclose}\ \isacommand{by}\isamarkupfalse%
\ simp\isanewline
\ \ \isacommand{qed}\isamarkupfalse%
\isanewline
\isacommand{qed}\isamarkupfalse%
\endisatagproof
{\isafoldproof}%
\isadelimproof
\isanewline
\endisadelimproof
\isanewline
\isacommand{lemma}\isamarkupfalse%
\ infchain{\isacharunderscore}outruns{\isacharunderscore}all{\isacharcolon}\ \isanewline
\ \ \isakeyword{assumes}\ {\isachardoublequoteopen}finite\ {\isacharparenleft}UNIV\ {\isacharcolon}{\isacharcolon}\ {\isacharprime}a\ set{\isacharparenright}{\isachardoublequoteclose}\ \isanewline
\ \ \ \ \isakeyword{and}\ {\isachardoublequoteopen}{\isasymforall}i\ {\isacharcolon}{\isacharcolon}\ nat{\isachardot}\ {\isacharparenleft}{\isasymtau}\ {\isacharcircum}\ i{\isacharparenright}\ {\isacharparenleft}{\isacharbraceleft}{\isacharbraceright}{\isacharcolon}{\isacharcolon}\ {\isacharprime}a\ set{\isacharparenright}\ {\isasymsubset}\ {\isacharparenleft}{\isasymtau}\ {\isacharcircum}\ i\ {\isacharplus}\ {\isacharparenleft}{\isadigit{1}}\ {\isacharcolon}{\isacharcolon}\ nat{\isacharparenright}{\isacharparenright}\ {\isacharbraceleft}{\isacharbraceright}{\isachardoublequoteclose}\isanewline
\ \ \isakeyword{shows}\ {\isachardoublequoteopen}{\isasymforall}j\ {\isacharcolon}{\isacharcolon}\ nat{\isachardot}\ {\isasymexists}i\ {\isacharcolon}{\isacharcolon}\ nat{\isachardot}\ j\ {\isacharless}\ card\ {\isacharparenleft}{\isacharparenleft}{\isasymtau}\ {\isacharcircum}\ i{\isacharparenright}\ {\isacharbraceleft}{\isacharbraceright}{\isacharparenright}{\isachardoublequoteclose}\isanewline
\prf\isanewline\isanewline
\isacommand{lemma}\isamarkupfalse%
\ no{\isacharunderscore}infinite{\isacharunderscore}subset{\isacharunderscore}chain{\isacharcolon}\ \isanewline
\ \ \ \isakeyword{assumes}\ {\isachardoublequoteopen}finite\ {\isacharparenleft}UNIV\ {\isacharcolon}{\isacharcolon}\ {\isacharprime}a\ set{\isacharparenright}{\isachardoublequoteclose}\isanewline
\ \ \ \ \isakeyword{and}\ \ \ \ {\isachardoublequoteopen}monotone\ {\isacharparenleft}{\isasymtau}\ {\isacharcolon}{\isacharcolon}\ {\isacharparenleft}{\isacharprime}a\ set\ {\isasymRightarrow}\ {\isacharprime}a\ set{\isacharparenright}{\isacharparenright}{\isachardoublequoteclose}\isanewline
\ \ \ \ \isakeyword{and}\ \ \ \ {\isachardoublequoteopen}{\isasymforall}i\ {\isacharcolon}{\isacharcolon}\ nat{\isachardot}\ {\isacharparenleft}{\isacharparenleft}{\isasymtau}\ {\isacharcolon}{\isacharcolon}\ {\isacharprime}a\ set\ {\isasymRightarrow}\ {\isacharprime}a\ set{\isacharparenright}\ {\isacharcircum}\ i{\isacharparenright}\ {\isacharbraceleft}{\isacharbraceright}\ {\isasymsubset}\ {\isacharparenleft}{\isasymtau}\ {\isacharcircum}\ i\ {\isacharplus}\ {\isacharparenleft}{\isadigit{1}}\ {\isacharcolon}{\isacharcolon}\ nat{\isacharparenright}{\isacharparenright}\ {\isacharparenleft}{\isacharbraceleft}{\isacharbraceright}\ {\isacharcolon}{\isacharcolon}\ {\isacharprime}a\ set{\isacharparenright}{\isachardoublequoteclose}\ \isanewline
\ \ \isakeyword{shows}\ \ \ {\isachardoublequoteopen}False{\isachardoublequoteclose}\isanewline
\prf\isanewline\isanewline
\isacommand{lemma}\isamarkupfalse%
\ finite{\isacharunderscore}fixp{\isacharcolon}\ \isanewline
\ \ \isakeyword{assumes}\ {\isachardoublequoteopen}finite{\isacharparenleft}UNIV\ {\isacharcolon}{\isacharcolon}\ {\isacharprime}a\ set{\isacharparenright}{\isachardoublequoteclose}\ \isanewline
\ \ \ \ \ \ \isakeyword{and}\ {\isachardoublequoteopen}monotone\ {\isacharparenleft}{\isasymtau}\ {\isacharcolon}{\isacharcolon}\ {\isacharparenleft}{\isacharprime}a\ set\ {\isasymRightarrow}\ {\isacharprime}a\ set{\isacharparenright}{\isacharparenright}{\isachardoublequoteclose}\isanewline
\ \ \ \ \isakeyword{shows}\ {\isachardoublequoteopen}{\isasymexists}\ i{\isachardot}\ {\isacharparenleft}{\isasymtau}\ {\isacharcircum}\ i{\isacharparenright}\ {\isacharparenleft}{\isacharbraceleft}{\isacharbraceright}{\isacharparenright}\ {\isacharequal}\ {\isacharparenleft}{\isasymtau}\ {\isacharcircum}{\isacharparenleft}i\ {\isacharplus}\ {\isadigit{1}}{\isacharparenright}{\isacharparenright}{\isacharparenleft}{\isacharbraceleft}{\isacharbraceright}{\isacharparenright}{\isachardoublequoteclose}\isanewline
\prf\isanewline\isanewline
\isacommand{lemma}\isamarkupfalse%
\ predtrans{\isacharunderscore}UNIV{\isacharcolon}\ \isanewline
\ \ \isakeyword{assumes}\ {\isachardoublequoteopen}monotone\ {\isasymtau}{\isachardoublequoteclose}\isanewline
\ \ \isakeyword{shows}\ {\isachardoublequoteopen}{\isasymforall}\ i{\isachardot}\ {\isacharparenleft}{\isasymtau}\ {\isacharcircum}\ i{\isacharparenright}\ {\isacharparenleft}UNIV{\isacharparenright}\ {\isasymsupseteq}\ {\isacharparenleft}{\isasymtau}\ {\isacharcircum}{\isacharparenleft}i\ {\isacharplus}\ {\isadigit{1}}{\isacharparenright}{\isacharparenright}{\isacharparenleft}UNIV{\isacharparenright}{\isachardoublequoteclose}\isanewline
\prf\isanewline\isanewline
\isacommand{lemma}\isamarkupfalse%
\ down{\isacharunderscore}chain{\isacharunderscore}reaches{\isacharunderscore}empty{\isacharcolon}\isanewline
\ \ \isakeyword{assumes}\ {\isachardoublequoteopen}finite\ {\isacharparenleft}UNIV\ {\isacharcolon}{\isacharcolon}\ {\isacharprime}a\ set{\isacharparenright}{\isachardoublequoteclose}\ \isakeyword{and}\ {\isachardoublequoteopen}monotone\ {\isacharparenleft}{\isasymtau}\ {\isacharcolon}{\isacharcolon}\ {\isacharprime}a\ set\ {\isasymRightarrow}\ {\isacharprime}a\ set{\isacharparenright}{\isachardoublequoteclose}\isanewline
\ \ \ \isakeyword{and}\ {\isachardoublequoteopen}{\isacharparenleft}{\isasymforall}i\ {\isacharcolon}{\isacharcolon}\ nat{\isachardot}\ {\isacharparenleft}{\isacharparenleft}{\isasymtau}\ {\isacharcolon}{\isacharcolon}\ {\isacharprime}a\ set\ {\isasymRightarrow}\ {\isacharprime}a\ set{\isacharparenright}\ {\isacharcircum}\ i\ {\isacharplus}\ {\isacharparenleft}{\isadigit{1}}\ {\isacharcolon}{\isacharcolon}\ nat{\isacharparenright}{\isacharparenright}\ UNIV\ {\isasymsubset}\ {\isacharparenleft}{\isasymtau}\ {\isacharcircum}\ i{\isacharparenright}\ UNIV{\isacharparenright}{\isachardoublequoteclose}\isanewline
\ \isakeyword{shows}\ {\isachardoublequoteopen}{\isasymexists}\ {\isacharparenleft}j\ {\isacharcolon}{\isacharcolon}\ nat{\isacharparenright}{\isachardot}\ {\isacharparenleft}{\isasymtau}\ {\isacharcircum}\ j{\isacharparenright}\ UNIV\ {\isacharequal}\ {\isacharbraceleft}{\isacharbraceright}{\isachardoublequoteclose}\isanewline
\prf\isanewline\isanewline
\isacommand{lemma}\isamarkupfalse%
\ lfp{\isacharunderscore}loop{\isacharcolon}\ \isanewline
\ \ \isakeyword{assumes}\ {\isachardoublequoteopen}finite\ {\isacharparenleft}UNIV\ {\isacharcolon}{\isacharcolon}\ {\isacharprime}b\ set{\isacharparenright}{\isachardoublequoteclose}\ \isakeyword{and}\ {\isachardoublequoteopen}monotone\ {\isacharparenleft}{\isasymtau}\ {\isacharcolon}{\isacharcolon}\ {\isacharparenleft}{\isacharprime}b\ set\ {\isasymRightarrow}\ {\isacharprime}b\ set{\isacharparenright}{\isacharparenright}{\isachardoublequoteclose}\isanewline
\ \ \isakeyword{shows}\ {\isachardoublequoteopen}{\isasymexists}\ n\ {\isachardot}\ lfp\ {\isasymtau}\ \ {\isacharequal}\ {\isacharparenleft}{\isasymtau}\ {\isacharcircum}\ n{\isacharparenright}\ {\isacharbraceleft}{\isacharbraceright}{\isachardoublequoteclose}\isanewline
\prf\isanewline
\isanewline
\isacommand{lemma}\isamarkupfalse%
\ gfp{\isacharunderscore}loop{\isacharcolon}\ \isanewline
\ \ \isakeyword{assumes}\ {\isachardoublequoteopen}finite\ {\isacharparenleft}UNIV\ {\isacharcolon}{\isacharcolon}\ {\isacharprime}b\ set{\isacharparenright}{\isachardoublequoteclose}\isanewline
\ \ \ \isakeyword{and}\ {\isachardoublequoteopen}monotone\ {\isacharparenleft}{\isasymtau}\ {\isacharcolon}{\isacharcolon}\ {\isacharparenleft}{\isacharprime}b\ set\ {\isasymRightarrow}\ {\isacharprime}b\ set{\isacharparenright}{\isacharparenright}{\isachardoublequoteclose}\isanewline
\ \ \ \ \isakeyword{shows}\ {\isachardoublequoteopen}{\isasymexists}\ n\ {\isachardot}\ gfp\ {\isasymtau}\ \ {\isacharequal}\ {\isacharparenleft}{\isasymtau}\ {\isacharcircum}\ n{\isacharparenright}{\isacharparenleft}UNIV\ {\isacharcolon}{\isacharcolon}\ {\isacharprime}b\ set{\isacharparenright}{\isachardoublequoteclose}\isanewline
\prf\isanewline
\isanewline
\isacommand{class}\isamarkupfalse%
\ state\ {\isacharequal}\ \isanewline
\ \ \isakeyword{fixes}\ state{\isacharunderscore}transition\ {\isacharcolon}{\isacharcolon}\ {\isachardoublequoteopen}{\isacharbrackleft}{\isacharprime}a\ {\isacharcolon}{\isacharcolon}\ type{\isacharcomma}\ {\isacharprime}a{\isacharbrackright}\ {\isasymRightarrow}\ bool{\isachardoublequoteclose}\ \ {\isacharparenleft}{\isachardoublequoteopen}{\isacharparenleft}{\isacharunderscore}\ {\isasymrightarrow}\isactrlsub i\ {\isacharunderscore}{\isacharparenright}{\isachardoublequoteclose}\ {\isadigit{5}}{\isadigit{0}}{\isacharparenright}\isanewline
\ \ \ \ \isanewline
\isacommand{definition}\isamarkupfalse%
\ AX\ \isakeyword{where}\ {\isachardoublequoteopen}AX\ f\ {\isasymequiv}\ {\isacharbraceleft}s{\isachardot}\ {\isacharbraceleft}f{\isadigit{0}}{\isachardot}\ s\ {\isasymrightarrow}\isactrlsub i\ f{\isadigit{0}}{\isacharbraceright}\ {\isasymsubseteq}\ f{\isacharbraceright}{\isachardoublequoteclose}\isanewline
\isacommand{definition}\isamarkupfalse%
\ EX{\isacharprime}\ \isakeyword{where}\ {\isachardoublequoteopen}EX{\isacharprime}\ f\ {\isasymequiv}\ {\isacharbraceleft}s\ {\isachardot}\ {\isasymexists}\ f{\isadigit{0}}\ {\isasymin}\ f{\isachardot}\ s\ {\isasymrightarrow}\isactrlsub i\ f{\isadigit{0}}\ {\isacharbraceright}{\isachardoublequoteclose}\isanewline
\isanewline
\isacommand{definition}\isamarkupfalse%
\ AF\ \isakeyword{where}\ {\isachardoublequoteopen}AF\ f\ {\isasymequiv}\ lfp\ {\isacharparenleft}{\isasymlambda}\ Z{\isachardot}\ f\ {\isasymunion}\ AX\ Z{\isacharparenright}{\isachardoublequoteclose}\isanewline
\isacommand{definition}\isamarkupfalse%
\ EF\ \isakeyword{where}\ {\isachardoublequoteopen}EF\ f\ {\isasymequiv}\ lfp\ {\isacharparenleft}{\isasymlambda}\ Z{\isachardot}\ f\ {\isasymunion}\ EX{\isacharprime}\ Z{\isacharparenright}{\isachardoublequoteclose}\isanewline
\isacommand{definition}\isamarkupfalse%
\ AG\ \isakeyword{where}\ {\isachardoublequoteopen}AG\ f\ {\isasymequiv}\ gfp\ {\isacharparenleft}{\isasymlambda}\ Z{\isachardot}\ f\ {\isasyminter}\ AX\ Z{\isacharparenright}{\isachardoublequoteclose}\isanewline
\isacommand{definition}\isamarkupfalse%
\ EG\ \isakeyword{where}\ {\isachardoublequoteopen}EG\ f\ {\isasymequiv}\ gfp\ {\isacharparenleft}{\isasymlambda}\ Z{\isachardot}\ f\ {\isasyminter}\ EX{\isacharprime}\ Z{\isacharparenright}{\isachardoublequoteclose}\isanewline
\isacommand{definition}\isamarkupfalse%
\ AU\ \isakeyword{where}\ {\isachardoublequoteopen}AU\ f{\isadigit{1}}\ f{\isadigit{2}}\ {\isasymequiv}\ lfp{\isacharparenleft}{\isasymlambda}\ Z{\isachardot}\ f{\isadigit{2}}\ {\isasymunion}\ {\isacharparenleft}f{\isadigit{1}}\ {\isasyminter}\ AX\ Z{\isacharparenright}{\isacharparenright}{\isachardoublequoteclose}\isanewline
\isacommand{definition}\isamarkupfalse%
\ EU\ \isakeyword{where}\ {\isachardoublequoteopen}EU\ f{\isadigit{1}}\ f{\isadigit{2}}\ {\isasymequiv}\ lfp{\isacharparenleft}{\isasymlambda}\ Z{\isachardot}\ f{\isadigit{2}}\ {\isasymunion}\ {\isacharparenleft}f{\isadigit{1}}\ {\isasyminter}\ EX{\isacharprime}\ Z{\isacharparenright}{\isacharparenright}{\isachardoublequoteclose}\isanewline
\isacommand{definition}\isamarkupfalse%
\ AR\ \isakeyword{where}\ {\isachardoublequoteopen}AR\ f{\isadigit{1}}\ f{\isadigit{2}}\ {\isasymequiv}\ gfp{\isacharparenleft}{\isasymlambda}\ Z{\isachardot}\ f{\isadigit{2}}\ {\isasyminter}\ {\isacharparenleft}f{\isadigit{1}}\ {\isasymunion}\ AX\ Z{\isacharparenright}{\isacharparenright}{\isachardoublequoteclose}\isanewline
\isacommand{definition}\isamarkupfalse%
\ ER\ \isakeyword{where}\ {\isachardoublequoteopen}ER\ f{\isadigit{1}}\ f{\isadigit{2}}\ {\isasymequiv}\ gfp{\isacharparenleft}{\isasymlambda}\ Z{\isachardot}\ f{\isadigit{2}}\ {\isasyminter}\ {\isacharparenleft}f{\isadigit{1}}\ {\isasymunion}\ EX{\isacharprime}\ Z{\isacharparenright}{\isacharparenright}{\isachardoublequoteclose}\isanewline
\isanewline
\isacommand{datatype}\isamarkupfalse%
\ {\isacharprime}a\ kripke\ {\isacharequal}\ 
\ \ Kripke\ {\isachardoublequoteopen}{\isacharprime}a\ set{\isachardoublequoteclose}\ {\isachardoublequoteopen}{\isacharprime}a\ set{\isachardoublequoteclose}\isanewline
\isanewline
\isacommand{primrec}\isamarkupfalse%
\ states\ \isakeyword{where}\ {\isachardoublequoteopen}states\ {\isacharparenleft}Kripke\ S\ I{\isacharparenright}\ {\isacharequal}\ S{\isachardoublequoteclose}\ \isanewline
\isacommand{primrec}\isamarkupfalse%
\ init\ \isakeyword{where}\ {\isachardoublequoteopen}init\ {\isacharparenleft}Kripke\ S\ I{\isacharparenright}\ {\isacharequal}\ I{\isachardoublequoteclose}\ \isanewline
\isanewline
\isacommand{definition}\isamarkupfalse%
\ check\ {\isacharparenleft}{\isachardoublequoteopen}{\isacharunderscore}\ {\isasymturnstile}\ {\isacharunderscore}{\isachardoublequoteclose}\ {\isadigit{5}}{\isadigit{0}}{\isacharparenright}\isanewline
\ \isakeyword{where}\ {\isachardoublequoteopen}M\ {\isasymturnstile}\ f\ {\isasymequiv}\ {\isacharparenleft}init\ M{\isacharparenright}\ {\isasymsubseteq}\ {\isacharbraceleft}s\ {\isasymin}\ {\isacharparenleft}states\ M{\isacharparenright}{\isachardot}\ s\ {\isasymin}\ f\ {\isacharbraceright}{\isachardoublequoteclose}\isanewline
\isanewline
\isacommand{definition}\isamarkupfalse%
\ state{\isacharunderscore}transition{\isacharunderscore}refl\ {\isacharparenleft}{\isachardoublequoteopen}{\isacharparenleft}{\isacharunderscore}\ {\isasymrightarrow}\isactrlsub i{\isacharasterisk}\ {\isacharunderscore}{\isacharparenright}{\isachardoublequoteclose}\ {\isadigit{5}}{\isadigit{0}}{\isacharparenright}\isanewline
\isakeyword{where}\ {\isachardoublequoteopen}s\ {\isasymrightarrow}\isactrlsub i{\isacharasterisk}\ s{\isacharprime}\ {\isasymequiv}\ {\isacharparenleft}{\isacharparenleft}s{\isacharcomma}s{\isacharprime}{\isacharparenright}\ {\isasymin}\ {\isacharbraceleft}{\isacharparenleft}x{\isacharcomma}y{\isacharparenright}{\isachardot}\ state{\isacharunderscore}transition\ x\ y{\isacharbraceright}\isactrlsup {\isacharasterisk}{\isacharparenright}{\isachardoublequoteclose}\isanewline
\ \ \isanewline
\isacommand{lemma}\isamarkupfalse%
\ EX{\isacharunderscore}step{\isacharcolon}\ \isakeyword{assumes}\ {\isachardoublequoteopen}x\ \ {\isasymrightarrow}\isactrlsub i\ y{\isachardoublequoteclose}\ \isakeyword{and}\ {\isachardoublequoteopen}y\ {\isasymin}\ f{\isachardoublequoteclose}\ \isakeyword{shows}\ {\isachardoublequoteopen}x\ {\isasymin}\ EX{\isacharprime}\ f{\isachardoublequoteclose}\isanewline
\prf\isanewline\isanewline
\isacommand{lemma}\isamarkupfalse%
\ EF{\isacharunderscore}step{\isacharcolon}\ \isakeyword{assumes}\ {\isachardoublequoteopen}x\ \ {\isasymrightarrow}\isactrlsub i\ y{\isachardoublequoteclose}\ \isakeyword{and}\ {\isachardoublequoteopen}y\ {\isasymin}\ f{\isachardoublequoteclose}\ \isakeyword{shows}\ {\isachardoublequoteopen}x\ {\isasymin}\ EF\ f{\isachardoublequoteclose}\isanewline
\prf\isanewline\isanewline
\isacommand{lemma}\isamarkupfalse%
\ EF{\isacharunderscore}step{\isacharunderscore}step{\isacharcolon}\ \isakeyword{assumes}\ {\isachardoublequoteopen}x\ \ {\isasymrightarrow}\isactrlsub i\ y{\isachardoublequoteclose}\ \isakeyword{and}\ {\isachardoublequoteopen}y\ {\isasymin}\ EF\ f{\isachardoublequoteclose}\ \isakeyword{shows}\ \ {\isachardoublequoteopen}x\ {\isasymin}\ EF\ f{\isachardoublequoteclose}\isanewline
\prf\isanewline\isanewline
\isacommand{lemma}\isamarkupfalse%
\ EF{\isacharunderscore}step{\isacharunderscore}star{\isacharcolon}\ {\isachardoublequoteopen}{\isasymlbrakk}\ x\ \ {\isasymrightarrow}\isactrlsub i{\isacharasterisk}\ y{\isacharsemicolon}\ y\ {\isasymin}\ f\ {\isasymrbrakk}\ {\isasymLongrightarrow}\ x\ {\isasymin}\ EF\ f{\isachardoublequoteclose}\isanewline
\prf\isanewline\isanewline
\isacommand{lemma}\isamarkupfalse%
\ EF{\isacharunderscore}induct{\isacharcolon}\ {\isachardoublequoteopen}{\isacharparenleft}a{\isacharcolon}{\isacharcolon}{\isacharprime}a{\isacharcolon}{\isacharcolon}state{\isacharparenright}\ {\isasymin}\ EF\ {\isacharparenleft}f\ {\isacharcolon}{\isacharcolon}\ {\isacharprime}a\ {\isacharcolon}{\isacharcolon}\ state\ set{\isacharparenright}\ {\isasymLongrightarrow}\isanewline
\ \ \ \ mono\ \ {\isacharparenleft}{\isasymlambda}\ Z{\isachardot}\ {\isacharparenleft}f{\isacharcolon}{\isacharcolon}{\isacharprime}a{\isacharcolon}{\isacharcolon}state\ set{\isacharparenright}\ {\isasymunion}\ EX{\isacharprime}\ Z{\isacharparenright}\ {\isasymLongrightarrow}\isanewline
\ \ \ \ {\isacharparenleft}{\isasymAnd}x{\isacharcolon}{\isacharcolon}{\isacharprime}a{\isacharcolon}{\isacharcolon}state{\isachardot}\isanewline
\ \ \ \ \ \ \ \ x\ {\isasymin}\ {\isacharparenleft}{\isacharparenleft}{\isasymlambda}\ Z{\isachardot}\ {\isacharparenleft}f{\isacharcolon}{\isacharcolon}{\isacharprime}a{\isacharcolon}{\isacharcolon}state\ set{\isacharparenright}\ {\isasymunion}\ EX{\isacharprime}\ Z{\isacharparenright}{\isacharparenleft}EF\ f\ {\isasyminter}\ {\isacharbraceleft}x{\isacharcolon}{\isacharcolon}{\isacharprime}a{\isacharcolon}{\isacharcolon}state{\isachardot}\ {\isacharparenleft}P{\isacharcolon}{\isacharcolon}{\isacharprime}a{\isacharcolon}{\isacharcolon}state\ {\isasymRightarrow}\ bool{\isacharparenright}\ x{\isacharbraceright}{\isacharparenright}{\isacharparenright}\ {\isasymLongrightarrow}\ P\ x{\isacharparenright}\ {\isasymLongrightarrow}\isanewline
\ \ \ \ P\ a{\isachardoublequoteclose}\isanewline
\prf\isanewline\isanewline
\isacommand{lemma}\isamarkupfalse%
\ EF{\isacharunderscore}step{\isacharunderscore}star{\isacharunderscore}rev{\isacharbrackleft}rule{\isacharunderscore}format{\isacharbrackright}{\isacharcolon}\ {\isachardoublequoteopen}x\ {\isasymin}\ EF\ s\ {\isasymLongrightarrow}\ \ {\isacharparenleft}{\isasymexists}\ y\ {\isasymin}\ s{\isachardot}\ \ x\ \ {\isasymrightarrow}\isactrlsub i{\isacharasterisk}\ y{\isacharparenright}{\isachardoublequoteclose}\isanewline
\prf\isanewline\isanewline
\isacommand{lemma}\isamarkupfalse%
\ EF{\isacharunderscore}step{\isacharunderscore}inv{\isacharcolon}\ {\isachardoublequoteopen}{\isacharparenleft}I\ {\isasymsubseteq}\ {\isacharbraceleft}sa{\isacharcolon}{\isacharcolon}{\isacharprime}s\ {\isacharcolon}{\isacharcolon}\ state{\isachardot}\ {\isacharparenleft}{\isasymexists}i{\isacharcolon}{\isacharcolon}{\isacharprime}s{\isasymin}I{\isachardot}\ i\ {\isasymrightarrow}\isactrlsub i{\isacharasterisk}\ sa{\isacharparenright}\ {\isasymand}\ sa\ {\isasymin}\ EF\ s{\isacharbraceright}{\isacharparenright}\ \ \isanewline
\ \ \ \ \ \ \ \ \ {\isasymLongrightarrow}\ {\isasymforall}\ x\ {\isasymin}\ I{\isachardot}\ {\isasymexists}\ y\ {\isasymin}\ s{\isachardot}\ x\ {\isasymrightarrow}\isactrlsub i{\isacharasterisk}\ y{\isachardoublequoteclose}\isanewline
\prf\isanewline\isanewline
\isacommand{lemma}\isamarkupfalse%
\ AG{\isacharunderscore}in{\isacharunderscore}lem{\isacharcolon}\ \ \ {\isachardoublequoteopen}x\ {\isasymin}\ AG\ s\ {\isasymLongrightarrow}\ x\ {\isasymin}\ s{\isachardoublequoteclose}\ \ \isanewline
\prf\isanewline\isanewline
\isacommand{lemma}\isamarkupfalse%
\ AG{\isacharunderscore}step{\isacharcolon}\ {\isachardoublequoteopen}y\ {\isasymrightarrow}\isactrlsub i\ z\ {\isasymLongrightarrow}\ y\ {\isasymin}\ AG\ s\ {\isasymLongrightarrow}\ z\ {\isasymin}\ AG\ s{\isachardoublequoteclose}\ \ \isanewline
\prf\isanewline\isanewline
\isacommand{lemma}\isamarkupfalse%
\ AG{\isacharunderscore}all{\isacharunderscore}s{\isacharcolon}\ {\isachardoublequoteopen}\ x\ {\isasymrightarrow}\isactrlsub i{\isacharasterisk}\ y\ {\isasymLongrightarrow}\ x\ {\isasymin}\ AG\ s\ {\isasymLongrightarrow}\ y\ {\isasymin}\ AG\ s{\isachardoublequoteclose}\isanewline
\prf\isanewline\isanewline
\isacommand{lemma}\isamarkupfalse%
\ AG{\isacharunderscore}imp{\isacharunderscore}notnotEF{\isacharcolon}\ \isanewline
{\isachardoublequoteopen}I\ {\isasymnoteq}\ {\isacharbraceleft}{\isacharbraceright}\ {\isasymLongrightarrow}\ {\isacharparenleft}{\isacharparenleft}Kripke\ {\isacharbraceleft}s\ {\isacharcolon}{\isacharcolon}\ {\isacharparenleft}{\isacharprime}s\ {\isacharcolon}{\isacharcolon}\ state{\isacharparenright}{\isachardot}\ {\isasymexists}\ i\ {\isasymin}\ I{\isachardot}\ {\isacharparenleft}i\ {\isasymrightarrow}\isactrlsub i{\isacharasterisk}\ s{\isacharparenright}{\isacharbraceright}\ {\isacharparenleft}I\ {\isacharcolon}{\isacharcolon}\ {\isacharparenleft}{\isacharprime}s\ {\isacharcolon}{\isacharcolon}\ state{\isacharparenright}set{\isacharparenright}\ \ {\isasymturnstile}\ AG\ s{\isacharparenright}{\isacharparenright}\ {\isasymLongrightarrow}\ \isanewline
\ {\isacharparenleft}{\isasymnot}{\isacharparenleft}Kripke\ {\isacharbraceleft}s\ {\isacharcolon}{\isacharcolon}\ {\isacharparenleft}{\isacharprime}s\ {\isacharcolon}{\isacharcolon}\ state{\isacharparenright}{\isachardot}\ {\isasymexists}\ i\ {\isasymin}\ I{\isachardot}\ {\isacharparenleft}i\ {\isasymrightarrow}\isactrlsub i{\isacharasterisk}\ s{\isacharparenright}{\isacharbraceright}\ {\isacharparenleft}I\ {\isacharcolon}{\isacharcolon}\ {\isacharparenleft}{\isacharprime}s\ {\isacharcolon}{\isacharcolon}\ state{\isacharparenright}set{\isacharparenright}\ \ {\isasymturnstile}\ EF\ {\isacharparenleft}{\isacharminus}\ s{\isacharparenright}{\isacharparenright}{\isacharparenright}{\isachardoublequoteclose}\isanewline
\prf\isanewline\isanewline
\endisadelimtheory
\isatagtheory
\isacommand{end}\isamarkupfalse%
\endisatagtheory
{\isafoldtheory}%
\isadelimtheory
\endisadelimtheory
\end{isabellebody}%

\subsection{Insider Framework}
\begin{isabellebody}%
\setisabellecontext{AirInsider}%
\isadelimtheory
\endisadelimtheory
\isatagtheory
\isacommand{theory}\isamarkupfalse%
\ AirInsider\isanewline
\isakeyword{imports}\ MC\isanewline
\isakeyword{begin}%
\endisatagtheory
{\isafoldtheory}%
\isadelimtheory
\isanewline
\endisadelimtheory
\isacommand{datatype}\isamarkupfalse%
\ action\ {\isacharequal}\ get\ {\isacharbar}\ move\ {\isacharbar}\ eval\ {\isacharbar}put\isanewline
\isanewline
\isacommand{typedecl}\isamarkupfalse%
\ actor\ \isanewline
\isacommand{consts}\isamarkupfalse%
\ Actor\ {\isacharcolon}{\isacharcolon}\ {\isachardoublequoteopen}string\ {\isasymRightarrow}\ actor{\isachardoublequoteclose}\ \isanewline
\isanewline
\isacommand{type{\isacharunderscore}synonym}\isamarkupfalse%
\ identity\ {\isacharequal}\ string\isanewline
\isacommand{type{\isacharunderscore}synonym}\isamarkupfalse%
\ \ policy\ {\isacharequal}\ {\isachardoublequoteopen}{\isacharparenleft}{\isacharparenleft}actor\ {\isasymRightarrow}\ bool{\isacharparenright}\ {\isacharasterisk}\ action\ set{\isacharparenright}{\isachardoublequoteclose}\isanewline
\isanewline
\isacommand{datatype}\isamarkupfalse%
\ location\ {\isacharequal}\ Location\ nat\isanewline
\isanewline
\isacommand{datatype}\isamarkupfalse%
\ igraph\ {\isacharequal}\ Lgraph\ {\isachardoublequoteopen}{\isacharparenleft}location\ {\isacharasterisk}\ location{\isacharparenright}set{\isachardoublequoteclose}\ {\isachardoublequoteopen}location\ {\isasymRightarrow}\ identity\ list{\isachardoublequoteclose}\isanewline
\ \ \ \ \ \ \ \ \ \ \ \ \ \ \ \ \ \ \ \ \ \ \ \ \ {\isachardoublequoteopen}actor\ {\isasymRightarrow}\ {\isacharparenleft}string\ list\ {\isacharasterisk}\ string\ list{\isacharparenright}{\isachardoublequoteclose}\ \ {\isachardoublequoteopen}location\ {\isasymRightarrow}\ string\ list{\isachardoublequoteclose}\isanewline
\isacommand{datatype}\isamarkupfalse%
\ infrastructure\ {\isacharequal}\ \isanewline
\ \ \ \ \ \ \ \ \ Infrastructure\ {\isachardoublequoteopen}igraph{\isachardoublequoteclose}\ \isanewline
\ \ \ \ \ \ \ \ \ \ \ \ \ \ \ \ \ \ \ \ \ \ \ \ {\isachardoublequoteopen}{\isacharbrackleft}igraph{\isacharcomma}\ location{\isacharbrackright}\ {\isasymRightarrow}\ policy\ set{\isachardoublequoteclose}\ \isanewline
\ \ \ \ \ \ \ \ \ \ \ \ \ \ \ \ \ \ \ \ \ \ \ \isanewline
\isacommand{primrec}\isamarkupfalse%
\ loc\ {\isacharcolon}{\isacharcolon}\ {\isachardoublequoteopen}location\ {\isasymRightarrow}\ nat{\isachardoublequoteclose}\isanewline
\isakeyword{where}\ \ {\isachardoublequoteopen}loc{\isacharparenleft}Location\ n{\isacharparenright}\ {\isacharequal}\ n{\isachardoublequoteclose}\isanewline
\isacommand{primrec}\isamarkupfalse%
\ gra\ {\isacharcolon}{\isacharcolon}\ {\isachardoublequoteopen}igraph\ {\isasymRightarrow}\ {\isacharparenleft}location\ {\isacharasterisk}\ location{\isacharparenright}set{\isachardoublequoteclose}\isanewline
\isakeyword{where}\ \ {\isachardoublequoteopen}gra{\isacharparenleft}Lgraph\ g\ a\ c\ l{\isacharparenright}\ {\isacharequal}\ g{\isachardoublequoteclose}\isanewline
\isacommand{primrec}\isamarkupfalse%
\ agra\ {\isacharcolon}{\isacharcolon}\ {\isachardoublequoteopen}igraph\ {\isasymRightarrow}\ {\isacharparenleft}location\ {\isasymRightarrow}\ identity\ list{\isacharparenright}{\isachardoublequoteclose}\isanewline
\isakeyword{where}\ \ {\isachardoublequoteopen}agra{\isacharparenleft}Lgraph\ g\ a\ c\ l{\isacharparenright}\ {\isacharequal}\ a{\isachardoublequoteclose}\isanewline
\isacommand{primrec}\isamarkupfalse%
\ cgra\ {\isacharcolon}{\isacharcolon}\ {\isachardoublequoteopen}igraph\ {\isasymRightarrow}\ {\isacharparenleft}actor\ {\isasymRightarrow}\ string\ list\ {\isacharasterisk}\ string\ list{\isacharparenright}{\isachardoublequoteclose}\isanewline
\isakeyword{where}\ \ {\isachardoublequoteopen}cgra{\isacharparenleft}Lgraph\ g\ a\ c\ l{\isacharparenright}\ {\isacharequal}\ c{\isachardoublequoteclose}\isanewline
\isacommand{primrec}\isamarkupfalse%
\ lgra\ {\isacharcolon}{\isacharcolon}\ {\isachardoublequoteopen}igraph\ {\isasymRightarrow}\ {\isacharparenleft}location\ {\isasymRightarrow}\ string\ list{\isacharparenright}{\isachardoublequoteclose}\isanewline
\isakeyword{where}\ \ {\isachardoublequoteopen}lgra{\isacharparenleft}Lgraph\ g\ a\ c\ l{\isacharparenright}\ {\isacharequal}\ l{\isachardoublequoteclose}\isanewline
\isanewline
\isacommand{definition}\isamarkupfalse%
\ nodes\ {\isacharcolon}{\isacharcolon}\ {\isachardoublequoteopen}igraph\ {\isasymRightarrow}\ location\ set{\isachardoublequoteclose}\ \isanewline
\isakeyword{where}\ {\isachardoublequoteopen}nodes\ g\ {\isacharequal}{\isacharequal}\ {\isacharbraceleft}\ x{\isachardot}\ {\isacharparenleft}{\isacharquery}\ y{\isachardot}\ {\isacharparenleft}{\isacharparenleft}x{\isacharcomma}y{\isacharparenright}{\isacharcolon}\ gra\ g{\isacharparenright}\ {\isacharbar}\ {\isacharparenleft}{\isacharparenleft}y{\isacharcomma}x{\isacharparenright}{\isacharcolon}\ gra\ g{\isacharparenright}{\isacharparenright}{\isacharbraceright}{\isachardoublequoteclose}\isanewline
\isanewline
\isacommand{definition}\isamarkupfalse%
\ actors{\isacharunderscore}graph\ {\isacharcolon}{\isacharcolon}\ {\isachardoublequoteopen}igraph\ {\isasymRightarrow}\ identity\ set{\isachardoublequoteclose}\ \ \isanewline
\isakeyword{where}\ \ {\isachardoublequoteopen}actors{\isacharunderscore}graph\ g\ {\isacharequal}{\isacharequal}\ {\isacharbraceleft}x{\isachardot}\ {\isacharquery}\ y{\isachardot}\ y\ {\isacharcolon}\ nodes\ g\ {\isasymand}\ x\ {\isasymin}\ set{\isacharparenleft}agra\ g\ y{\isacharparenright}{\isacharbraceright}{\isachardoublequoteclose}\isanewline
\isanewline
\isacommand{primrec}\isamarkupfalse%
\ graphI\ {\isacharcolon}{\isacharcolon}\ {\isachardoublequoteopen}infrastructure\ {\isasymRightarrow}\ igraph{\isachardoublequoteclose}\isanewline
\isakeyword{where}\ {\isachardoublequoteopen}graphI\ {\isacharparenleft}Infrastructure\ g\ d{\isacharparenright}\ {\isacharequal}\ g{\isachardoublequoteclose}\isanewline
\isacommand{primrec}\isamarkupfalse%
\ delta\ {\isacharcolon}{\isacharcolon}\ {\isachardoublequoteopen}{\isacharbrackleft}infrastructure{\isacharcomma}\ igraph{\isacharcomma}\ location{\isacharbrackright}\ {\isasymRightarrow}\ policy\ set{\isachardoublequoteclose}\isanewline
\isakeyword{where}\ {\isachardoublequoteopen}delta\ {\isacharparenleft}Infrastructure\ g\ d{\isacharparenright}\ {\isacharequal}\ d{\isachardoublequoteclose}\isanewline
\isacommand{primrec}\isamarkupfalse%
\ tspace\ {\isacharcolon}{\isacharcolon}\ {\isachardoublequoteopen}{\isacharbrackleft}infrastructure{\isacharcomma}\ actor\ {\isacharbrackright}\ {\isasymRightarrow}\ string\ list\ {\isacharasterisk}\ string\ list{\isachardoublequoteclose}\isanewline
\ \ \isakeyword{where}\ {\isachardoublequoteopen}tspace\ {\isacharparenleft}Infrastructure\ g\ d{\isacharparenright}\ {\isacharequal}\ cgra\ g{\isachardoublequoteclose}\isanewline
\isacommand{primrec}\isamarkupfalse%
\ lspace\ {\isacharcolon}{\isacharcolon}\ {\isachardoublequoteopen}{\isacharbrackleft}infrastructure{\isacharcomma}\ location\ {\isacharbrackright}\ {\isasymRightarrow}\ string\ list{\isachardoublequoteclose}\isanewline
\isakeyword{where}\ {\isachardoublequoteopen}lspace\ {\isacharparenleft}Infrastructure\ g\ d{\isacharparenright}\ {\isacharequal}\ lgra\ g{\isachardoublequoteclose}\isanewline
\isanewline
\isacommand{definition}\isamarkupfalse%
\ credentials\ {\isacharcolon}{\isacharcolon}\ {\isachardoublequoteopen}string\ list\ {\isacharasterisk}\ string\ list\ {\isasymRightarrow}\ string\ set{\isachardoublequoteclose}\isanewline
\ \ \isakeyword{where}\ \ {\isachardoublequoteopen}credentials\ lxl\ {\isasymequiv}\ set\ {\isacharparenleft}fst\ lxl{\isacharparenright}{\isachardoublequoteclose}\isanewline
\isacommand{definition}\isamarkupfalse%
\ has\ {\isacharcolon}{\isacharcolon}\ {\isachardoublequoteopen}{\isacharbrackleft}igraph{\isacharcomma}\ actor\ {\isacharasterisk}\ string{\isacharbrackright}\ {\isasymRightarrow}\ bool{\isachardoublequoteclose}\isanewline
\ \ \isakeyword{where}\ {\isachardoublequoteopen}has\ G\ ac\ {\isasymequiv}\ snd\ ac\ {\isasymin}\ credentials{\isacharparenleft}cgra\ G\ {\isacharparenleft}fst\ ac{\isacharparenright}{\isacharparenright}{\isachardoublequoteclose}\isanewline
\isacommand{definition}\isamarkupfalse%
\ roles\ {\isacharcolon}{\isacharcolon}\ {\isachardoublequoteopen}string\ list\ {\isacharasterisk}\ string\ list\ {\isasymRightarrow}\ string\ set{\isachardoublequoteclose}\isanewline
\ \ \isakeyword{where}\ \ {\isachardoublequoteopen}roles\ lxl\ {\isasymequiv}\ set\ {\isacharparenleft}snd\ lxl{\isacharparenright}{\isachardoublequoteclose}\isanewline
\isacommand{definition}\isamarkupfalse%
\ role\ {\isacharcolon}{\isacharcolon}\ {\isachardoublequoteopen}{\isacharbrackleft}igraph{\isacharcomma}\ actor\ {\isacharasterisk}\ string{\isacharbrackright}\ {\isasymRightarrow}\ bool{\isachardoublequoteclose}\isanewline
\ \ \isakeyword{where}\ {\isachardoublequoteopen}role\ G\ ac\ {\isasymequiv}\ snd\ ac\ {\isasymin}\ roles{\isacharparenleft}cgra\ G\ {\isacharparenleft}fst\ ac{\isacharparenright}{\isacharparenright}{\isachardoublequoteclose}\isanewline
\isacommand{definition}\isamarkupfalse%
\ isin\ {\isacharcolon}{\isacharcolon}\ {\isachardoublequoteopen}{\isacharbrackleft}igraph{\isacharcomma}location{\isacharcomma}\ string{\isacharbrackright}\ {\isasymRightarrow}\ bool{\isachardoublequoteclose}\ \isanewline
\ \ \isakeyword{where}\ {\isachardoublequoteopen}isin\ G\ l\ s\ {\isasymequiv}\ s\ {\isasymin}\ set{\isacharparenleft}lgra\ G\ l{\isacharparenright}{\isachardoublequoteclose}\isanewline
\ \ \isanewline
\isacommand{datatype}\isamarkupfalse%
\ psy{\isacharunderscore}states\ {\isacharequal}\ happy\ {\isacharbar}\ depressed\ {\isacharbar}\ disgruntled\ {\isacharbar}\ angry\ {\isacharbar}\ stressed\isanewline
\isacommand{datatype}\isamarkupfalse%
\ motivations\ {\isacharequal}\ financial\ {\isacharbar}\ political\ {\isacharbar}\ revenge\ {\isacharbar}\ curious\ {\isacharbar}\ competitive{\isacharunderscore}advantage\ {\isacharbar}\ power\ {\isacharbar}\ peer{\isacharunderscore}recognition\isanewline
\isanewline
\isacommand{datatype}\isamarkupfalse%
\ actor{\isacharunderscore}state\ {\isacharequal}\ Actor{\isacharunderscore}state\ {\isachardoublequoteopen}psy{\isacharunderscore}states{\isachardoublequoteclose}\ {\isachardoublequoteopen}motivations\ set{\isachardoublequoteclose}\isanewline
\isacommand{primrec}\isamarkupfalse%
\ motivation\ {\isacharcolon}{\isacharcolon}\ {\isachardoublequoteopen}actor{\isacharunderscore}state\ {\isasymRightarrow}\ motivations\ set{\isachardoublequoteclose}\ \isanewline
\isakeyword{where}\ {\isachardoublequoteopen}motivation\ \ {\isacharparenleft}Actor{\isacharunderscore}state\ p\ m{\isacharparenright}\ {\isacharequal}\ \ m{\isachardoublequoteclose}\isanewline
\isacommand{primrec}\isamarkupfalse%
\ psy{\isacharunderscore}state\ {\isacharcolon}{\isacharcolon}\ {\isachardoublequoteopen}actor{\isacharunderscore}state\ {\isasymRightarrow}\ psy{\isacharunderscore}states{\isachardoublequoteclose}\ \isanewline
\isakeyword{where}\ {\isachardoublequoteopen}psy{\isacharunderscore}state\ \ {\isacharparenleft}Actor{\isacharunderscore}state\ p\ m{\isacharparenright}\ {\isacharequal}\ p{\isachardoublequoteclose}\isanewline
\isanewline
\isacommand{definition}\isamarkupfalse%
\ tipping{\isacharunderscore}point\ {\isacharcolon}{\isacharcolon}\ {\isachardoublequoteopen}actor{\isacharunderscore}state\ {\isasymRightarrow}\ bool{\isachardoublequoteclose}\ \isakeyword{where}\isanewline
\ \ {\isachardoublequoteopen}tipping{\isacharunderscore}point\ a\ {\isasymequiv}\ {\isacharparenleft}{\isacharparenleft}motivation\ a\ {\isasymnoteq}\ {\isacharbraceleft}{\isacharbraceright}{\isacharparenright}\ {\isasymand}\ {\isacharparenleft}happy\ {\isasymnoteq}\ psy{\isacharunderscore}state\ a{\isacharparenright}{\isacharparenright}{\isachardoublequoteclose}\isanewline
\isanewline
\isacommand{definition}\isamarkupfalse%
\ UasI\ {\isacharcolon}{\isacharcolon}\ \ {\isachardoublequoteopen}{\isacharbrackleft}identity{\isacharcomma}\ identity{\isacharbrackright}\ {\isasymRightarrow}\ bool\ {\isachardoublequoteclose}\ \isanewline
\isakeyword{where}\ {\isachardoublequoteopen}UasI\ a\ b\ {\isasymequiv}\ {\isacharparenleft}Actor\ a\ {\isacharequal}\ Actor\ b{\isacharparenright}\ {\isasymand}\ {\isacharparenleft}{\isasymforall}\ x\ y{\isachardot}\ x\ {\isasymnoteq}\ a\ {\isasymand}\ y\ {\isasymnoteq}\ a\ {\isasymand}\ Actor\ x\ {\isacharequal}\ Actor\ y\ {\isasymlongrightarrow}\ x\ {\isacharequal}\ y{\isacharparenright}{\isachardoublequoteclose}\isanewline
\isanewline
\isacommand{definition}\isamarkupfalse%
\ Insider\ {\isacharcolon}{\isacharcolon}\ {\isachardoublequoteopen}{\isacharbrackleft}identity{\isacharcomma}\ identity\ set{\isacharcomma}\ identity\ {\isasymRightarrow}\ actor{\isacharunderscore}state{\isacharbrackright}\ {\isasymRightarrow}\ bool{\isachardoublequoteclose}\ \isanewline
\isakeyword{where}\ {\isachardoublequoteopen}Insider\ a\ C\ as\ {\isasymequiv}\ {\isacharparenleft}tipping{\isacharunderscore}point\ {\isacharparenleft}as\ a{\isacharparenright}\ {\isasymlongrightarrow}\ {\isacharparenleft}{\isasymforall}\ b{\isasymin}C{\isachardot}\ UasI\ a\ b{\isacharparenright}{\isacharparenright}{\isachardoublequoteclose}\isanewline
\isanewline
\isacommand{definition}\isamarkupfalse%
\ atI\ {\isacharcolon}{\isacharcolon}\ {\isachardoublequoteopen}{\isacharbrackleft}identity{\isacharcomma}\ igraph{\isacharcomma}\ location{\isacharbrackright}\ {\isasymRightarrow}\ bool{\isachardoublequoteclose}\ {\isacharparenleft}{\isachardoublequoteopen}{\isacharunderscore}\ {\isacharat}\isactrlbsub {\isacharparenleft}{\isacharunderscore}{\isacharparenright}\isactrlesub \ {\isacharunderscore}{\isachardoublequoteclose}\ {\isadigit{5}}{\isadigit{0}}{\isacharparenright}\isanewline
\isakeyword{where}\ {\isachardoublequoteopen}a\ {\isacharat}\isactrlbsub G\isactrlesub \ l\ {\isasymequiv}\ a\ {\isasymin}\ set{\isacharparenleft}agra\ G\ l{\isacharparenright}{\isachardoublequoteclose}\isanewline
\isanewline
\isacommand{definition}\isamarkupfalse%
\ enables\ {\isacharcolon}{\isacharcolon}\ {\isachardoublequoteopen}{\isacharbrackleft}infrastructure{\isacharcomma}\ location{\isacharcomma}\ actor{\isacharcomma}\ action{\isacharbrackright}\ {\isasymRightarrow}\ bool{\isachardoublequoteclose}\isanewline
\isakeyword{where}\isanewline
{\isachardoublequoteopen}enables\ I\ l\ a\ a{\isacharprime}\ {\isasymequiv}\ \ {\isacharparenleft}{\isasymexists}\ {\isacharparenleft}p{\isacharcomma}e{\isacharparenright}\ {\isasymin}\ delta\ I\ {\isacharparenleft}graphI\ I{\isacharparenright}\ l{\isachardot}\ a{\isacharprime}\ {\isasymin}\ e\ {\isasymand}\ p\ a{\isacharparenright}{\isachardoublequoteclose}\isanewline
\isanewline
\isacommand{primrec}\isamarkupfalse%
\ nodup\ {\isacharcolon}{\isacharcolon}\ {\isachardoublequoteopen}{\isacharbrackleft}{\isacharprime}a{\isacharcomma}\ {\isacharprime}a\ list{\isacharbrackright}\ {\isasymRightarrow}\ bool{\isachardoublequoteclose}\isanewline
\ \ \isakeyword{where}\ \isanewline
\ \ \ \ nodup{\isacharunderscore}nil{\isacharcolon}\ {\isachardoublequoteopen}nodup\ a\ {\isacharbrackleft}{\isacharbrackright}\ {\isacharequal}\ True{\isachardoublequoteclose}\ {\isacharbar}\isanewline
\ \ \ \ nodup{\isacharunderscore}step{\isacharcolon}\ {\isachardoublequoteopen}nodup\ a\ {\isacharparenleft}x\ {\isacharhash}\ ls{\isacharparenright}\ {\isacharequal}\ {\isacharparenleft}if\ x\ {\isacharequal}\ a\ then\ {\isacharparenleft}a\ {\isasymnotin}\ {\isacharparenleft}set\ ls{\isacharparenright}{\isacharparenright}\ else\ nodup\ a\ ls{\isacharparenright}{\isachardoublequoteclose}\isanewline
\isanewline
\isacommand{definition}\isamarkupfalse%
\ move{\isacharunderscore}graph{\isacharunderscore}a\ {\isacharcolon}{\isacharcolon}\ {\isachardoublequoteopen}{\isacharbrackleft}identity{\isacharcomma}\ location{\isacharcomma}\ location{\isacharcomma}\ igraph{\isacharbrackright}\ {\isasymRightarrow}\ igraph{\isachardoublequoteclose}\isanewline
\isakeyword{where}\ {\isachardoublequoteopen}move{\isacharunderscore}graph{\isacharunderscore}a\ n\ l\ l{\isacharprime}\ g\ {\isasymequiv}\ Lgraph\ {\isacharparenleft}gra\ g{\isacharparenright}\ \isanewline
\ \ \ \ \ \ \ \ \ \ \ \ \ \ \ \ \ \ \ \ {\isacharparenleft}if\ n\ {\isasymin}\ set\ {\isacharparenleft}{\isacharparenleft}agra\ g{\isacharparenright}\ l{\isacharparenright}\ {\isacharampersand}\ \ n\ {\isasymnotin}\ set\ {\isacharparenleft}{\isacharparenleft}agra\ g{\isacharparenright}\ l{\isacharprime}{\isacharparenright}\ then\ \isanewline
\ \ \ \ \ \ \ \ \ \ \ \ \ \ \ \ \ \ \ \ \ {\isacharparenleft}{\isacharparenleft}agra\ g{\isacharparenright}{\isacharparenleft}l\ {\isacharcolon}{\isacharequal}\ del\ n\ {\isacharparenleft}agra\ g\ l{\isacharparenright}{\isacharparenright}{\isacharparenright}{\isacharparenleft}l{\isacharprime}\ {\isacharcolon}{\isacharequal}\ {\isacharparenleft}n\ {\isacharhash}\ {\isacharparenleft}agra\ g\ l{\isacharprime}{\isacharparenright}{\isacharparenright}{\isacharparenright}\isanewline
\ \ \ \ \ \ \ \ \ \ \ \ \ \ \ \ \ \ \ \ \ else\ {\isacharparenleft}agra\ g{\isacharparenright}{\isacharparenright}{\isacharparenleft}cgra\ g{\isacharparenright}{\isacharparenleft}lgra\ g{\isacharparenright}{\isachardoublequoteclose}\isanewline
\isanewline
\isacommand{inductive}\isamarkupfalse%
\ state{\isacharunderscore}transition{\isacharunderscore}in\ {\isacharcolon}{\isacharcolon}\ {\isachardoublequoteopen}{\isacharbrackleft}infrastructure{\isacharcomma}\ infrastructure{\isacharbrackright}\ {\isasymRightarrow}\ bool{\isachardoublequoteclose}\ {\isacharparenleft}{\isachardoublequoteopen}{\isacharparenleft}{\isacharunderscore}\ {\isasymrightarrow}\isactrlsub n\ {\isacharunderscore}{\isacharparenright}{\isachardoublequoteclose}\ {\isadigit{5}}{\isadigit{0}}{\isacharparenright}\isanewline
\isakeyword{where}\isanewline
\ \ move{\isacharcolon}\ {\isachardoublequoteopen}{\isasymlbrakk}\ G\ {\isacharequal}\ graphI\ I{\isacharsemicolon}\ a\ {\isacharat}\isactrlbsub G\isactrlesub \ l{\isacharsemicolon}\ l\ {\isasymin}\ nodes\ G{\isacharsemicolon}\ l{\isacharprime}\ {\isasymin}\ nodes\ G{\isacharsemicolon}\isanewline
\ \ \ \ \ \ \ \ \ \ {\isacharparenleft}a{\isacharparenright}\ {\isasymin}\ actors{\isacharunderscore}graph{\isacharparenleft}graphI\ I{\isacharparenright}{\isacharsemicolon}\ enables\ I\ l{\isacharprime}\ {\isacharparenleft}Actor\ a{\isacharparenright}\ move{\isacharsemicolon}\isanewline
\ \ \ \ \ \ \ \ \ I{\isacharprime}\ {\isacharequal}\ Infrastructure\ {\isacharparenleft}move{\isacharunderscore}graph{\isacharunderscore}a\ a\ l\ l{\isacharprime}\ {\isacharparenleft}graphI\ I{\isacharparenright}{\isacharparenright}{\isacharparenleft}delta\ I{\isacharparenright}\ {\isasymrbrakk}\ {\isasymLongrightarrow}\ I\ {\isasymrightarrow}\isactrlsub n\ I{\isacharprime}{\isachardoublequoteclose}\ \isanewline
{\isacharbar}\ get\ {\isacharcolon}\ {\isachardoublequoteopen}{\isasymlbrakk}\ G\ {\isacharequal}\ graphI\ I{\isacharsemicolon}\ a\ {\isacharat}\isactrlbsub G\isactrlesub \ l{\isacharsemicolon}\ a{\isacharprime}\ {\isacharat}\isactrlbsub G\isactrlesub \ l{\isacharsemicolon}\ has\ G\ {\isacharparenleft}Actor\ a{\isacharcomma}\ z{\isacharparenright}{\isacharsemicolon}\isanewline
\ \ \ \ \ \ \ \ enables\ I\ l\ {\isacharparenleft}Actor\ a{\isacharparenright}\ get{\isacharsemicolon}\isanewline
\ \ \ \ \ \ \ \ I{\isacharprime}\ {\isacharequal}\ Infrastructure\ \isanewline
\ \ \ \ \ \ \ \ \ \ \ \ \ \ \ \ \ \ \ {\isacharparenleft}Lgraph\ {\isacharparenleft}gra\ G{\isacharparenright}{\isacharparenleft}agra\ G{\isacharparenright}\isanewline
\ \ \ \ \ \ \ \ \ \ \ \ \ \ \ \ \ \ \ \ \ \ \ \ \ \ \ {\isacharparenleft}{\isacharparenleft}cgra\ G{\isacharparenright}{\isacharparenleft}Actor\ a{\isacharprime}\ {\isacharcolon}{\isacharequal}\ \isanewline
\ \ \ \ \ \ \ \ \ \ \ \ \ \ \ \ \ \ \ \ \ \ \ \ \ \ \ \ \ \ \ \ {\isacharparenleft}z\ {\isacharhash}\ {\isacharparenleft}fst{\isacharparenleft}cgra\ G\ {\isacharparenleft}Actor\ a{\isacharprime}{\isacharparenright}{\isacharparenright}{\isacharparenright}{\isacharcomma}\ snd{\isacharparenleft}cgra\ G\ {\isacharparenleft}Actor\ a{\isacharprime}{\isacharparenright}{\isacharparenright}{\isacharparenright}{\isacharparenright}{\isacharparenright}\isanewline
\ \ \ \ \ \ \ \ \ \ \ \ \ \ \ \ \ \ \ \ \ \ \ \ \ \ \ {\isacharparenleft}lgra\ G{\isacharparenright}{\isacharparenright}\isanewline
\ \ \ \ \ \ \ \ \ \ \ \ \ \ \ \ \ \ \ {\isacharparenleft}delta\ I{\isacharparenright}\isanewline
\ \ \ \ \ \ \ \ \ {\isasymrbrakk}\ {\isasymLongrightarrow}\ I\ {\isasymrightarrow}\isactrlsub n\ I{\isacharprime}{\isachardoublequoteclose}\isanewline
{\isacharbar}\ put\ {\isacharcolon}\ {\isachardoublequoteopen}{\isasymlbrakk}\ G\ {\isacharequal}\ graphI\ I{\isacharsemicolon}\ a\ {\isacharat}\isactrlbsub G\isactrlesub \ l{\isacharsemicolon}\ enables\ I\ l\ {\isacharparenleft}Actor\ a{\isacharparenright}\ put{\isacharsemicolon}\isanewline
\ \ \ \ \ \ \ \ I{\isacharprime}\ {\isacharequal}\ Infrastructure\ \isanewline
\ \ \ \ \ \ \ \ \ \ \ \ \ \ \ \ \ \ {\isacharparenleft}Lgraph\ {\isacharparenleft}gra\ G{\isacharparenright}{\isacharparenleft}agra\ G{\isacharparenright}{\isacharparenleft}cgra\ G{\isacharparenright}\isanewline
\ \ \ \ \ \ \ \ \ \ \ \ \ \ \ \ \ \ \ \ \ \ \ \ \ \ {\isacharparenleft}{\isacharparenleft}lgra\ G{\isacharparenright}{\isacharparenleft}l\ {\isacharcolon}{\isacharequal}\ {\isacharbrackleft}z{\isacharbrackright}{\isacharparenright}{\isacharparenright}{\isacharparenright}\isanewline
\ \ \ \ \ \ \ \ \ \ \ \ \ \ \ \ \ \ \ {\isacharparenleft}delta\ I{\isacharparenright}\ {\isasymrbrakk}\isanewline
\ \ \ \ \ \ \ \ \ {\isasymLongrightarrow}\ I\ {\isasymrightarrow}\isactrlsub n\ I{\isacharprime}{\isachardoublequoteclose}\ \ \isanewline
{\isacharbar}\ put{\isacharunderscore}remote\ {\isacharcolon}\ {\isachardoublequoteopen}{\isasymlbrakk}\ G\ {\isacharequal}\ graphI\ I{\isacharsemicolon}\ enables\ I\ l\ {\isacharparenleft}Actor\ a{\isacharparenright}\ put{\isacharsemicolon}\isanewline
\ \ \ \ \ \ \ \ I{\isacharprime}\ {\isacharequal}\ Infrastructure\ \isanewline
\ \ \ \ \ \ \ \ \ \ \ \ \ \ \ \ \ \ {\isacharparenleft}Lgraph\ {\isacharparenleft}gra\ G{\isacharparenright}{\isacharparenleft}agra\ G{\isacharparenright}{\isacharparenleft}cgra\ G{\isacharparenright}\isanewline
\ \ \ \ \ \ \ \ \ \ \ \ \ \ \ \ \ \ \ \ \ \ \ \ \ \ \ \ {\isacharparenleft}{\isacharparenleft}lgra\ G{\isacharparenright}{\isacharparenleft}l\ {\isacharcolon}{\isacharequal}\ {\isacharbrackleft}z{\isacharbrackright}{\isacharparenright}{\isacharparenright}{\isacharparenright}\isanewline
\ \ \ \ \ \ \ \ \ \ \ \ \ \ \ \ \ \ \ \ {\isacharparenleft}delta\ I{\isacharparenright}\ {\isasymrbrakk}\isanewline
\ \ \ \ \ \ \ \ \ {\isasymLongrightarrow}\ I\ {\isasymrightarrow}\isactrlsub n\ I{\isacharprime}{\isachardoublequoteclose}\isanewline
\ \ \isanewline
\isanewline
\isacommand{instantiation}\isamarkupfalse%
\ {\isachardoublequoteopen}infrastructure{\isachardoublequoteclose}\ {\isacharcolon}{\isacharcolon}\ state\isanewline
\isakeyword{begin}\isanewline
\isacommand{definition}\isamarkupfalse%
\ \isanewline
\ \ \ state{\isacharunderscore}transition{\isacharunderscore}infra{\isacharunderscore}def{\isacharcolon}\ {\isachardoublequoteopen}{\isacharparenleft}i\ {\isasymrightarrow}\isactrlsub i\ i{\isacharprime}{\isacharparenright}\ {\isacharequal}\ \ {\isacharparenleft}i\ {\isasymrightarrow}\isactrlsub n\ {\isacharparenleft}i{\isacharprime}\ {\isacharcolon}{\isacharcolon}\ infrastructure{\isacharparenright}{\isacharparenright}{\isachardoublequoteclose}\isanewline
\isanewline
\isacommand{instance}\isamarkupfalse%
\isanewline
\isadelimproof
\ \ %
\endisadelimproof
\isatagproof
\isacommand{by}\isamarkupfalse%
\ {\isacharparenleft}rule\ MC{\isachardot}class{\isachardot}MC{\isachardot}state{\isachardot}of{\isacharunderscore}class{\isachardot}intro{\isacharparenright}%
\endisatagproof
{\isafoldproof}%
\isadelimproof
\isanewline
\isanewline
\endisadelimproof
\isacommand{definition}\isamarkupfalse%
\ state{\isacharunderscore}transition{\isacharunderscore}in{\isacharunderscore}refl\ {\isacharparenleft}{\isachardoublequoteopen}{\isacharparenleft}{\isacharunderscore}\ {\isasymrightarrow}\isactrlsub n{\isacharasterisk}\ {\isacharunderscore}{\isacharparenright}{\isachardoublequoteclose}\ {\isadigit{5}}{\isadigit{0}}{\isacharparenright}\isanewline
\isakeyword{where}\ {\isachardoublequoteopen}s\ {\isasymrightarrow}\isactrlsub n{\isacharasterisk}\ s{\isacharprime}\ {\isasymequiv}\ {\isacharparenleft}{\isacharparenleft}s{\isacharcomma}s{\isacharprime}{\isacharparenright}\ {\isasymin}\ {\isacharbraceleft}{\isacharparenleft}x{\isacharcomma}y{\isacharparenright}{\isachardot}\ state{\isacharunderscore}transition{\isacharunderscore}in\ x\ y{\isacharbraceright}\isactrlsup {\isacharasterisk}{\isacharparenright}{\isachardoublequoteclose}\isanewline
\isanewline
\isacommand{lemma}\isamarkupfalse%
\ move{\isacharunderscore}graph{\isacharunderscore}eq{\isacharcolon}\ {\isachardoublequoteopen}move{\isacharunderscore}graph{\isacharunderscore}a\ a\ l\ l\ g\ {\isacharequal}\ g{\isachardoublequoteclose}\ \ \isanewline
\isadelimproof
\ \ %
\endisadelimproof
\isatagproof
\isacommand{by}\isamarkupfalse%
\ {\isacharparenleft}simp\ add{\isacharcolon}\ move{\isacharunderscore}graph{\isacharunderscore}a{\isacharunderscore}def{\isacharcomma}\ case{\isacharunderscore}tac\ g{\isacharcomma}\ force{\isacharparenright}%
\endisatagproof
{\isafoldproof}%
\isadelimproof
\isanewline
\endisadelimproof
\isanewline
\isacommand{lemma}\isamarkupfalse%
\ delta{\isacharunderscore}invariant{\isacharcolon}\ {\isachardoublequoteopen}{\isasymforall}\ z\ z{\isacharprime}{\isachardot}\ z\ {\isasymrightarrow}\isactrlsub n\ z{\isacharprime}\ {\isasymlongrightarrow}\ \ delta{\isacharparenleft}z{\isacharparenright}\ {\isacharequal}\ delta{\isacharparenleft}z{\isacharprime}{\isacharparenright}{\isachardoublequoteclose}\ \ \ \ \isanewline
\isadelimproof
\ \ %
\endisadelimproof
\isatagproof
\isacommand{by}\isamarkupfalse%
\ {\isacharparenleft}clarify{\isacharcomma}\ erule\ state{\isacharunderscore}transition{\isacharunderscore}in{\isachardot}cases{\isacharcomma}\ simp{\isacharplus}{\isacharparenright}%
\endisatagproof
{\isafoldproof}%
\isadelimproof
\isanewline
\endisadelimproof
\isanewline
\isacommand{lemma}\isamarkupfalse%
\ init{\isacharunderscore}state{\isacharunderscore}policy{\isacharcolon}\ {\isachardoublequoteopen}{\isasymlbrakk}\ {\isacharparenleft}x{\isacharcomma}y{\isacharparenright}\ {\isasymin}\ {\isacharbraceleft}{\isacharparenleft}x{\isacharcolon}{\isacharcolon}infrastructure{\isacharcomma}\ y{\isacharcolon}{\isacharcolon}infrastructure{\isacharparenright}{\isachardot}\ x\ {\isasymrightarrow}\isactrlsub n\ y{\isacharbraceright}\isactrlsup {\isacharasterisk}\ {\isasymrbrakk}\ {\isasymLongrightarrow}\ \isanewline
\ \ \ \ \ \ \ \ \ \ \ \ \ \ \ \ \ \ \ \ \ \ \ \ \ \ delta{\isacharparenleft}x{\isacharparenright}\ {\isacharequal}\ delta{\isacharparenleft}y{\isacharparenright}{\isachardoublequoteclose}\ \ \isanewline
\isadelimproof
\endisadelimproof
\isatagproof
\isacommand{proof}\isamarkupfalse%
\ {\isacharminus}\isanewline
\ \ \isacommand{have}\isamarkupfalse%
\ ind{\isacharcolon}\ {\isachardoublequoteopen}{\isacharparenleft}x{\isacharcomma}y{\isacharparenright}\ {\isasymin}\ {\isacharbraceleft}{\isacharparenleft}x{\isacharcolon}{\isacharcolon}infrastructure{\isacharcomma}\ y{\isacharcolon}{\isacharcolon}infrastructure{\isacharparenright}{\isachardot}\ x\ {\isasymrightarrow}\isactrlsub n\ y{\isacharbraceright}\isactrlsup {\isacharasterisk}\isanewline
\ \ \ \ \ \ \ \ \ \ \ \ \ {\isasymlongrightarrow}\ delta{\isacharparenleft}x{\isacharparenright}\ {\isacharequal}\ delta{\isacharparenleft}y{\isacharparenright}{\isachardoublequoteclose}\isanewline
\ \ \isacommand{proof}\isamarkupfalse%
\ {\isacharparenleft}insert\ assms{\isacharcomma}\ erule\ rtrancl{\isachardot}induct{\isacharparenright}\isanewline
\ \ \ \ \isacommand{show}\isamarkupfalse%
\ {\isachardoublequoteopen}{\isacharparenleft}{\isasymAnd}\ a{\isacharcolon}{\isacharcolon}infrastructure{\isachardot}\isanewline
\ \ \ \ \ \ \ {\isacharparenleft}{\isasymforall}{\isacharparenleft}z{\isacharcolon}{\isacharcolon}infrastructure{\isacharparenright}{\isacharparenleft}z{\isacharprime}{\isacharcolon}{\isacharcolon}infrastructure{\isacharparenright}{\isachardot}\ {\isacharparenleft}z\ {\isasymrightarrow}\isactrlsub n\ z{\isacharprime}{\isacharparenright}\ {\isasymlongrightarrow}\ {\isacharparenleft}delta\ z\ {\isacharequal}\ delta\ z{\isacharprime}{\isacharparenright}{\isacharparenright}\ {\isasymLongrightarrow}\isanewline
\ \ \ \ \ \ \ {\isacharparenleft}{\isacharparenleft}{\isacharparenleft}a{\isacharcomma}\ a{\isacharparenright}\ {\isasymin}\ {\isacharbraceleft}{\isacharparenleft}x\ {\isacharcolon}{\isacharcolon}infrastructure{\isacharcomma}\ y\ {\isacharcolon}{\isacharcolon}\ infrastructure{\isacharparenright}{\isachardot}\ x\ {\isasymrightarrow}\isactrlsub n\ y{\isacharbraceright}\isactrlsup {\isacharasterisk}{\isacharparenright}\ {\isasymlongrightarrow}\isanewline
\ \ \ \ \ \ \ {\isacharparenleft}delta\ a\ {\isacharequal}\ delta\ a{\isacharparenright}{\isacharparenright}{\isacharparenright}{\isachardoublequoteclose}\isanewline
\ \ \ \ \isacommand{by}\isamarkupfalse%
\ {\isacharparenleft}rule\ impI{\isacharcomma}\ rule\ refl{\isacharparenright}\isanewline
\isacommand{next}\isamarkupfalse%
\ \isacommand{fix}\isamarkupfalse%
\ a\ b\ c\isanewline
\ \ \isacommand{assume}\isamarkupfalse%
\ a{\isadigit{0}}{\isacharcolon}\ {\isachardoublequoteopen}{\isasymforall}{\isacharparenleft}z{\isacharcolon}{\isacharcolon}infrastructure{\isacharparenright}\ z{\isacharprime}{\isacharcolon}{\isacharcolon}infrastructure{\isachardot}\ z\ {\isasymrightarrow}\isactrlsub n\ z{\isacharprime}\ {\isasymlongrightarrow}\ delta\ z\ {\isacharequal}\ delta\ z{\isacharprime}{\isachardoublequoteclose}\isanewline
\ \ \ \ \ \isakeyword{and}\ a{\isadigit{1}}{\isacharcolon}\ {\isachardoublequoteopen}{\isacharparenleft}a{\isacharcomma}\ b{\isacharparenright}\ {\isasymin}\ {\isacharbraceleft}{\isacharparenleft}x{\isacharcolon}{\isacharcolon}infrastructure{\isacharcomma}\ y{\isacharcolon}{\isacharcolon}infrastructure{\isacharparenright}{\isachardot}\ x\ {\isasymrightarrow}\isactrlsub n\ y{\isacharbraceright}\isactrlsup {\isacharasterisk}{\isachardoublequoteclose}\isanewline
\ \ \ \ \ \isakeyword{and}\ a{\isadigit{2}}{\isacharcolon}\ {\isachardoublequoteopen}{\isacharparenleft}a{\isacharcomma}\ b{\isacharparenright}\ {\isasymin}\ {\isacharbraceleft}{\isacharparenleft}x{\isacharcolon}{\isacharcolon}infrastructure{\isacharcomma}\ y{\isacharcolon}{\isacharcolon}infrastructure{\isacharparenright}{\isachardot}\ x\ {\isasymrightarrow}\isactrlsub n\ y{\isacharbraceright}\isactrlsup {\isacharasterisk}\ {\isasymlongrightarrow}\isanewline
\ \ \ \ \ \ \ \ \ delta\ a\ {\isacharequal}\ delta\ b{\isachardoublequoteclose}\isanewline
\ \ \ \ \ \isakeyword{and}\ a{\isadigit{3}}{\isacharcolon}\ {\isachardoublequoteopen}{\isacharparenleft}b{\isacharcomma}\ c{\isacharparenright}\ {\isasymin}\ {\isacharbraceleft}{\isacharparenleft}x{\isacharcolon}{\isacharcolon}infrastructure{\isacharcomma}\ y{\isacharcolon}{\isacharcolon}infrastructure{\isacharparenright}{\isachardot}\ x\ {\isasymrightarrow}\isactrlsub n\ y{\isacharbraceright}{\isachardoublequoteclose}\isanewline
\ \ \ \ \ \isacommand{show}\isamarkupfalse%
\ {\isachardoublequoteopen}{\isacharparenleft}a{\isacharcomma}\ c{\isacharparenright}\ {\isasymin}\ {\isacharbraceleft}{\isacharparenleft}x{\isacharcolon}{\isacharcolon}infrastructure{\isacharcomma}\ y{\isacharcolon}{\isacharcolon}infrastructure{\isacharparenright}{\isachardot}\ x\ {\isasymrightarrow}\isactrlsub n\ y{\isacharbraceright}\isactrlsup {\isacharasterisk}\ {\isasymlongrightarrow}\isanewline
\ \ \ \ \ \ \ delta\ a\ {\isacharequal}\ delta\ c{\isachardoublequoteclose}\isanewline
\ \ \isacommand{proof}\isamarkupfalse%
\ {\isacharminus}\isanewline
\ \ \ \ \isacommand{have}\isamarkupfalse%
\ a{\isadigit{4}}{\isacharcolon}\ {\isachardoublequoteopen}delta\ b\ {\isacharequal}\ delta\ c{\isachardoublequoteclose}\ \isacommand{using}\isamarkupfalse%
\ a{\isadigit{0}}\ a{\isadigit{1}}\ a{\isadigit{2}}\ a{\isadigit{3}}\ \isacommand{by}\isamarkupfalse%
\ simp\isanewline
\ \ \ \ \isacommand{show}\isamarkupfalse%
\ {\isacharquery}thesis\ \isacommand{using}\isamarkupfalse%
\ a{\isadigit{0}}\ a{\isadigit{1}}\ a{\isadigit{2}}\ a{\isadigit{3}}\ \isacommand{by}\isamarkupfalse%
\ simp\isanewline
\ \ \isacommand{qed}\isamarkupfalse%
\isanewline
\isacommand{qed}\isamarkupfalse%
\isanewline
\isacommand{show}\isamarkupfalse%
\ {\isacharquery}thesis\ \isanewline
\ \ \isacommand{by}\isamarkupfalse%
\ {\isacharparenleft}insert\ ind{\isacharcomma}\ insert\ assms{\isacharparenleft}{\isadigit{2}}{\isacharparenright}{\isacharcomma}\ simp{\isacharparenright}\isanewline
\isacommand{qed}\isamarkupfalse%
\endisatagproof
{\isafoldproof}%
\isadelimproof
\isanewline
\endisadelimproof
\isanewline
\isacommand{lemma}\isamarkupfalse%
\ same{\isacharunderscore}nodes{\isacharcolon}\ {\isachardoublequoteopen}{\isacharparenleft}I{\isacharcomma}\ y{\isacharparenright}\ {\isasymin}\ {\isacharbraceleft}{\isacharparenleft}x{\isacharcolon}{\isacharcolon}infrastructure{\isacharcomma}\ y{\isacharcolon}{\isacharcolon}infrastructure{\isacharparenright}{\isachardot}\ x\ {\isasymrightarrow}\isactrlsub n\ y{\isacharbraceright}\isactrlsup {\isacharasterisk}\ \isanewline
\ \ \ \ \ \ \ \ \ \ \ \ \ \ \ \ \ \ \ {\isasymLongrightarrow}\ nodes{\isacharparenleft}graphI\ y{\isacharparenright}\ {\isacharequal}\ nodes{\isacharparenleft}graphI\ I{\isacharparenright}{\isachardoublequoteclose}\ \ \isanewline
\prf\isanewline\isanewline
\isacommand{lemma}\isamarkupfalse%
\ same{\isacharunderscore}actors{\isacharcolon}\ {\isachardoublequoteopen}{\isacharparenleft}I{\isacharcomma}\ y{\isacharparenright}\ {\isasymin}\ {\isacharbraceleft}{\isacharparenleft}x{\isacharcolon}{\isacharcolon}infrastructure{\isacharcomma}\ y{\isacharcolon}{\isacharcolon}infrastructure{\isacharparenright}{\isachardot}\ x\ {\isasymrightarrow}\isactrlsub n\ y{\isacharbraceright}\isactrlsup {\isacharasterisk}\ \isanewline
\ \ \ \ \ \ \ \ \ \ \ \ \ \ {\isasymLongrightarrow}\ actors{\isacharunderscore}graph{\isacharparenleft}graphI\ I{\isacharparenright}\ {\isacharequal}\ actors{\isacharunderscore}graph{\isacharparenleft}graphI\ y{\isacharparenright}{\isachardoublequoteclose}\isanewline
\prf\isanewline\isanewline
\isacommand{end}\isamarkupfalse%
\isanewline
\isadelimtheory
\endisadelimtheory
\isatagtheory
\isacommand{end}\isamarkupfalse%
\endisatagtheory
{\isafoldtheory}%
\isadelimtheory
\endisadelimtheory
\end{isabellebody}%

\subsection{Airplane}
\begin{isabellebody}%
\setisabellecontext{Airplane}%
\isadelimtheory
\endisadelimtheory
\isatagtheory
\isacommand{theory}\isamarkupfalse%
\ Airplane\isanewline
\isakeyword{imports}\ AirInsider\isanewline
\isakeyword{begin}%
\endisatagtheory
{\isafoldtheory}%
\isadelimtheory
\isanewline
\endisadelimtheory
\isacommand{datatype}\isamarkupfalse%
\ doorstate\ {\isacharequal}\ locked\ {\isacharbar}\ norm\ {\isacharbar}\ unlocked\isanewline
\isacommand{datatype}\isamarkupfalse%
\ position\ {\isacharequal}\ air\ {\isacharbar}\ airport\ {\isacharbar}\ ground\isanewline
\isanewline
\isacommand{locale}\isamarkupfalse%
\ airplane\ {\isacharequal}\isanewline
\isakeyword{fixes}\ airplane{\isacharunderscore}actors\ {\isacharcolon}{\isacharcolon}\ {\isachardoublequoteopen}identity\ set{\isachardoublequoteclose}\isanewline
\isakeyword{defines}\ airplane{\isacharunderscore}actors{\isacharunderscore}def{\isacharcolon}\ {\isachardoublequoteopen}airplane{\isacharunderscore}actors\ {\isasymequiv}\ {\isacharbraceleft}{\isacharprime}{\isacharprime}Bob{\isacharprime}{\isacharprime}{\isacharcomma}\ {\isacharprime}{\isacharprime}Charly{\isacharprime}{\isacharprime}{\isacharcomma}\ {\isacharprime}{\isacharprime}Alice{\isacharprime}{\isacharprime}{\isacharbraceright}{\isachardoublequoteclose}\isanewline
\isakeyword{fixes}\ airplane{\isacharunderscore}locations\ {\isacharcolon}{\isacharcolon}\ {\isachardoublequoteopen}location\ set{\isachardoublequoteclose}\isanewline
\isakeyword{defines}\ airplane{\isacharunderscore}locations{\isacharunderscore}def{\isacharcolon}\ \isanewline
{\isachardoublequoteopen}airplane{\isacharunderscore}locations\ {\isasymequiv}\ {\isacharbraceleft}Location\ {\isadigit{0}}{\isacharcomma}\ Location\ {\isadigit{1}}{\isacharcomma}\ Location\ {\isadigit{2}}{\isacharbraceright}{\isachardoublequoteclose}\isanewline
\isakeyword{fixes}\ cockpit\ {\isacharcolon}{\isacharcolon}\ {\isachardoublequoteopen}location{\isachardoublequoteclose}\isanewline
\isakeyword{defines}\ cockpit{\isacharunderscore}def{\isacharcolon}\ {\isachardoublequoteopen}cockpit\ {\isasymequiv}\ Location\ {\isadigit{2}}{\isachardoublequoteclose}\ \isanewline
\isakeyword{fixes}\ door\ {\isacharcolon}{\isacharcolon}\ {\isachardoublequoteopen}location{\isachardoublequoteclose}\isanewline
\isakeyword{defines}\ door{\isacharunderscore}def{\isacharcolon}\ {\isachardoublequoteopen}door\ {\isasymequiv}\ Location\ {\isadigit{1}}{\isachardoublequoteclose}\ \isanewline
\isakeyword{fixes}\ cabin\ {\isacharcolon}{\isacharcolon}\ {\isachardoublequoteopen}location{\isachardoublequoteclose}\isanewline
\isakeyword{defines}\ cabin{\isacharunderscore}def{\isacharcolon}\ {\isachardoublequoteopen}cabin\ {\isasymequiv}\ Location\ {\isadigit{0}}{\isachardoublequoteclose}\ \isanewline
\isanewline
\isakeyword{fixes}\ global{\isacharunderscore}policy\ {\isacharcolon}{\isacharcolon}\ {\isachardoublequoteopen}{\isacharbrackleft}infrastructure{\isacharcomma}\ identity{\isacharbrackright}\ {\isasymRightarrow}\ bool{\isachardoublequoteclose}\isanewline
\isakeyword{defines}\ global{\isacharunderscore}policy{\isacharunderscore}def{\isacharcolon}\ {\isachardoublequoteopen}global{\isacharunderscore}policy\ I\ a\ {\isasymequiv}\ a\ {\isasymnotin}\ airplane{\isacharunderscore}actors\ \isanewline
\ \ \ \ \ \ \ \ \ \ \ \ \ \ \ \ \ {\isasymlongrightarrow}\ {\isasymnot}{\isacharparenleft}enables\ I\ cockpit\ {\isacharparenleft}Actor\ a{\isacharparenright}\ put{\isacharparenright}{\isachardoublequoteclose}\isanewline
\isanewline
\isakeyword{fixes}\ ex{\isacharunderscore}creds\ {\isacharcolon}{\isacharcolon}\ {\isachardoublequoteopen}actor\ \ {\isasymRightarrow}\ {\isacharparenleft}string\ list\ {\isacharasterisk}\ string\ list{\isacharparenright}{\isachardoublequoteclose}\isanewline
\isakeyword{defines}\ ex{\isacharunderscore}creds{\isacharunderscore}def{\isacharcolon}\ {\isachardoublequoteopen}ex{\isacharunderscore}creds\ {\isasymequiv}\ \isanewline
\ \ \ \ \ \ \ \ {\isacharparenleft}{\isasymlambda}\ x{\isachardot}{\isacharparenleft}if\ x\ {\isacharequal}\ Actor\ {\isacharprime}{\isacharprime}Bob{\isacharprime}{\isacharprime}\ \isanewline
\ \ \ \ \ \ \ \ \ \ \ \ \ \ then\ {\isacharparenleft}{\isacharbrackleft}{\isacharprime}{\isacharprime}PIN{\isacharprime}{\isacharprime}{\isacharbrackright}{\isacharcomma}\ {\isacharbrackleft}{\isacharprime}{\isacharprime}pilot{\isacharprime}{\isacharprime}{\isacharbrackright}{\isacharparenright}\ \ \ \ \ \ \ \ \isanewline
\ \ \ \ \ \ \ \ \ \ \ \ \ \ else\ {\isacharparenleft}if\ x\ {\isacharequal}\ Actor\ {\isacharprime}{\isacharprime}Charly{\isacharprime}{\isacharprime}\ \isanewline
\ \ \ \ \ \ \ \ \ \ \ \ \ \ \ \ \ \ \ \ then\ {\isacharparenleft}{\isacharbrackleft}{\isacharprime}{\isacharprime}PIN{\isacharprime}{\isacharprime}{\isacharbrackright}{\isacharcomma}{\isacharbrackleft}{\isacharprime}{\isacharprime}copilot{\isacharprime}{\isacharprime}{\isacharbrackright}{\isacharparenright}\isanewline
\ \ \ \ \ \ \ \ \ \ \ \ \ \ \ \ \ \ \ \ else\ {\isacharparenleft}if\ x\ {\isacharequal}\ Actor\ {\isacharprime}{\isacharprime}Alice{\isacharprime}{\isacharprime}\ \ \isanewline
\ \ \ \ \ \ \ \ \ \ \ \ \ \ \ \ \ \ \ \ \ \ \ \ \ \ then\ {\isacharparenleft}{\isacharbrackleft}{\isacharprime}{\isacharprime}PIN{\isacharprime}{\isacharprime}{\isacharbrackright}{\isacharcomma}{\isacharbrackleft}{\isacharprime}{\isacharprime}flightattendant{\isacharprime}{\isacharprime}{\isacharbrackright}{\isacharparenright}\isanewline
\ \ \ \ \ \ \ \ \ \ \ \ \ \ \ \ \ \ \ \ \ \ \ \ \ \ \ \ \ \ \ \ else\ {\isacharparenleft}{\isacharbrackleft}{\isacharbrackright}{\isacharcomma}{\isacharbrackleft}{\isacharbrackright}{\isacharparenright}{\isacharparenright}{\isacharparenright}{\isacharparenright}{\isacharparenright}{\isachardoublequoteclose}\isanewline
\isanewline
\isakeyword{fixes}\ ex{\isacharunderscore}locs\ {\isacharcolon}{\isacharcolon}\ {\isachardoublequoteopen}location\ {\isasymRightarrow}\ string\ list{\isachardoublequoteclose}\isanewline
\isakeyword{defines}\ ex{\isacharunderscore}locs{\isacharunderscore}def{\isacharcolon}\ {\isachardoublequoteopen}ex{\isacharunderscore}locs\ {\isasymequiv}\ \ {\isacharparenleft}{\isasymlambda}\ x{\isachardot}\ if\ x\ {\isacharequal}\ door\ then\ {\isacharbrackleft}{\isacharprime}{\isacharprime}norm{\isacharprime}{\isacharprime}{\isacharbrackright}\ else\ \isanewline
\ \ \ \ \ \ \ \ \ \ \ \ \ \ \ \ \ \ \ \ \ \ \ \ \ \ \ \ \ \ \ \ \ \ \ \ \ \ \ {\isacharparenleft}if\ x\ {\isacharequal}\ cockpit\ then\ {\isacharbrackleft}{\isacharprime}{\isacharprime}air{\isacharprime}{\isacharprime}{\isacharbrackright}\ else\ {\isacharbrackleft}{\isacharbrackright}{\isacharparenright}{\isacharparenright}{\isachardoublequoteclose}\isanewline
\ \ \isanewline
\isanewline
\isakeyword{fixes}\ ex{\isacharunderscore}locs{\isacharprime}\ {\isacharcolon}{\isacharcolon}\ {\isachardoublequoteopen}location\ {\isasymRightarrow}\ string\ list{\isachardoublequoteclose}\isanewline
\isakeyword{defines}\ ex{\isacharunderscore}locs{\isacharprime}{\isacharunderscore}def{\isacharcolon}\ {\isachardoublequoteopen}ex{\isacharunderscore}locs{\isacharprime}\ {\isasymequiv}\ \ {\isacharparenleft}{\isasymlambda}\ x{\isachardot}\ if\ x\ {\isacharequal}\ door\ then\ {\isacharbrackleft}{\isacharprime}{\isacharprime}locked{\isacharprime}{\isacharprime}{\isacharbrackright}\ else\isanewline
\ \ \ \ \ \ \ \ \ \ \ \ \ \ \ \ \ \ \ \ \ \ \ \ \ \ \ \ \ \ \ \ \ \ \ \ \ \ \ \ \ {\isacharparenleft}if\ x\ {\isacharequal}\ cockpit\ then\ {\isacharbrackleft}{\isacharprime}{\isacharprime}air{\isacharprime}{\isacharprime}{\isacharbrackright}\ else\ {\isacharbrackleft}{\isacharbrackright}{\isacharparenright}{\isacharparenright}{\isachardoublequoteclose}\isanewline
\isakeyword{fixes}\ ex{\isacharunderscore}graph\ {\isacharcolon}{\isacharcolon}\ {\isachardoublequoteopen}igraph{\isachardoublequoteclose}\isanewline
\isakeyword{defines}\ ex{\isacharunderscore}graph{\isacharunderscore}def{\isacharcolon}\ {\isachardoublequoteopen}ex{\isacharunderscore}graph\ {\isasymequiv}\ Lgraph\isanewline
\ \ \ \ \ \ {\isacharbraceleft}{\isacharparenleft}cockpit{\isacharcomma}\ door{\isacharparenright}{\isacharcomma}{\isacharparenleft}door{\isacharcomma}cabin{\isacharparenright}{\isacharbraceright}\isanewline
\ \ \ \ \ \ {\isacharparenleft}{\isasymlambda}\ x{\isachardot}\ if\ x\ {\isacharequal}\ cockpit\ then\ {\isacharbrackleft}{\isacharprime}{\isacharprime}Bob{\isacharprime}{\isacharprime}{\isacharcomma}\ {\isacharprime}{\isacharprime}Charly{\isacharprime}{\isacharprime}{\isacharbrackright}\ \isanewline
\ \ \ \ \ \ \ \ \ \ \ \ else\ {\isacharparenleft}if\ x\ {\isacharequal}\ door\ then\ {\isacharbrackleft}{\isacharbrackright}\ \isanewline
\ \ \ \ \ \ \ \ \ \ \ \ \ \ \ \ \ \ else\ {\isacharparenleft}if\ x\ {\isacharequal}\ cabin\ then\ {\isacharbrackleft}{\isacharprime}{\isacharprime}Alice{\isacharprime}{\isacharprime}{\isacharbrackright}\ else\ {\isacharbrackleft}{\isacharbrackright}{\isacharparenright}{\isacharparenright}{\isacharparenright}\isanewline
\ \ \ \ \ \ ex{\isacharunderscore}creds\ ex{\isacharunderscore}locs{\isachardoublequoteclose}\isanewline
\isanewline
\isakeyword{fixes}\ aid{\isacharunderscore}graph\ {\isacharcolon}{\isacharcolon}\ {\isachardoublequoteopen}igraph{\isachardoublequoteclose}\isanewline
\isakeyword{defines}\ aid{\isacharunderscore}graph{\isacharunderscore}def{\isacharcolon}\ {\isachardoublequoteopen}aid{\isacharunderscore}graph\ {\isasymequiv}\ \ Lgraph\isanewline
\ \ \ \ \ \ {\isacharbraceleft}{\isacharparenleft}cockpit{\isacharcomma}\ door{\isacharparenright}{\isacharcomma}{\isacharparenleft}door{\isacharcomma}cabin{\isacharparenright}{\isacharbraceright}\isanewline
\ \ \ \ \ \ {\isacharparenleft}{\isasymlambda}\ x{\isachardot}\ if\ x\ {\isacharequal}\ cockpit\ then\ {\isacharbrackleft}{\isacharprime}{\isacharprime}Charly{\isacharprime}{\isacharprime}{\isacharbrackright}\ \isanewline
\ \ \ \ \ \ \ \ \ \ \ \ else\ {\isacharparenleft}if\ x\ {\isacharequal}\ door\ then\ {\isacharbrackleft}{\isacharbrackright}\ \isanewline
\ \ \ \ \ \ \ \ \ \ \ \ \ \ \ \ \ \ else\ {\isacharparenleft}if\ x\ {\isacharequal}\ cabin\ then\ {\isacharbrackleft}{\isacharprime}{\isacharprime}Bob{\isacharprime}{\isacharprime}{\isacharcomma}\ {\isacharprime}{\isacharprime}Alice{\isacharprime}{\isacharprime}{\isacharbrackright}\ else\ {\isacharbrackleft}{\isacharbrackright}{\isacharparenright}{\isacharparenright}{\isacharparenright}\isanewline
\ \ \ \ \ \ ex{\isacharunderscore}creds\ ex{\isacharunderscore}locs{\isacharprime}{\isachardoublequoteclose}\isanewline
\ \ \isanewline
\isakeyword{fixes}\ aid{\isacharunderscore}graph{\isadigit{0}}\ {\isacharcolon}{\isacharcolon}\ {\isachardoublequoteopen}igraph{\isachardoublequoteclose}\isanewline
\isakeyword{defines}\ aid{\isacharunderscore}graph{\isadigit{0}}{\isacharunderscore}def{\isacharcolon}\ {\isachardoublequoteopen}aid{\isacharunderscore}graph{\isadigit{0}}\ {\isasymequiv}\ \ Lgraph\isanewline
\ \ \ \ \ \ {\isacharbraceleft}{\isacharparenleft}cockpit{\isacharcomma}\ door{\isacharparenright}{\isacharcomma}{\isacharparenleft}door{\isacharcomma}cabin{\isacharparenright}{\isacharbraceright}\isanewline
\ \ \ \ \ \ {\isacharparenleft}{\isasymlambda}\ x{\isachardot}\ if\ x\ {\isacharequal}\ cockpit\ then\ {\isacharbrackleft}{\isacharprime}{\isacharprime}Charly{\isacharprime}{\isacharprime}{\isacharbrackright}\ \isanewline
\ \ \ \ \ \ \ \ \ \ \ \ else\ {\isacharparenleft}if\ x\ {\isacharequal}\ door\ then\ {\isacharbrackleft}{\isacharprime}{\isacharprime}Bob{\isacharprime}{\isacharprime}{\isacharbrackright}\ \isanewline
\ \ \ \ \ \ \ \ \ \ \ \ \ \ \ \ \ \ else\ {\isacharparenleft}if\ x\ {\isacharequal}\ cabin\ then\ {\isacharbrackleft}{\isacharprime}{\isacharprime}Alice{\isacharprime}{\isacharprime}{\isacharbrackright}\ else\ {\isacharbrackleft}{\isacharbrackright}{\isacharparenright}{\isacharparenright}{\isacharparenright}\isanewline
\ \ \ \ \ \ \ \ ex{\isacharunderscore}creds\ ex{\isacharunderscore}locs{\isachardoublequoteclose}\isanewline
\isanewline
\isakeyword{fixes}\ agid{\isacharunderscore}graph\ {\isacharcolon}{\isacharcolon}\ {\isachardoublequoteopen}igraph{\isachardoublequoteclose}\isanewline
\isakeyword{defines}\ agid{\isacharunderscore}graph{\isacharunderscore}def{\isacharcolon}\ {\isachardoublequoteopen}agid{\isacharunderscore}graph\ {\isasymequiv}\ \ Lgraph\isanewline
\ \ \ \ \ \ {\isacharbraceleft}{\isacharparenleft}cockpit{\isacharcomma}\ door{\isacharparenright}{\isacharcomma}{\isacharparenleft}door{\isacharcomma}cabin{\isacharparenright}{\isacharbraceright}\isanewline
\ \ \ \ \ \ {\isacharparenleft}{\isasymlambda}\ x{\isachardot}\ if\ x\ {\isacharequal}\ cockpit\ then\ {\isacharbrackleft}{\isacharprime}{\isacharprime}Charly{\isacharprime}{\isacharprime}{\isacharbrackright}\ \isanewline
\ \ \ \ \ \ \ \ \ \ \ \ else\ {\isacharparenleft}if\ x\ {\isacharequal}\ door\ then\ {\isacharbrackleft}{\isacharbrackright}\ \isanewline
\ \ \ \ \ \ \ \ \ \ \ \ \ \ \ \ \ \ else\ {\isacharparenleft}if\ x\ {\isacharequal}\ cabin\ then\ {\isacharbrackleft}{\isacharprime}{\isacharprime}Bob{\isacharprime}{\isacharprime}{\isacharcomma}\ {\isacharprime}{\isacharprime}Alice{\isacharprime}{\isacharprime}{\isacharbrackright}\ else\ {\isacharbrackleft}{\isacharbrackright}{\isacharparenright}{\isacharparenright}{\isacharparenright}\isanewline
\ \ \ \ \ \ ex{\isacharunderscore}creds\ ex{\isacharunderscore}locs{\isachardoublequoteclose}\isanewline
\ \ \isanewline
\isakeyword{fixes}\ local{\isacharunderscore}policies\ {\isacharcolon}{\isacharcolon}\ {\isachardoublequoteopen}{\isacharbrackleft}igraph{\isacharcomma}\ \ location{\isacharbrackright}\ {\isasymRightarrow}\ policy\ set{\isachardoublequoteclose}\isanewline
\isakeyword{defines}\ local{\isacharunderscore}policies{\isacharunderscore}def{\isacharcolon}\ {\isachardoublequoteopen}local{\isacharunderscore}policies\ G\ {\isasymequiv}\ \ \isanewline
\ \ \ {\isacharparenleft}{\isasymlambda}\ y{\isachardot}\ if\ y\ {\isacharequal}\ cockpit\ then\isanewline
\ \ \ \ \ \ \ \ \ \ \ \ \ {\isacharbraceleft}{\isacharparenleft}{\isasymlambda}\ x{\isachardot}\ {\isacharparenleft}{\isacharquery}\ n{\isachardot}\ {\isacharparenleft}n\ {\isacharat}\isactrlbsub G\isactrlesub \ cockpit{\isacharparenright}\ {\isasymand}\ Actor\ n\ {\isacharequal}\ x{\isacharparenright}{\isacharcomma}\ {\isacharbraceleft}put{\isacharbraceright}{\isacharparenright}{\isacharcomma}\isanewline
\ \ \ \ \ \ \ \ \ \ \ \ \ \ {\isacharparenleft}{\isasymlambda}\ x{\isachardot}\ {\isacharparenleft}{\isacharquery}\ n{\isachardot}\ {\isacharparenleft}n\ {\isacharat}\isactrlbsub G\isactrlesub \ cabin{\isacharparenright}\ {\isasymand}\ Actor\ n\ {\isacharequal}\ x\ {\isasymand}\ has\ G\ {\isacharparenleft}x{\isacharcomma}\ {\isacharprime}{\isacharprime}PIN{\isacharprime}{\isacharprime}{\isacharparenright}\isanewline
\ \ \ \ \ \ \ \ \ \ \ \ \ \ \ \ \ \ \ \ {\isasymand}\ isin\ G\ door\ {\isacharprime}{\isacharprime}norm{\isacharprime}{\isacharprime}{\isacharparenright}{\isacharcomma}{\isacharbraceleft}move{\isacharbraceright}{\isacharparenright}\isanewline
\ \ \ \ \ \ \ \ \ \ \ \ \ {\isacharbraceright}\isanewline
\ \ \ \ \ \ \ \ \ else\ {\isacharparenleft}if\ y\ {\isacharequal}\ door\ then\ {\isacharbraceleft}{\isacharparenleft}{\isasymlambda}\ x{\isachardot}\ True{\isacharcomma}\ {\isacharbraceleft}move{\isacharbraceright}{\isacharparenright}{\isacharcomma}\isanewline
\ \ \ \ \ \ \ \ \ \ \ \ \ \ \ \ \ \ \ \ \ \ \ {\isacharparenleft}{\isasymlambda}\ x{\isachardot}\ {\isacharparenleft}{\isacharquery}\ n{\isachardot}\ {\isacharparenleft}n\ {\isacharat}\isactrlbsub G\isactrlesub \ cockpit{\isacharparenright}\ {\isasymand}\ Actor\ n\ {\isacharequal}\ x{\isacharparenright}{\isacharcomma}\ {\isacharbraceleft}put{\isacharbraceright}{\isacharparenright}{\isacharbraceright}\isanewline
\ \ \ \ \ \ \ \ \ \ \ \ \ \ \ else\ {\isacharparenleft}if\ y\ {\isacharequal}\ cabin\ then\ {\isacharbraceleft}{\isacharparenleft}{\isasymlambda}\ x{\isachardot}\ True{\isacharcomma}\ {\isacharbraceleft}move{\isacharbraceright}{\isacharparenright}{\isacharbraceright}\ \isanewline
\ \ \ \ \ \ \ \ \ \ \ \ \ \ \ \ \ \ \ \ \ else\ {\isacharbraceleft}{\isacharbraceright}{\isacharparenright}{\isacharparenright}{\isacharparenright}{\isachardoublequoteclose}\isanewline
\isanewline
\isakeyword{fixes}\ local{\isacharunderscore}policies{\isacharunderscore}four{\isacharunderscore}eyes\ {\isacharcolon}{\isacharcolon}\ {\isachardoublequoteopen}{\isacharbrackleft}igraph{\isacharcomma}\ location{\isacharbrackright}\ {\isasymRightarrow}\ policy\ set{\isachardoublequoteclose}\isanewline
\isakeyword{defines}\ local{\isacharunderscore}policies{\isacharunderscore}four{\isacharunderscore}eyes{\isacharunderscore}def{\isacharcolon}\ {\isachardoublequoteopen}local{\isacharunderscore}policies{\isacharunderscore}four{\isacharunderscore}eyes\ G\ {\isasymequiv}\ \ \isanewline
\ \ \ {\isacharparenleft}{\isasymlambda}\ y{\isachardot}\ if\ y\ {\isacharequal}\ cockpit\ then\isanewline
\ \ \ \ \ \ \ \ \ \ \ \ \ {\isacharbraceleft}{\isacharparenleft}{\isasymlambda}\ x{\isachardot}\ \ {\isacharparenleft}{\isacharquery}\ n{\isachardot}\ {\isacharparenleft}n\ {\isacharat}\isactrlbsub G\isactrlesub \ cockpit{\isacharparenright}\ {\isasymand}\ Actor\ n\ {\isacharequal}\ x{\isacharparenright}\ {\isasymand}\isanewline
\ \ \ \ \ \ \ \ \ \ \ \ \ \ \ \ \ \ \ \ \ {\isadigit{2}}\ {\isasymle}\ length{\isacharparenleft}agra\ G\ y{\isacharparenright}\ {\isasymand}\ {\isacharparenleft}{\isasymforall}\ h\ {\isasymin}\ set{\isacharparenleft}agra\ G\ y{\isacharparenright}{\isachardot}\ h\ {\isasymin}\ airplane{\isacharunderscore}actors{\isacharparenright}{\isacharcomma}\ {\isacharbraceleft}put{\isacharbraceright}{\isacharparenright}{\isacharcomma}\isanewline
\ \ \ \ \ \ \ \ \ \ \ \ \ \ {\isacharparenleft}{\isasymlambda}\ x{\isachardot}\ {\isacharparenleft}{\isacharquery}\ n{\isachardot}\ {\isacharparenleft}n\ {\isacharat}\isactrlbsub G\isactrlesub \ cabin{\isacharparenright}\ {\isasymand}\ Actor\ n\ {\isacharequal}\ x\ {\isasymand}\ has\ G\ {\isacharparenleft}x{\isacharcomma}\ {\isacharprime}{\isacharprime}PIN{\isacharprime}{\isacharprime}{\isacharparenright}\ {\isasymand}\ \isanewline
\ \ \ \ \ \ \ \ \ \ \ \ \ \ \ \ \ \ \ \ \ \ \ \ \ \ \ isin\ G\ door\ {\isacharprime}{\isacharprime}norm{\isacharprime}{\isacharprime}\ {\isacharparenright}{\isacharcomma}{\isacharbraceleft}move{\isacharbraceright}{\isacharparenright}\isanewline
\ \ \ \ \ \ \ \ \ \ \ \ \ {\isacharbraceright}\isanewline
\ \ \ \ \ \ \ \ \ else\ {\isacharparenleft}if\ y\ {\isacharequal}\ door\ then\ \isanewline
\ \ \ \ \ \ \ \ \ \ \ \ \ \ {\isacharbraceleft}{\isacharparenleft}{\isasymlambda}\ x{\isachardot}\ \ {\isacharparenleft}{\isacharparenleft}{\isacharquery}\ n{\isachardot}\ {\isacharparenleft}n\ {\isacharat}\isactrlbsub G\isactrlesub \ cockpit{\isacharparenright}\ {\isasymand}\ Actor\ n\ {\isacharequal}\ x{\isacharparenright}\ {\isasymand}\ {\isadigit{3}}\ {\isasymle}\ length{\isacharparenleft}agra\ G\ cockpit{\isacharparenright}{\isacharparenright}{\isacharcomma}\ {\isacharbraceleft}move{\isacharbraceright}{\isacharparenright}{\isacharbraceright}\isanewline
\ \ \ \ \ \ \ \ \ \ \ \ \ \ \ else\ {\isacharparenleft}if\ y\ {\isacharequal}\ cabin\ then\ \isanewline
\ \ \ \ \ \ \ \ \ \ \ \ \ \ \ \ \ \ \ \ \ {\isacharbraceleft}{\isacharparenleft}{\isasymlambda}\ x{\isachardot}\ {\isacharparenleft}{\isacharparenleft}{\isacharquery}\ n{\isachardot}\ {\isacharparenleft}n\ {\isacharat}\isactrlbsub G\isactrlesub \ door{\isacharparenright}\ {\isasymand}\ Actor\ n\ {\isacharequal}\ x{\isacharparenright}{\isacharparenright}{\isacharcomma}\ {\isacharbraceleft}move{\isacharbraceright}{\isacharparenright}{\isacharbraceright}\ \isanewline
\ \ \ \ \ \ \ \ \ \ \ \ \ \ \ \ \ \ \ \ \ \ \ \ \ \ \ else\ {\isacharbraceleft}{\isacharbraceright}{\isacharparenright}{\isacharparenright}{\isacharparenright}{\isachardoublequoteclose}\isanewline
\isanewline
\isakeyword{fixes}\ Airplane{\isacharunderscore}scenario\ {\isacharcolon}{\isacharcolon}\ {\isachardoublequoteopen}infrastructure{\isachardoublequoteclose}\ {\isacharparenleft}\isakeyword{structure}{\isacharparenright}\isanewline
\isakeyword{defines}\ Airplane{\isacharunderscore}scenario{\isacharunderscore}def{\isacharcolon}\isanewline
{\isachardoublequoteopen}Airplane{\isacharunderscore}scenario\ {\isasymequiv}\ Infrastructure\ ex{\isacharunderscore}graph\ local{\isacharunderscore}policies{\isachardoublequoteclose}\isanewline
\isanewline
\isakeyword{fixes}\ Airplane{\isacharunderscore}in{\isacharunderscore}danger\ {\isacharcolon}{\isacharcolon}\ {\isachardoublequoteopen}infrastructure{\isachardoublequoteclose}\isanewline
\isakeyword{defines}\ Airplane{\isacharunderscore}in{\isacharunderscore}danger{\isacharunderscore}def{\isacharcolon}\isanewline
{\isachardoublequoteopen}Airplane{\isacharunderscore}in{\isacharunderscore}danger\ {\isasymequiv}\ Infrastructure\ aid{\isacharunderscore}graph\ local{\isacharunderscore}policies{\isachardoublequoteclose}\isanewline
\isanewline
\isakeyword{fixes}\ Airplane{\isacharunderscore}getting{\isacharunderscore}in{\isacharunderscore}danger{\isadigit{0}}\ {\isacharcolon}{\isacharcolon}\ {\isachardoublequoteopen}infrastructure{\isachardoublequoteclose}\isanewline
\isakeyword{defines}\ Airplane{\isacharunderscore}getting{\isacharunderscore}in{\isacharunderscore}danger{\isadigit{0}}{\isacharunderscore}def{\isacharcolon}\isanewline
{\isachardoublequoteopen}Airplane{\isacharunderscore}getting{\isacharunderscore}in{\isacharunderscore}danger{\isadigit{0}}\ {\isasymequiv}\ Infrastructure\ aid{\isacharunderscore}graph{\isadigit{0}}\ local{\isacharunderscore}policies{\isachardoublequoteclose}\isanewline
\isanewline
\isakeyword{fixes}\ Airplane{\isacharunderscore}getting{\isacharunderscore}in{\isacharunderscore}danger\ {\isacharcolon}{\isacharcolon}\ {\isachardoublequoteopen}infrastructure{\isachardoublequoteclose}\isanewline
\isakeyword{defines}\ Airplane{\isacharunderscore}getting{\isacharunderscore}in{\isacharunderscore}danger{\isacharunderscore}def{\isacharcolon}\isanewline
{\isachardoublequoteopen}Airplane{\isacharunderscore}getting{\isacharunderscore}in{\isacharunderscore}danger\ {\isasymequiv}\ Infrastructure\ agid{\isacharunderscore}graph\ local{\isacharunderscore}policies{\isachardoublequoteclose}\isanewline
\isanewline
\isakeyword{fixes}\ Air{\isacharunderscore}states\isanewline
\isakeyword{defines}\ Air{\isacharunderscore}states{\isacharunderscore}def{\isacharcolon}\ {\isachardoublequoteopen}Air{\isacharunderscore}states\ {\isasymequiv}\ {\isacharbraceleft}\ I{\isachardot}\ Airplane{\isacharunderscore}scenario\ {\isasymrightarrow}\isactrlsub n{\isacharasterisk}\ I\ {\isacharbraceright}{\isachardoublequoteclose}\isanewline
\isanewline
\isakeyword{fixes}\ Air{\isacharunderscore}Kripke\isanewline
\isakeyword{defines}\ {\isachardoublequoteopen}Air{\isacharunderscore}Kripke\ {\isasymequiv}\ Kripke\ Air{\isacharunderscore}states\ {\isacharbraceleft}Airplane{\isacharunderscore}scenario{\isacharbraceright}{\isachardoublequoteclose}\isanewline
\isanewline
\isakeyword{fixes}\ Airplane{\isacharunderscore}not{\isacharunderscore}in{\isacharunderscore}danger\ {\isacharcolon}{\isacharcolon}\ {\isachardoublequoteopen}infrastructure{\isachardoublequoteclose}\isanewline
\isakeyword{defines}\ Airplane{\isacharunderscore}not{\isacharunderscore}in{\isacharunderscore}danger{\isacharunderscore}def{\isacharcolon}\isanewline
{\isachardoublequoteopen}Airplane{\isacharunderscore}not{\isacharunderscore}in{\isacharunderscore}danger\ {\isasymequiv}\ Infrastructure\ aid{\isacharunderscore}graph\ local{\isacharunderscore}policies{\isacharunderscore}four{\isacharunderscore}eyes{\isachardoublequoteclose}\isanewline
\isanewline
\isakeyword{fixes}\ Airplane{\isacharunderscore}not{\isacharunderscore}in{\isacharunderscore}danger{\isacharunderscore}init\ {\isacharcolon}{\isacharcolon}\ {\isachardoublequoteopen}infrastructure{\isachardoublequoteclose}\isanewline
\isakeyword{defines}\ Airplane{\isacharunderscore}not{\isacharunderscore}in{\isacharunderscore}danger{\isacharunderscore}init{\isacharunderscore}def{\isacharcolon}\isanewline
{\isachardoublequoteopen}Airplane{\isacharunderscore}not{\isacharunderscore}in{\isacharunderscore}danger{\isacharunderscore}init\ {\isasymequiv}\ Infrastructure\ ex{\isacharunderscore}graph\ local{\isacharunderscore}policies{\isacharunderscore}four{\isacharunderscore}eyes{\isachardoublequoteclose}\isanewline
\isanewline
\isakeyword{fixes}\ Air{\isacharunderscore}tp{\isacharunderscore}states\isanewline
\isakeyword{defines}\ Air{\isacharunderscore}tp{\isacharunderscore}states{\isacharunderscore}def{\isacharcolon}\ {\isachardoublequoteopen}Air{\isacharunderscore}tp{\isacharunderscore}states\ {\isasymequiv}\ {\isacharbraceleft}\ I{\isachardot}\ Airplane{\isacharunderscore}not{\isacharunderscore}in{\isacharunderscore}danger{\isacharunderscore}init\ {\isasymrightarrow}\isactrlsub n{\isacharasterisk}\ I\ {\isacharbraceright}{\isachardoublequoteclose}\isanewline
\isanewline
\isakeyword{fixes}\ Air{\isacharunderscore}tp{\isacharunderscore}Kripke\isanewline
\isakeyword{defines}\ {\isachardoublequoteopen}Air{\isacharunderscore}tp{\isacharunderscore}Kripke\ {\isasymequiv}\ Kripke\ Air{\isacharunderscore}tp{\isacharunderscore}states\ {\isacharbraceleft}Airplane{\isacharunderscore}not{\isacharunderscore}in{\isacharunderscore}danger{\isacharunderscore}init{\isacharbraceright}{\isachardoublequoteclose}\isanewline
\isanewline
\isakeyword{fixes}\ Safety\ {\isacharcolon}{\isacharcolon}\ {\isachardoublequoteopen}{\isacharbrackleft}infrastructure{\isacharcomma}\ identity{\isacharbrackright}\ {\isasymRightarrow}\ bool{\isachardoublequoteclose}\isanewline
\isakeyword{defines}\ Safety{\isacharunderscore}def{\isacharcolon}\ {\isachardoublequoteopen}Safety\ I\ a\ {\isasymequiv}\ a\ {\isasymin}\ airplane{\isacharunderscore}actors\ \ \isanewline
\ \ \ \ \ \ \ \ \ \ \ \ \ \ \ \ \ \ \ \ \ \ \ {\isasymlongrightarrow}\ {\isacharparenleft}enables\ I\ cockpit\ {\isacharparenleft}Actor\ a{\isacharparenright}\ move{\isacharparenright}{\isachardoublequoteclose}\isanewline
\isanewline
\isakeyword{fixes}\ Security\ {\isacharcolon}{\isacharcolon}\ {\isachardoublequoteopen}{\isacharbrackleft}infrastructure{\isacharcomma}\ identity{\isacharbrackright}\ {\isasymRightarrow}\ bool{\isachardoublequoteclose}\isanewline
\isakeyword{defines}\ Security{\isacharunderscore}def{\isacharcolon}\ {\isachardoublequoteopen}Security\ I\ a\ {\isasymequiv}\ \ {\isacharparenleft}isin\ {\isacharparenleft}graphI\ I{\isacharparenright}\ door\ {\isacharprime}{\isacharprime}locked{\isacharprime}{\isacharprime}{\isacharparenright}\ \isanewline
\ \ \ \ \ \ \ \ \ \ \ \ \ \ \ \ \ \ \ \ \ \ \ {\isasymlongrightarrow}\ {\isasymnot}{\isacharparenleft}enables\ I\ cockpit\ {\isacharparenleft}Actor\ a{\isacharparenright}\ move{\isacharparenright}{\isachardoublequoteclose}\isanewline
\isanewline
\isakeyword{fixes}\ foe{\isacharunderscore}control\ {\isacharcolon}{\isacharcolon}\ {\isachardoublequoteopen}{\isacharbrackleft}location{\isacharcomma}\ action{\isacharbrackright}\ {\isasymRightarrow}\ bool{\isachardoublequoteclose}\isanewline
\isakeyword{defines}\ foe{\isacharunderscore}control{\isacharunderscore}def{\isacharcolon}\ {\isachardoublequoteopen}foe{\isacharunderscore}control\ l\ c\ {\isasymequiv}\ \ \isanewline
\ \ \ {\isacharparenleft}{\isacharbang}\ I{\isacharcolon}{\isacharcolon}\ infrastructure{\isachardot}\ {\isacharparenleft}{\isacharquery}\ x\ {\isacharcolon}{\isacharcolon}\ identity{\isachardot}\ \isanewline
\ \ \ \ \ \ \ \ x\ {\isacharat}\isactrlbsub graphI\ I\isactrlesub \ l\ {\isasymand}\ Actor\ x\ {\isasymnoteq}\ Actor\ {\isacharprime}{\isacharprime}Eve{\isacharprime}{\isacharprime}{\isacharparenright}\isanewline
\ \ \ \ \ \ \ \ \ \ \ \ \ {\isasymlongrightarrow}\ {\isasymnot}{\isacharparenleft}enables\ I\ l\ {\isacharparenleft}Actor\ {\isacharprime}{\isacharprime}Eve{\isacharprime}{\isacharprime}{\isacharparenright}\ c{\isacharparenright}{\isacharparenright}{\isachardoublequoteclose}\isanewline
\isanewline
\isakeyword{fixes}\ astate{\isacharcolon}{\isacharcolon}\ {\isachardoublequoteopen}identity\ {\isasymRightarrow}\ actor{\isacharunderscore}state{\isachardoublequoteclose}\isanewline
\isakeyword{defines}\ astate{\isacharunderscore}def{\isacharcolon}\ {\isachardoublequoteopen}astate\ x\ {\isasymequiv}\ \ {\isacharparenleft}case\ x\ of\ \isanewline
\ \ \ \ \ \ \ \ \ \ \ {\isacharprime}{\isacharprime}Eve{\isacharprime}{\isacharprime}\ {\isasymRightarrow}\ Actor{\isacharunderscore}state\ depressed\ {\isacharbraceleft}revenge{\isacharcomma}\ peer{\isacharunderscore}recognition{\isacharbraceright}\isanewline
\ \ \ \ \ \ \ \ \ \ {\isacharbar}\ {\isacharunderscore}\ {\isasymRightarrow}\ Actor{\isacharunderscore}state\ happy\ {\isacharbraceleft}{\isacharbraceright}{\isacharparenright}{\isachardoublequoteclose}\isanewline
\isanewline
\isakeyword{assumes}\ Eve{\isacharunderscore}precipitating{\isacharunderscore}event{\isacharcolon}\ {\isachardoublequoteopen}tipping{\isacharunderscore}point\ {\isacharparenleft}astate\ {\isacharprime}{\isacharprime}Eve{\isacharprime}{\isacharprime}{\isacharparenright}{\isachardoublequoteclose}\isanewline
\isakeyword{assumes}\ Insider{\isacharunderscore}Eve{\isacharcolon}\ {\isachardoublequoteopen}Insider\ {\isacharprime}{\isacharprime}Eve{\isacharprime}{\isacharprime}\ {\isacharbraceleft}{\isacharprime}{\isacharprime}Charly{\isacharprime}{\isacharprime}{\isacharbraceright}\ astate{\isachardoublequoteclose}\isanewline
\isakeyword{assumes}\ cockpit{\isacharunderscore}foe{\isacharunderscore}control{\isacharcolon}\ {\isachardoublequoteopen}foe{\isacharunderscore}control\ cockpit\ put{\isachardoublequoteclose}\isanewline
\isanewline
\isakeyword{begin}\isanewline
\isanewline
\isacommand{lemma}\isamarkupfalse%
\ Safety{\isacharcolon}\ {\isachardoublequoteopen}Safety\ Airplane{\isacharunderscore}scenario\ {\isacharparenleft}{\isacharprime}{\isacharprime}Alice{\isacharprime}{\isacharprime}{\isacharparenright}{\isachardoublequoteclose}\isanewline
\prf\isanewline\isanewline
\isacommand{lemma}\isamarkupfalse%
\ Security{\isacharcolon}\ {\isachardoublequoteopen}Security\ Airplane{\isacharunderscore}scenario\ s{\isachardoublequoteclose}\isanewline
\prf\isanewline\isanewline
\isacommand{lemma}\isamarkupfalse%
\ step{\isadigit{0}}r{\isacharcolon}\ {\isachardoublequoteopen}Airplane{\isacharunderscore}scenario\ {\isasymrightarrow}\isactrlsub n{\isacharasterisk}\ Airplane{\isacharunderscore}getting{\isacharunderscore}in{\isacharunderscore}danger{\isadigit{0}}{\isachardoublequoteclose}\isanewline
\prf\isanewline\isanewline
\isacommand{lemma}\isamarkupfalse%
\ step{\isadigit{1}}r{\isacharcolon}\ {\isachardoublequoteopen}Airplane{\isacharunderscore}getting{\isacharunderscore}in{\isacharunderscore}danger{\isadigit{0}}\ {\isasymrightarrow}\isactrlsub n{\isacharasterisk}\ Airplane{\isacharunderscore}getting{\isacharunderscore}in{\isacharunderscore}danger{\isachardoublequoteclose}\isanewline
\prf\isanewline\isanewline
\isacommand{lemma}\isamarkupfalse%
\ step{\isadigit{2}}r{\isacharcolon}\ {\isachardoublequoteopen}Airplane{\isacharunderscore}getting{\isacharunderscore}in{\isacharunderscore}danger\ {\isasymrightarrow}\isactrlsub n{\isacharasterisk}\ Airplane{\isacharunderscore}in{\isacharunderscore}danger{\isachardoublequoteclose}\isanewline
\prf\isanewline\isanewline
\isacommand{theorem}\isamarkupfalse%
\ step{\isacharunderscore}allr{\isacharcolon}\ \ {\isachardoublequoteopen}Airplane{\isacharunderscore}scenario\ {\isasymrightarrow}\isactrlsub n{\isacharasterisk}\ Airplane{\isacharunderscore}in{\isacharunderscore}danger{\isachardoublequoteclose}\isanewline
\prf\isanewline\isanewline
\isacommand{theorem}\isamarkupfalse%
\ aid{\isacharunderscore}attack{\isacharcolon}\ {\isachardoublequoteopen}Air{\isacharunderscore}Kripke\ {\isasymturnstile}\ EF\ {\isacharparenleft}{\isacharbraceleft}x{\isachardot}\ {\isasymnot}\ global{\isacharunderscore}policy\ x\ {\isacharprime}{\isacharprime}Eve{\isacharprime}{\isacharprime}{\isacharbraceright}{\isacharparenright}{\isachardoublequoteclose}\isanewline
\isadelimproof
\endisadelimproof
\isatagproof
\isacommand{proof}\isamarkupfalse%
\ {\isacharparenleft}simp\ add{\isacharcolon}\ check{\isacharunderscore}def\ Air{\isacharunderscore}Kripke{\isacharunderscore}def{\isacharcomma}\ rule\ conjI{\isacharparenright}\isanewline
\ \ \isacommand{show}\isamarkupfalse%
\ {\isachardoublequoteopen}Airplane{\isacharunderscore}scenario\ {\isasymin}\ Air{\isacharunderscore}states{\isachardoublequoteclose}\ \isanewline
\ \ \ \ \isacommand{by}\isamarkupfalse%
\ {\isacharparenleft}simp\ add{\isacharcolon}\ Air{\isacharunderscore}states{\isacharunderscore}def\ state{\isacharunderscore}transition{\isacharunderscore}in{\isacharunderscore}refl{\isacharunderscore}def{\isacharparenright}\isanewline
\isacommand{next}\isamarkupfalse%
\ \isacommand{show}\isamarkupfalse%
\ {\isachardoublequoteopen}Airplane{\isacharunderscore}scenario\ {\isasymin}\ EF\ {\isacharbraceleft}x{\isacharcolon}{\isacharcolon}infrastructure{\isachardot}\ {\isasymnot}\ global{\isacharunderscore}policy\ x\ {\isacharprime}{\isacharprime}Eve{\isacharprime}{\isacharprime}{\isacharbraceright}{\isachardoublequoteclose}\isanewline
\ \ \isacommand{by}\isamarkupfalse%
\ {\isacharparenleft}rule\ EF{\isacharunderscore}lem{\isadigit{2}}b{\isacharcomma}\ subst\ EF{\isacharunderscore}lem{\isadigit{0}}{\isadigit{0}}{\isadigit{0}}{\isacharcomma}\ rule\ EX{\isacharunderscore}lem{\isadigit{0}}r{\isacharcomma}\ subst\ EF{\isacharunderscore}lem{\isadigit{0}}{\isadigit{0}}{\isadigit{0}}{\isacharcomma}\ rule\ EX{\isacharunderscore}step{\isacharcomma}\isanewline
\ \ \ \ \ unfold\ state{\isacharunderscore}transition{\isacharunderscore}infra{\isacharunderscore}def{\isacharcomma}\ rule\ step{\isadigit{0}}{\isacharcomma}\ rule\ EX{\isacharunderscore}lem{\isadigit{0}}r{\isacharcomma}\isanewline
\ \ \ \ \ rule{\isacharunderscore}tac\ y\ {\isacharequal}\ {\isachardoublequoteopen}Airplane{\isacharunderscore}getting{\isacharunderscore}in{\isacharunderscore}danger{\isachardoublequoteclose}\ \isakeyword{in}\ EX{\isacharunderscore}step{\isacharcomma}\isanewline
\ \ \ \ \ unfold\ state{\isacharunderscore}transition{\isacharunderscore}infra{\isacharunderscore}def{\isacharcomma}\ rule\ step{\isadigit{1}}{\isacharcomma}\ subst\ EF{\isacharunderscore}lem{\isadigit{0}}{\isadigit{0}}{\isadigit{0}}{\isacharcomma}\ rule\ EX{\isacharunderscore}lem{\isadigit{0}}l{\isacharcomma}\isanewline
\ \ \ \ \ rule{\isacharunderscore}tac\ y\ {\isacharequal}\ {\isachardoublequoteopen}Airplane{\isacharunderscore}in{\isacharunderscore}danger{\isachardoublequoteclose}\ \isakeyword{in}\ EX{\isacharunderscore}step{\isacharcomma}\ unfold\ state{\isacharunderscore}transition{\isacharunderscore}infra{\isacharunderscore}def{\isacharcomma}\isanewline
\ \ \ \ \ rule\ step{\isadigit{2}}{\isacharcomma}\ rule\ CollectI{\isacharcomma}\ rule\ ex{\isacharunderscore}inv{\isadigit{4}}{\isacharparenright}\isanewline
\isacommand{qed}\isamarkupfalse%
\endisatagproof
{\isafoldproof}%
\isadelimproof
\ \isanewline
\endisadelimproof
\ \ \ \ \isanewline
\isacommand{lemma}\isamarkupfalse%
\ \ actors{\isacharunderscore}unique{\isacharunderscore}loc{\isacharunderscore}base{\isacharcolon}\ \isanewline
\ \ \isakeyword{assumes}\ {\isachardoublequoteopen}I\ {\isasymrightarrow}\isactrlsub n\ I{\isacharprime}{\isachardoublequoteclose}\isanewline
\ \ \ \ \ \ \isakeyword{and}\ {\isachardoublequoteopen}{\isacharparenleft}{\isasymforall}\ l\ l{\isacharprime}{\isachardot}\ a\ {\isacharat}\isactrlbsub graphI\ I\isactrlesub \ l\ {\isasymand}\ a\ {\isacharat}\isactrlbsub graphI\ I\isactrlesub \ l{\isacharprime}\ {\isasymlongrightarrow}\ l\ {\isacharequal}\ l{\isacharprime}{\isacharparenright}{\isasymand}\isanewline
\ \ \ \ \ \ \ \ \ \ \ {\isacharparenleft}{\isasymforall}\ l{\isachardot}\ nodup\ a\ {\isacharparenleft}agra\ {\isacharparenleft}graphI\ I{\isacharparenright}\ l{\isacharparenright}{\isacharparenright}{\isachardoublequoteclose}\isanewline
\ \ \ \ \isakeyword{shows}\ {\isachardoublequoteopen}{\isacharparenleft}{\isasymforall}\ l\ l{\isacharprime}{\isachardot}\ a\ {\isacharat}\isactrlbsub graphI\ I{\isacharprime}\isactrlesub \ l\ {\isasymand}\ a\ {\isacharat}\isactrlbsub graphI\ I{\isacharprime}\isactrlesub \ l{\isacharprime}\ \ {\isasymlongrightarrow}\ l\ {\isacharequal}\ l{\isacharprime}{\isacharparenright}\ {\isasymand}\isanewline
\ \ \ \ \ \ \ \ \ \ \ {\isacharparenleft}{\isasymforall}\ l{\isachardot}\ nodup\ a\ {\isacharparenleft}agra\ {\isacharparenleft}graphI\ I{\isacharprime}{\isacharparenright}\ l{\isacharparenright}{\isacharparenright}{\isachardoublequoteclose}\isanewline
\prf\isanewline\isanewline
\isacommand{lemma}\isamarkupfalse%
\ actors{\isacharunderscore}unique{\isacharunderscore}loc{\isacharunderscore}step{\isacharcolon}\ \isanewline
\ \ \isakeyword{assumes}\ {\isachardoublequoteopen}{\isacharparenleft}I{\isacharcomma}\ I{\isacharprime}{\isacharparenright}\ {\isasymin}\ {\isacharbraceleft}{\isacharparenleft}x{\isacharcolon}{\isacharcolon}infrastructure{\isacharcomma}\ y{\isacharcolon}{\isacharcolon}infrastructure{\isacharparenright}{\isachardot}\ x\ {\isasymrightarrow}\isactrlsub n\ y{\isacharbraceright}\isactrlsup {\isacharasterisk}{\isachardoublequoteclose}\ \isanewline
\ \ \ \ \ \ \isakeyword{and}\ {\isachardoublequoteopen}{\isasymforall}\ a{\isachardot}\ {\isacharparenleft}{\isasymforall}\ l\ l{\isacharprime}{\isachardot}\ a\ {\isacharat}\isactrlbsub graphI\ I\isactrlesub \ l\ {\isasymand}\ a\ {\isacharat}\isactrlbsub graphI\ I\isactrlesub \ l{\isacharprime}\ {\isasymlongrightarrow}\ l\ {\isacharequal}\ l{\isacharprime}{\isacharparenright}{\isasymand}\isanewline
\ \ \ \ \ \ \ \ \ \ {\isacharparenleft}{\isasymforall}\ l{\isachardot}\ nodup\ a\ {\isacharparenleft}agra\ {\isacharparenleft}graphI\ I{\isacharparenright}\ l{\isacharparenright}{\isacharparenright}{\isachardoublequoteclose}\ \isanewline
\ \ \ \ \isakeyword{shows}\ \ \ {\isachardoublequoteopen}{\isasymforall}\ a{\isachardot}\ {\isacharparenleft}{\isasymforall}\ l\ l{\isacharprime}{\isachardot}\ a\ {\isacharat}\isactrlbsub graphI\ I{\isacharprime}\isactrlesub \ l\ {\isasymand}\ a\ {\isacharat}\isactrlbsub graphI\ I{\isacharprime}\isactrlesub \ l{\isacharprime}\ \ {\isasymlongrightarrow}\ l\ {\isacharequal}\ l{\isacharprime}{\isacharparenright}\ {\isasymand}\isanewline
\ \ \ \ \ \ \ \ \ \ {\isacharparenleft}{\isasymforall}\ l{\isachardot}\ nodup\ a\ {\isacharparenleft}agra\ {\isacharparenleft}graphI\ I{\isacharprime}{\isacharparenright}\ l{\isacharparenright}{\isacharparenright}{\isachardoublequoteclose}\isanewline
\prf\isanewline\isanewline
\isacommand{lemma}\isamarkupfalse%
\ two{\isacharunderscore}person{\isacharunderscore}inv{\isacharcolon}\ \isanewline
\ \ \isakeyword{fixes}\ z\ z{\isacharprime}\ \isanewline
\ \ \isakeyword{assumes}\ {\isachardoublequoteopen}{\isacharparenleft}{\isadigit{2}}{\isacharcolon}{\isacharcolon}nat{\isacharparenright}\ {\isasymle}\ length\ {\isacharparenleft}agra\ {\isacharparenleft}graphI\ z{\isacharparenright}\ cockpit{\isacharparenright}{\isachardoublequoteclose}\isanewline
\ \ \ \ \ \ \isakeyword{and}\ {\isachardoublequoteopen}nodes{\isacharparenleft}graphI\ z{\isacharparenright}\ {\isacharequal}\ nodes{\isacharparenleft}graphI\ Airplane{\isacharunderscore}not{\isacharunderscore}in{\isacharunderscore}danger{\isacharunderscore}init{\isacharparenright}{\isachardoublequoteclose}\isanewline
\ \ \ \ \ \ \isakeyword{and}\ {\isachardoublequoteopen}delta{\isacharparenleft}z{\isacharparenright}\ {\isacharequal}\ delta{\isacharparenleft}Airplane{\isacharunderscore}not{\isacharunderscore}in{\isacharunderscore}danger{\isacharunderscore}init{\isacharparenright}{\isachardoublequoteclose}\isanewline
\ \ \ \ \ \ \isakeyword{and}\ {\isachardoublequoteopen}{\isacharparenleft}Airplane{\isacharunderscore}not{\isacharunderscore}in{\isacharunderscore}danger{\isacharunderscore}init{\isacharcomma}z{\isacharparenright}\ {\isasymin}\ {\isacharbraceleft}{\isacharparenleft}x{\isacharcolon}{\isacharcolon}infrastructure{\isacharcomma}\ y{\isacharcolon}{\isacharcolon}infrastructure{\isacharparenright}{\isachardot}\ x\ {\isasymrightarrow}\isactrlsub n\ y{\isacharbraceright}\isactrlsup {\isacharasterisk}{\isachardoublequoteclose}\isanewline
\ \ \ \ \ \ \isakeyword{and}\ {\isachardoublequoteopen}z\ {\isasymrightarrow}\isactrlsub n\ z{\isacharprime}{\isachardoublequoteclose}\isanewline
\ \ \ \ \isakeyword{shows}\ {\isachardoublequoteopen}{\isacharparenleft}{\isadigit{2}}{\isacharcolon}{\isacharcolon}nat{\isacharparenright}\ {\isasymle}\ length\ {\isacharparenleft}agra\ {\isacharparenleft}graphI\ z{\isacharprime}{\isacharparenright}\ cockpit{\isacharparenright}{\isachardoublequoteclose}\isanewline
\prf\isanewline\isanewline
\isacommand{lemma}\isamarkupfalse%
\ airplane{\isacharunderscore}actors{\isacharunderscore}inv{\isacharcolon}\ \isanewline
\ \ \isakeyword{assumes}\ {\isachardoublequoteopen}{\isacharparenleft}Airplane{\isacharunderscore}not{\isacharunderscore}in{\isacharunderscore}danger{\isacharunderscore}init{\isacharcomma}z{\isacharparenright}\ {\isasymin}\ {\isacharbraceleft}{\isacharparenleft}x{\isacharcolon}{\isacharcolon}infrastructure{\isacharcomma}\ y{\isacharcolon}{\isacharcolon}infrastructure{\isacharparenright}{\isachardot}\ x\ {\isasymrightarrow}\isactrlsub n\ y{\isacharbraceright}\isactrlsup {\isacharasterisk}{\isachardoublequoteclose}\ \isanewline
\ \ \ \ \isakeyword{shows}\ {\isachardoublequoteopen}{\isasymforall}h{\isacharcolon}{\isacharcolon}char\ list{\isasymin}set\ {\isacharparenleft}agra\ {\isacharparenleft}graphI\ z{\isacharparenright}\ cockpit{\isacharparenright}{\isachardot}\ h\ {\isasymin}\ airplane{\isacharunderscore}actors{\isachardoublequoteclose}\ \ \ \ \ \isanewline
\prf\isanewline\isanewline
\isacommand{lemma}\isamarkupfalse%
\ Eve{\isacharunderscore}not{\isacharunderscore}in{\isacharunderscore}cockpit{\isacharcolon}\ {\isachardoublequoteopen}{\isacharparenleft}Airplane{\isacharunderscore}not{\isacharunderscore}in{\isacharunderscore}danger{\isacharunderscore}init{\isacharcomma}\ I{\isacharparenright}\isanewline
\ \ \ \ \ \ \ {\isasymin}\ {\isacharbraceleft}{\isacharparenleft}x{\isacharcolon}{\isacharcolon}infrastructure{\isacharcomma}\ y{\isacharcolon}{\isacharcolon}infrastructure{\isacharparenright}{\isachardot}\ x\ {\isasymrightarrow}\isactrlsub n\ y{\isacharbraceright}\isactrlsup {\isacharasterisk}\ {\isasymLongrightarrow}\isanewline
\ \ \ \ \ \ \ x\ {\isasymin}\ set\ {\isacharparenleft}agra\ {\isacharparenleft}graphI\ I{\isacharparenright}\ cockpit{\isacharparenright}\ {\isasymLongrightarrow}\ x\ {\isasymnoteq}\ {\isacharprime}{\isacharprime}Eve{\isacharprime}{\isacharprime}{\isachardoublequoteclose}\isanewline
\prf\isanewline\isanewline
\isacommand{lemma}\isamarkupfalse%
\ tp{\isacharunderscore}imp{\isacharunderscore}control{\isacharcolon}\isanewline
\ \ \isakeyword{assumes}\ {\isachardoublequoteopen}{\isacharparenleft}Airplane{\isacharunderscore}not{\isacharunderscore}in{\isacharunderscore}danger{\isacharunderscore}init{\isacharcomma}I{\isacharparenright}\ {\isasymin}\ {\isacharbraceleft}{\isacharparenleft}x{\isacharcolon}{\isacharcolon}infrastructure{\isacharcomma}\ y{\isacharcolon}{\isacharcolon}infrastructure{\isacharparenright}{\isachardot}\ x\ {\isasymrightarrow}\isactrlsub n\ y{\isacharbraceright}\isactrlsup {\isacharasterisk}{\isachardoublequoteclose}\isanewline
\ \ \isakeyword{shows}\ {\isachardoublequoteopen}{\isacharparenleft}{\isacharquery}\ x\ {\isacharcolon}{\isacharcolon}\ identity{\isachardot}\ \ x\ {\isacharat}\isactrlbsub graphI\ I\isactrlesub \ cockpit\ {\isasymand}\ Actor\ x\ {\isasymnoteq}\ Actor\ {\isacharprime}{\isacharprime}Eve{\isacharprime}{\isacharprime}{\isacharparenright}{\isachardoublequoteclose}\isanewline
\prf\isanewline\isanewline
\isacommand{lemma}\isamarkupfalse%
\ Fend{\isacharunderscore}{\isadigit{2}}{\isacharcolon}\ \ \ \ {\isachardoublequoteopen}{\isacharparenleft}Airplane{\isacharunderscore}not{\isacharunderscore}in{\isacharunderscore}danger{\isacharunderscore}init{\isacharcomma}I{\isacharparenright}\ {\isasymin}\ {\isacharbraceleft}{\isacharparenleft}x{\isacharcolon}{\isacharcolon}infrastructure{\isacharcomma}\ y{\isacharcolon}{\isacharcolon}infrastructure{\isacharparenright}{\isachardot}\ x\ {\isasymrightarrow}\isactrlsub n\ y{\isacharbraceright}\isactrlsup {\isacharasterisk}\ {\isasymLongrightarrow}\isanewline
\ \ \ \ \ \ \ \ \ {\isasymnot}\ enables\ I\ cockpit\ {\isacharparenleft}Actor\ {\isacharprime}{\isacharprime}Eve{\isacharprime}{\isacharprime}{\isacharparenright}\ put{\isachardoublequoteclose}\isanewline
\isadelimproof
\ \ %
\endisadelimproof
\isatagproof
\isacommand{by}\isamarkupfalse%
\ {\isacharparenleft}insert\ cockpit{\isacharunderscore}foe{\isacharunderscore}control{\isacharcomma}\ simp\ add{\isacharcolon}\ foe{\isacharunderscore}control{\isacharunderscore}def{\isacharcomma}\ drule{\isacharunderscore}tac\ x\ {\isacharequal}\ I\ \isakeyword{in}\ spec{\isacharcomma}\isanewline
\ \ \ \ \ \ erule\ mp{\isacharcomma}\ erule\ tp{\isacharunderscore}imp{\isacharunderscore}control{\isacharparenright}%
\endisatagproof
{\isafoldproof}%
\isadelimproof
\isanewline
\endisadelimproof
\isanewline
\isacommand{theorem}\isamarkupfalse%
\ Four{\isacharunderscore}eyes{\isacharunderscore}no{\isacharunderscore}danger{\isacharcolon}\ {\isachardoublequoteopen}Air{\isacharunderscore}tp{\isacharunderscore}Kripke\ {\isasymturnstile}\ AG\ {\isacharparenleft}{\isacharbraceleft}x{\isachardot}\ global{\isacharunderscore}policy\ x\ {\isacharprime}{\isacharprime}Eve{\isacharprime}{\isacharprime}{\isacharbraceright}{\isacharparenright}{\isachardoublequoteclose}\isanewline
\isadelimproof
\endisadelimproof
\isatagproof
\isacommand{proof}\isamarkupfalse%
\ {\isacharparenleft}simp\ add{\isacharcolon}\ Air{\isacharunderscore}tp{\isacharunderscore}Kripke{\isacharunderscore}def\ check{\isacharunderscore}def{\isacharcomma}\ rule\ conjI{\isacharparenright}\isanewline
\ \ \isacommand{show}\isamarkupfalse%
\ {\isachardoublequoteopen}Airplane{\isacharunderscore}not{\isacharunderscore}in{\isacharunderscore}danger{\isacharunderscore}init\ {\isasymin}\ Air{\isacharunderscore}tp{\isacharunderscore}states{\isachardoublequoteclose}\isanewline
\ \ \ \ \isacommand{by}\isamarkupfalse%
\ {\isacharparenleft}simp\ add{\isacharcolon}\ Airplane{\isacharunderscore}not{\isacharunderscore}in{\isacharunderscore}danger{\isacharunderscore}init{\isacharunderscore}def\ Air{\isacharunderscore}tp{\isacharunderscore}states{\isacharunderscore}def\isanewline
\ \ \ \ \ \ \ \ \ \ \ \ \ \ \ \ \ \ \ \ state{\isacharunderscore}transition{\isacharunderscore}in{\isacharunderscore}refl{\isacharunderscore}def{\isacharparenright}\isanewline
\isacommand{next}\isamarkupfalse%
\ \isacommand{show}\isamarkupfalse%
\ {\isachardoublequoteopen}Airplane{\isacharunderscore}not{\isacharunderscore}in{\isacharunderscore}danger{\isacharunderscore}init\ {\isasymin}\ AG\ {\isacharbraceleft}x{\isacharcolon}{\isacharcolon}infrastructure{\isachardot}\ global{\isacharunderscore}policy\ x\ {\isacharprime}{\isacharprime}Eve{\isacharprime}{\isacharprime}{\isacharbraceright}{\isachardoublequoteclose}\isanewline
\ \ \isacommand{proof}\isamarkupfalse%
\ {\isacharparenleft}unfold\ AG{\isacharunderscore}def{\isacharcomma}\ simp\ add{\isacharcolon}\ gfp{\isacharunderscore}def{\isacharcomma}\isanewline
\ \ \ rule{\isacharunderscore}tac\ x\ {\isacharequal}\ {\isachardoublequoteopen}{\isacharbraceleft}{\isacharparenleft}x\ {\isacharcolon}{\isacharcolon}\ infrastructure{\isacharparenright}\ {\isasymin}\ states\ Air{\isacharunderscore}tp{\isacharunderscore}Kripke{\isachardot}\ {\isachartilde}{\isacharparenleft}{\isacharprime}{\isacharprime}Eve{\isacharprime}{\isacharprime}\ {\isacharat}\isactrlbsub graphI\ x\isactrlesub \ cockpit{\isacharparenright}{\isacharbraceright}{\isachardoublequoteclose}\ \isakeyword{in}\ exI{\isacharcomma}\isanewline
\ \ \ rule\ conjI{\isacharparenright}\isanewline
\ \ \ \ \isacommand{show}\isamarkupfalse%
\ {\isachardoublequoteopen}{\isacharbraceleft}x{\isacharcolon}{\isacharcolon}infrastructure\ {\isasymin}\ states\ Air{\isacharunderscore}tp{\isacharunderscore}Kripke{\isachardot}\ {\isasymnot}\ {\isacharprime}{\isacharprime}Eve{\isacharprime}{\isacharprime}\ {\isacharat}\isactrlbsub graphI\ x\isactrlesub \ cockpit{\isacharbraceright}\isanewline
\ \ \ \ {\isasymsubseteq}\ {\isacharbraceleft}x{\isacharcolon}{\isacharcolon}infrastructure{\isachardot}\ global{\isacharunderscore}policy\ x\ {\isacharprime}{\isacharprime}Eve{\isacharprime}{\isacharprime}{\isacharbraceright}{\isachardoublequoteclose}\isanewline
\ \ \ \ \ \isacommand{by}\isamarkupfalse%
\ {\isacharparenleft}unfold\ global{\isacharunderscore}policy{\isacharunderscore}def{\isacharcomma}\ simp\ add{\isacharcolon}\ airplane{\isacharunderscore}actors{\isacharunderscore}def{\isacharcomma}\ rule\ subsetI{\isacharcomma}\isanewline
\ \ \ \ \ \ \ \ \ drule\ CollectD{\isacharcomma}\ rule\ CollectI{\isacharcomma}\ erule\ conjE{\isacharcomma}\isanewline
\ \ \ \ \ \ \ \ \ simp\ add{\isacharcolon}\ Air{\isacharunderscore}tp{\isacharunderscore}Kripke{\isacharunderscore}def\ Air{\isacharunderscore}tp{\isacharunderscore}states{\isacharunderscore}def\ state{\isacharunderscore}transition{\isacharunderscore}in{\isacharunderscore}refl{\isacharunderscore}def{\isacharcomma}\isanewline
\ \ \ \ \ \ \ \ \ erule\ Fend{\isacharunderscore}{\isadigit{2}}{\isacharparenright}\isanewline
\ \isacommand{next}\isamarkupfalse%
\ \isacommand{show}\isamarkupfalse%
\ {\isachardoublequoteopen}{\isacharbraceleft}x{\isacharcolon}{\isacharcolon}infrastructure\ {\isasymin}\ states\ Air{\isacharunderscore}tp{\isacharunderscore}Kripke{\isachardot}\ {\isasymnot}\ {\isacharprime}{\isacharprime}Eve{\isacharprime}{\isacharprime}\ {\isacharat}\isactrlbsub graphI\ x\isactrlesub \ cockpit{\isacharbraceright}\isanewline
\ \ \ \ {\isasymsubseteq}\ AX\ {\isacharbraceleft}x{\isacharcolon}{\isacharcolon}infrastructure\ {\isasymin}\ states\ Air{\isacharunderscore}tp{\isacharunderscore}Kripke{\isachardot}\ {\isasymnot}\ {\isacharprime}{\isacharprime}Eve{\isacharprime}{\isacharprime}\ {\isacharat}\isactrlbsub graphI\ x\isactrlesub \ cockpit{\isacharbraceright}\ {\isasymand}\isanewline
\ \ \ \ Airplane{\isacharunderscore}not{\isacharunderscore}in{\isacharunderscore}danger{\isacharunderscore}init\isanewline
\ \ \ \ {\isasymin}\ {\isacharbraceleft}x{\isacharcolon}{\isacharcolon}infrastructure\ {\isasymin}\ states\ Air{\isacharunderscore}tp{\isacharunderscore}Kripke{\isachardot}\ {\isasymnot}\ {\isacharprime}{\isacharprime}Eve{\isacharprime}{\isacharprime}\ {\isacharat}\isactrlbsub graphI\ x\isactrlesub \ cockpit{\isacharbraceright}{\isachardoublequoteclose}\isanewline
\ \ \ \isacommand{proof}\isamarkupfalse%
\isanewline
\ \ \ \ \ \isacommand{show}\isamarkupfalse%
\ {\isachardoublequoteopen}Airplane{\isacharunderscore}not{\isacharunderscore}in{\isacharunderscore}danger{\isacharunderscore}init\isanewline
\ \ \ \ \ \ \ \ \ \ {\isasymin}\ {\isacharbraceleft}x{\isacharcolon}{\isacharcolon}infrastructure\ {\isasymin}\ states\ Air{\isacharunderscore}tp{\isacharunderscore}Kripke{\isachardot}\ {\isasymnot}\ {\isacharprime}{\isacharprime}Eve{\isacharprime}{\isacharprime}\ {\isacharat}\isactrlbsub graphI\ x\isactrlesub \ cockpit{\isacharbraceright}{\isachardoublequoteclose}\isanewline
\ \ \ \ \ \ \isacommand{by}\isamarkupfalse%
\ {\isacharparenleft}simp\ add{\isacharcolon}\ Airplane{\isacharunderscore}not{\isacharunderscore}in{\isacharunderscore}danger{\isacharunderscore}init{\isacharunderscore}def\ Air{\isacharunderscore}tp{\isacharunderscore}Kripke{\isacharunderscore}def\ Air{\isacharunderscore}tp{\isacharunderscore}states{\isacharunderscore}def\isanewline
\ \ \ \ \ \ \ \ \ \ \ \ \ \ \ \ \ \ \ \ state{\isacharunderscore}transition{\isacharunderscore}refl{\isacharunderscore}def\ ex{\isacharunderscore}graph{\isacharunderscore}def\ atI{\isacharunderscore}def\ Air{\isacharunderscore}tp{\isacharunderscore}Kripke{\isacharunderscore}def\isanewline
\ \ \ \ \ \ \ \ \ \ \ \ \ \ \ \ \ \ \ \ state{\isacharunderscore}transition{\isacharunderscore}in{\isacharunderscore}refl{\isacharunderscore}def{\isacharparenright}\isanewline
\ \ \isacommand{next}\isamarkupfalse%
\ \isacommand{show}\isamarkupfalse%
\ {\isachardoublequoteopen}{\isacharbraceleft}x{\isacharcolon}{\isacharcolon}infrastructure\ {\isasymin}\ states\ Air{\isacharunderscore}tp{\isacharunderscore}Kripke{\isachardot}\ {\isasymnot}\ {\isacharprime}{\isacharprime}Eve{\isacharprime}{\isacharprime}\ {\isacharat}\isactrlbsub graphI\ x\isactrlesub \ cockpit{\isacharbraceright}\isanewline
\ \ \ \ {\isasymsubseteq}\ AX\ {\isacharbraceleft}x{\isacharcolon}{\isacharcolon}infrastructure\ {\isasymin}\ states\ Air{\isacharunderscore}tp{\isacharunderscore}Kripke{\isachardot}\ {\isasymnot}\ {\isacharprime}{\isacharprime}Eve{\isacharprime}{\isacharprime}\ {\isacharat}\isactrlbsub graphI\ x\isactrlesub \ cockpit{\isacharbraceright}{\isachardoublequoteclose}\isanewline
\ \ \ \ \isacommand{proof}\isamarkupfalse%
\ {\isacharparenleft}rule\ subsetI{\isacharcomma}\ simp\ add{\isacharcolon}\ AX{\isacharunderscore}def{\isacharcomma}\ rule\ subsetI{\isacharcomma}\ rule\ CollectI{\isacharcomma}\ rule\ conjI{\isacharparenright}\isanewline
\ \ \ \ \ \ \isacommand{show}\isamarkupfalse%
\ {\isachardoublequoteopen}{\isasymAnd}{\isacharparenleft}x{\isacharcolon}{\isacharcolon}infrastructure{\isacharparenright}\ xa{\isacharcolon}{\isacharcolon}infrastructure{\isachardot}\isanewline
\ \ \ \ \ \ \ x\ {\isasymin}\ states\ Air{\isacharunderscore}tp{\isacharunderscore}Kripke\ {\isasymand}\ {\isasymnot}\ {\isacharprime}{\isacharprime}Eve{\isacharprime}{\isacharprime}\ {\isacharat}\isactrlbsub graphI\ x\isactrlesub \ cockpit\ {\isasymLongrightarrow}\isanewline
\ \ \ \ \ \ \ xa\ {\isasymin}\ Collect\ {\isacharparenleft}state{\isacharunderscore}transition\ x{\isacharparenright}\ {\isasymLongrightarrow}\ xa\ {\isasymin}\ states\ Air{\isacharunderscore}tp{\isacharunderscore}Kripke{\isachardoublequoteclose}\isanewline
\ \ \ \ \ \ \ \ \isacommand{by}\isamarkupfalse%
\ {\isacharparenleft}simp\ add{\isacharcolon}\ \ Air{\isacharunderscore}tp{\isacharunderscore}Kripke{\isacharunderscore}def\ Air{\isacharunderscore}tp{\isacharunderscore}states{\isacharunderscore}def\ state{\isacharunderscore}transition{\isacharunderscore}in{\isacharunderscore}refl{\isacharunderscore}def{\isacharcomma}\isanewline
\ \ \ \ \ \ \ \ \ \ \ \ simp\ add{\isacharcolon}\ atI{\isacharunderscore}def{\isacharcomma}\ erule\ conjE{\isacharcomma}\isanewline
\ \ \ \ \ \ \ \ \ \ \ \ unfold\ state{\isacharunderscore}transition{\isacharunderscore}infra{\isacharunderscore}def\ state{\isacharunderscore}transition{\isacharunderscore}in{\isacharunderscore}refl{\isacharunderscore}def{\isacharcomma}\isanewline
\ \ \ \ \ \ \ \ \ \ \ \ erule\ rtrancl{\isacharunderscore}into{\isacharunderscore}rtrancl{\isacharcomma}\ rule\ CollectI{\isacharcomma}\ simp{\isacharparenright}\isanewline
\ \ \ \ \isacommand{next}\isamarkupfalse%
\ \isacommand{fix}\isamarkupfalse%
\ x\ xa\isanewline
\ \ \ \ \ \ \ \ \isacommand{assume}\isamarkupfalse%
\ a{\isadigit{0}}{\isacharcolon}\ {\isachardoublequoteopen}x\ {\isasymin}\ states\ Air{\isacharunderscore}tp{\isacharunderscore}Kripke\ {\isasymand}\ {\isasymnot}\ {\isacharprime}{\isacharprime}Eve{\isacharprime}{\isacharprime}\ {\isacharat}\isactrlbsub graphI\ x\isactrlesub \ cockpit{\isachardoublequoteclose}\isanewline
\ \ \ \ \ \ \ \ \ \isakeyword{and}\ a{\isadigit{1}}{\isacharcolon}\ {\isachardoublequoteopen}xa\ {\isasymin}\ Collect\ {\isacharparenleft}state{\isacharunderscore}transition\ x{\isacharparenright}{\isachardoublequoteclose}\isanewline
\ \ \ \ \ \ \ \ \isacommand{show}\isamarkupfalse%
\ {\isachardoublequoteopen}{\isasymnot}\ {\isacharprime}{\isacharprime}Eve{\isacharprime}{\isacharprime}\ {\isacharat}\isactrlbsub graphI\ xa\isactrlesub \ cockpit{\isachardoublequoteclose}\isanewline
\ \ \ \ \ \ \isacommand{proof}\isamarkupfalse%
\ {\isacharminus}\isanewline
\ \ \ \ \ \ \ \ \isacommand{have}\isamarkupfalse%
\ b{\isacharcolon}\ {\isachardoublequoteopen}{\isacharparenleft}Airplane{\isacharunderscore}not{\isacharunderscore}in{\isacharunderscore}danger{\isacharunderscore}init{\isacharcomma}\ xa{\isacharparenright}\isanewline
\ \ \ \ \ \ \ {\isasymin}\ {\isacharbraceleft}{\isacharparenleft}x{\isacharcolon}{\isacharcolon}infrastructure{\isacharcomma}\ y{\isacharcolon}{\isacharcolon}infrastructure{\isacharparenright}{\isachardot}\ x\ {\isasymrightarrow}\isactrlsub n\ y{\isacharbraceright}\isactrlsup {\isacharasterisk}{\isachardoublequoteclose}\isanewline
\ \ \ \ \ \ \ \ \isacommand{proof}\isamarkupfalse%
\ {\isacharparenleft}insert\ a{\isadigit{0}}\ a{\isadigit{1}}{\isacharcomma}\ rule\ rtrancl{\isacharunderscore}trans{\isacharparenright}\isanewline
\ \ \ \ \ \ \ \ \ \ \isacommand{show}\isamarkupfalse%
\ {\isachardoublequoteopen}x\ {\isasymin}\ states\ Air{\isacharunderscore}tp{\isacharunderscore}Kripke\ {\isasymand}\ {\isasymnot}\ {\isacharprime}{\isacharprime}Eve{\isacharprime}{\isacharprime}\ {\isacharat}\isactrlbsub graphI\ x\isactrlesub \ cockpit\ {\isasymLongrightarrow}\isanewline
\ \ \ \ \ \ \ \ \ \ \ \ \ \ \ \ xa\ {\isasymin}\ Collect\ {\isacharparenleft}state{\isacharunderscore}transition\ x{\isacharparenright}\ {\isasymLongrightarrow}\isanewline
\ \ \ \ \ \ \ \ \ \ \ \ \ \ \ \ {\isacharparenleft}x{\isacharcomma}\ xa{\isacharparenright}\ {\isasymin}\ {\isacharbraceleft}{\isacharparenleft}x{\isacharcolon}{\isacharcolon}infrastructure{\isacharcomma}\ y{\isacharcolon}{\isacharcolon}infrastructure{\isacharparenright}{\isachardot}\ x\ {\isasymrightarrow}\isactrlsub n\ y{\isacharbraceright}\isactrlsup {\isacharasterisk}{\isachardoublequoteclose}\isanewline
\ \ \ \ \ \ \ \ \ \ \ \ \isacommand{by}\isamarkupfalse%
\ {\isacharparenleft}unfold\ state{\isacharunderscore}transition{\isacharunderscore}infra{\isacharunderscore}def{\isacharcomma}\ force{\isacharparenright}\isanewline
\ \ \ \ \ \ \ \ \isacommand{next}\isamarkupfalse%
\ \isacommand{show}\isamarkupfalse%
\ {\isachardoublequoteopen}x\ {\isasymin}\ states\ Air{\isacharunderscore}tp{\isacharunderscore}Kripke\ {\isasymand}\ {\isasymnot}\ {\isacharprime}{\isacharprime}Eve{\isacharprime}{\isacharprime}\ {\isacharat}\isactrlbsub graphI\ x\isactrlesub \ cockpit\ {\isasymLongrightarrow}\isanewline
\ \ \ \ \ \ \ \ \ \ \ \ \ \ \ \ \ \ xa\ {\isasymin}\ Collect\ {\isacharparenleft}state{\isacharunderscore}transition\ x{\isacharparenright}\ {\isasymLongrightarrow}\isanewline
\ \ \ \ \ \ \ \ \ \ \ \ \ \ \ \ \ \ {\isacharparenleft}Airplane{\isacharunderscore}not{\isacharunderscore}in{\isacharunderscore}danger{\isacharunderscore}init{\isacharcomma}\ x{\isacharparenright}\ {\isasymin}\ {\isacharbraceleft}{\isacharparenleft}x{\isacharcolon}{\isacharcolon}infrastructure{\isacharcomma}\ y{\isacharcolon}{\isacharcolon}infrastructure{\isacharparenright}{\isachardot}\ x\ {\isasymrightarrow}\isactrlsub n\ y{\isacharbraceright}\isactrlsup {\isacharasterisk}{\isachardoublequoteclose}\isanewline
\ \ \ \ \ \ \ \ \ \ \ \ \isacommand{by}\isamarkupfalse%
\ {\isacharparenleft}erule\ conjE{\isacharcomma}\ simp\ add{\isacharcolon}\ Air{\isacharunderscore}tp{\isacharunderscore}Kripke{\isacharunderscore}def\ Air{\isacharunderscore}tp{\isacharunderscore}states{\isacharunderscore}def\ state{\isacharunderscore}transition{\isacharunderscore}in{\isacharunderscore}refl{\isacharunderscore}def{\isacharparenright}{\isacharplus}\isanewline
\ \ \ \ \ \ \ \ \isacommand{qed}\isamarkupfalse%
\isanewline
\ \ \ \ \ \ \ \ \isacommand{show}\isamarkupfalse%
\ {\isacharquery}thesis\ \isanewline
\ \ \ \ \ \ \ \ \ \isacommand{by}\isamarkupfalse%
\ {\isacharparenleft}insert\ a{\isadigit{0}}\ a{\isadigit{1}}\ b{\isacharcomma}\ rule{\isacharunderscore}tac\ P\ {\isacharequal}\ {\isachardoublequoteopen}{\isacharprime}{\isacharprime}Eve{\isacharprime}{\isacharprime}\ {\isacharat}\isactrlbsub graphI\ xa\isactrlesub \ cockpit{\isachardoublequoteclose}\ \isakeyword{in}\ notI{\isacharcomma}\ \isanewline
\ \ \ \ \ \ \ \ \ \ \ \ simp\ add{\isacharcolon}\ atI{\isacharunderscore}def{\isacharcomma}\ drule\ Eve{\isacharunderscore}not{\isacharunderscore}in{\isacharunderscore}cockpit{\isacharcomma}\ assumption{\isacharcomma}\ simp{\isacharparenright}\isanewline
\ \ \ \ \ \isacommand{qed}\isamarkupfalse%
\ \ \ \ \ \ \ \isanewline
\ \ \ \isacommand{qed}\isamarkupfalse%
\isanewline
\ \isacommand{qed}\isamarkupfalse%
\isanewline
\isacommand{qed}\isamarkupfalse%
\isanewline
\isacommand{qed}\isamarkupfalse%
\endisatagproof
{\isafoldproof}%
\isadelimproof
\isanewline
\endisadelimproof
\isanewline
\isacommand{end}\isamarkupfalse%
\isanewline
\isanewline
\isacommand{interpretation}\isamarkupfalse%
\ airplane\ airplane{\isacharunderscore}actors\ airplane{\isacharunderscore}locations\ cockpit\ door\ cabin\ global{\isacharunderscore}policy\ \isanewline
\ \ \ \ \ \ \ \ \ \ \ \ \ \ \ ex{\isacharunderscore}creds\ ex{\isacharunderscore}locs\ ex{\isacharunderscore}locs{\isacharprime}\ ex{\isacharunderscore}graph\ aid{\isacharunderscore}graph\ aid{\isacharunderscore}graph{\isadigit{0}}\ agid{\isacharunderscore}graph\ \isanewline
\ \ \ \ \ \ \ \ \ \ \ \ \ \ \ local{\isacharunderscore}policies\ local{\isacharunderscore}policies{\isacharunderscore}four{\isacharunderscore}eyes\ Airplane{\isacharunderscore}scenario\ Airplane{\isacharunderscore}in{\isacharunderscore}danger\isanewline
\ \ \ \ \ \ \ \ \ \ \ \ \ \ \ Airplane{\isacharunderscore}getting{\isacharunderscore}in{\isacharunderscore}danger{\isadigit{0}}\ Airplane{\isacharunderscore}getting{\isacharunderscore}in{\isacharunderscore}danger\ Air{\isacharunderscore}states\ Air{\isacharunderscore}Kripke\isanewline
\ \ \ \ \ \ \ \ \ \ \ \ \ \ \ Airplane{\isacharunderscore}not{\isacharunderscore}in{\isacharunderscore}danger\ Airplane{\isacharunderscore}not{\isacharunderscore}in{\isacharunderscore}danger{\isacharunderscore}init\ Air{\isacharunderscore}tp{\isacharunderscore}states\ \isanewline
\ \ \ \ \ \ \ \ \ \ \ \ \ \ \ Air{\isacharunderscore}tp{\isacharunderscore}Kripke\ Safety\ Security\ foe{\isacharunderscore}control\ astate\isanewline
\prf\isanewline
\isadelimtheory
\isanewline
\endisadelimtheory
\isatagtheory
\isacommand{end}\isamarkupfalse%
\endisatagtheory
{\isafoldtheory}%
\isadelimtheory
\endisadelimtheory
\end{isabellebody}%

\end{document}